\documentclass[11pt]{article}
\usepackage{jheppub}
\usepackage{tensor}

\usepackage{mathtools}
\usepackage{amsmath, amssymb} 
\usepackage{mathrsfs} 			
\usepackage{bm}				
\usepackage{color}
\usepackage{ytableau}

\usepackage{changes}
\definechangesauthor[name=Dan]{Dan}

\usepackage{hyperref}
\hypersetup{
	unicode,	
	colorlinks,
	citecolor=[rgb]{0,.3,.5}, 
	linkcolor=[rgb]{0,.3,.5},
	urlcolor=[rgb]{0,.3,.5}, 
	bookmarksopen=true,
	bookmarksopenlevel=\maxdimen,
}

\newcommand{\cA}{\mathcal{A}}
\newcommand{\cD}{\mathcal{D}}

\newcommand{\cG}{\mathcal{G}}

\newcommand{\cL}{\mathcal{L}}

\newcommand{\cV}{\mathcal{V}}
\newcommand{\cW}{\mathcal{W}}

\newcommand{\pa}{\partial}

\newcommand{\ha}{{\hat a}}
\newcommand{\hb}{{\hat b}}
\newcommand{\hc}{{\hat c}}
\newcommand{\hd}{{\hat d}}
\newcommand{\he}{{\hat e}}
\renewcommand{\hm}{{\hat m}}
\newcommand{\hn}{{\hat n}}
\newcommand{\hp}{{\hat p}}
\newcommand{\hq}{{\hat q}}

\newcommand{\hA}{{\hat A}}
\newcommand{\hB}{{\hat B}}
\newcommand{\hC}{{\hat C}}
\newcommand{\hD}{{\hat D}}
\newcommand{\hE}{{\hat E}}
\newcommand{\hF}{{\hat F}}
\newcommand{\hG}{{\hat G}}
\newcommand{\hH}{{\hat H}}
\newcommand{\hM}{{\hat M}}
\newcommand{\hN}{{\hat N}}
\newcommand{\hP}{{\hat P}}
\newcommand{\hR}{{\hat R}}
\newcommand{\hT}{\hat{T}}

\newcommand{\halpha}{{\hat\alpha}}
\newcommand{\hbeta}{{\hat\beta}}
\newcommand{\hgamma}{{\hat\gamma}}
\newcommand{\hdelta}{{\hat\delta}}

\newcommand{\hOmega}{\hat{\Omega}}

\newcommand{\eps}{\epsilon}

\newcommand{\eol}{\nonumber \\}
\newcommand{\rd}{{\rm d}}
\newcommand{\rT}{{\rm T}}

\newcommand{\ul}{\underline}

\newcommand{\um}{{\underline m}}
\newcommand{\un}{{\underline n}}
\newcommand{\up}{{\underline p}}
\newcommand{\uq}{{\underline q}}
\newcommand{\ur}{{\underline r}}
\newcommand{\us}{{\underline s}}

\newcommand{\dalpha}{{\dot{\alpha}}}
\newcommand{\dbeta}{{\dot{\beta}}}
\newcommand{\dgamma}{{\dot{\gamma}}}
\newcommand{\ddelta}{{\dot{\delta}}}

\newcommand{\veps}{\varepsilon}
\newcommand{\rep}[1]{\mathbf{#1}}

\numberwithin{equation}{section}

\title{Linearized Off-shell 4+7 Supergeometry of 11D Supergravity}

\author{Katrin Becker,}
\author{Daniel Butter,}
\author{and Anindya Sengupta}
\affiliation{
George P. and Cynthia Woods
Mitchell Institute for
Fundamental Physics and Astronomy, \\
Texas A\&{}M University.\\
College Station, TX 77843, USA}

\emailAdd{kbecker@physics.tamu.edu}
\emailAdd{dbutter@tamu.edu}
\emailAdd{anindya.sengupta@tamu.edu}

\preprint{MI-HET-784}

\arxivnumber{XXXX.XXXXX}

\abstract{
We describe the linearized supergeometry of eleven dimensional supergravity with four off-shell local supersymmetries. We start with a background Minkowski 11D, N=1 superspace, and an additional ingredient of a global, constant, $G_2$-structure which facilitates the definition of a $4|4+7$ background superspace. A bottom-up construction of linear fluctuations of the geometric constituents (such as supervielbein, spin connection, and the super 3-form of 11D supergravity) is given in terms of 4D, N=1 prepotential superfields. This is complemented by a top-down description of the linearized supergeometry of the $4|4+7$ superspace dealing directly with torsion, curvature, and Bianchi identities. Torsion constraints that (combined with the Bianchi identities) lead to the preceding prepotential expressions of the gauge fields are identified. All irreducible consequences of the torsion and 4-form Bianchi identities are systematically derived except for dimension 2 Bianchi identities of the 4-form, and dimension $\tfrac 52$ Bianchi identities of torsion, which set bosonic curls of components of one lower dimension to zero.
}

\begin{document}
\maketitle

\section{Introduction}
\label{S:Intro}

When a supersymmetric system possesses an off-shell formulation, this can offer crucial insight into its structure. Such formulations incorporate, in addition to the usual physical bosonic and fermionic fields, new auxiliary fields that do not typically propagate but exist to close the supersymmetry algebra off-shell -- that is, without imposing equations of motion. In the context of finding higher derivative corrections, this is especially useful because the problem of finding the supersymmetric action and the supersymmetry transformations are divorced: the transformations are fixed and only an invariant action must be sought.

For systems with four supercharges, such as the familiar 4D $N=1$ supersymmetry, off-shell approaches are straightforward. Even for systems with eight supercharges (e.g. 4D $N=2$), where the number of auxiliaries become infinite, techniques are available involving harmonic or projective superspace to tame this zoo. For 11D supergravity, the situation is quite different. While there exists an off-shell approach in pure spinor superspace
\cite{Berkovits:2000fe, Cederwall:2010tn}, it is even more technically challenging (see e.g. \cite{Berkovits:2018gbq} where the connection to 11D supergravity was first directly shown). Meanwhile, other techniques to construct higher derivative terms -- either via deformations of on-shell superspace \cite{Cederwall:2004cg} or by working directly at the component level \cite{Hyakutake:2006aq} -- have not been completely successful.

An alternative approach is to maintain only some fraction of the full supersymmetry by rewriting 11D supergravity in a lower dimensional superspace, specifically 4D $N=1$ superspace, while keeping additional parametric dependence on the other seven bosonic coordinates, which are spectators from the point of view of $N=1$ supersymmetry. We denote such a framework as a $4|4+7$ superspace for brevity; it is also convenient to think of the seven dimensional space as an ``internal space'' and the 4D $N=1$ superspace as ``external'', although we will neither truncate the theory nor expand on the internal space in harmonic functions.\footnote{This approach is inspired by the early rewriting of 10D super Yang-Mills in 4D $N=1$ superspace \cite{Marcus:1983wb}.} 

In recent years, two of us, along with numerous collaborators, have been exploring just how this works. The key point of departure is the 3-form in 11D; it descends to an abelian tensor hierarchy in 4D language, and the structure of such a hierarchy in 4D $N=1$ superspace is quite rigid \cite{Becker:2016xgv,Becker:2016rku} and \emph{just by itself} already correctly reproduces the internal sector of 11D supergravity \cite{Becker:2016edk}. A key missing feature is the additional gravitini supermultiplets, and taking these into account was shown in \cite{Becker:2017zwe} to correctly reproduce the bosonic part of the linearized component theory. A first step towards its non-linear completion was taken in \cite{Becker:2018phr} and the appropriate framework for this eventual completion was constructed in \cite{Becker:2020hym}. 

While this line of approach is promising, the resulting framework is cumbersome. Operationally, this is due to the additional gravitino superfields, which are encoded in curvature superfields of vanishing mass dimension. This fact means non-polynomial functions of this superfield play a role when constructing invariant superspace actions. Even more, the non-manifest supersymmetry is joined with a large number of other local symmetries. The upshot is that a larger number of component fields exist than one would normally expect, and the extra ones (aside from the auxiliary fields) turn out to be pure gauge degrees of freedom, set to zero by a Wess-Zumino condition or otherwise eaten by a gauge field. The framework is also more abstractly cumbersome because no trace of the higher Lorentz group remains; internal indices are purely world (curved) indices carried by various $p$-forms and even the internal metric is entirely encoded in a 3-form identifiable as a $G_2$ structure. While all of this leads to extremely natural $N=1$ multiplets fitting together into two elegant hierarchies, when taken together it deeply obfuscates the connection to the original 11D theory.

An alternative approach is to build on our earlier linearized work \cite{Becker:2017zwe}, maintaining (at least some of) the additional Lorentz symmetry and identifiable elements of the 11D theory. Indeed, this is the path we recently followed in \cite{Becker:2021oiz}, where employing the $N=1$ superfield constituents uncovered in \cite{Becker:2017zwe}, we identified the linearized components of 11D supergravity along with the auxiliary fields implied by $N=1$ supersymmetry. We worked purely at the component level, putting the various components of the $N=1$ superfields together into fields with the right transformation rules, until we recovered (on-shell) the known linearized transformations of 11D supergravity. The goal of the present paper is to explore this from a complementary perspective: that of a $4|4+7$ superspace for which those component fields are the natural geometric constituents.

Before delving more into that, we should remark on a surprising feature of the superspace we will be discussing (and the corresponding component theory given in \cite{Becker:2021oiz}): it involves the full 11D Lorentz group $SO(10,1)$. At first blush, in a Kaluza-Klein-like $4+7$ reformulation, one would expect the full Lorentz group to be broken to $SO(3,1) \times SO(7)$, say after fixing an upper triangular gauge for the vielbein. Moreover, since we are keeping only 1/8 of the supersymmetry, one might further expect that the $SO(7)$ should be broken to $G_2$, the subgroup respecting the selection of an $N=1$ subsector of the natural $N=8$ supersymmetry expected in 4D. 
The crux of the matter is that we are discussing a linearized supergeometry and must distinguish between background Lorentz transformations (which are manifestly broken when the background becomes rigid) and ones associated with the linearized fluctuation, which retain their 11D character. Naturally, this means that the formulation we are presenting is rather special to the linearized setting. Undoubtedly in trying to construct a non-linear version, we would need to specialize to an $SO(3,1) \times G_2$ Lorentz group, and this would presumably arise as some sort of gauge-fixing of the supergeometry discussed in \cite{Becker:2020hym}. We leave that question for future work.

The body of this paper is organized as follows. In section \ref{S:Sect2} we give a review of the prepotential superfields that encode dynamical fields of 11D supergravity along with auxiliary fields required for off-shell closure of four supersymmetries. We also construct partial invariants under the linearized transformations of these prepotentials. Section \ref{S:Sect3} gives an explicit, ``bottom up",  construction of the $4|4+7$ superspace. There are two key steps in this construction -- linearization around a background, and reduction to $4|4+7$ superspace -- both of which are explained in detail. The $4|4+7$ superspace is equipped with a frame, a spin connection, a super three form gauge field and its associated field strength, and seven extra gravitini. Components of all these constituents are defined in terms of the partial invariants introduced in section \ref{S:Sect2}. At this point, we change our perspective to ask the following supergeometric questions. Any generic superspace equipped with the same geometric ingredients as above is constrained to satisfy Bianchi identities. What additional data should be specified so that it is further constrained to match the superspace we built explicitly? In other words, what additional torsion constraints should be imposed so that a solution of the resulting Bianchi identities in terms of unconstrained prepotentials is given by our construction in section \ref{S:Sect3}? These questions are addressed in section \ref{S:Sect4}. We go all the way in analyzing the Bianchi identities to show which torsion or curvature components are determined fully in terms of lower dimensional components, and which components do not get constrained by Bianchi identities at all. We systematically derive derivative relations between these various torsion and curvature components as well.

\section{Linearized 4D $N=1$ prepotentials}
\label{S:Sect2}
The spectrum of 11D supergravity consists of a frame field $ e_\hm {}^\ha $, a $32$-component gravitino $\psi_\hm {}^\halpha $, and a 3-form gauge field $C_{\hm \hn \hp}$. The fields provide a realization
of the 11D $N=1$ supersymmetry algebra provided they are on-shell. The superspace formulation of this theory has also been long known \cite{Brink:1980az, Cremmer:1980ru}, which suffers from the same on-shell problem -- the combination of torsion constraints and superspace Bianchi identities imply equations of motion. A brief review of this on-shell $11|32$ superspace will be given in subsection \ref{S:Sect3.1}.

\begin{table}[ht]
	{\footnotesize
		\begin{align*}
			{\renewcommand{\arraystretch}{1.5} 
				\begin{array}{|c|c|c|c|c|c|c|c|}
					\hline
					\textrm{index} & ~~~~\textrm{range}~~~~ & \textrm{description} \\
					\hline
					\hm, \hn, \dots & 0, \cdots , 10 & \textrm{11D coordinate} \\
					\ha, \hb, \dots & 0, \cdots , 10 & \textrm{11D tangent} \\
					\halpha, \hbeta, \dots & 1, \cdots , 32 & \textrm{11D spinor} \\
					\hline
					m, n, \dots & 0, 1,2 , 3 & \textrm{4D coordinate} \\
					a, b, \dots & 0, 1,2 , 3 & \textrm{4D tangent}  \\
					~~\alpha, \beta , \dots , \dalpha,  \dbeta \dots & 1, 2 & \textrm{4D spinor} \\
					\hline
					\um,\un,\dots & 1,\cdots , 7 & ~~\textrm{7-component label} ~~\\
					\hline
					\hM, \hN, \cdots & (\hm, \hat \mu) & 11|32~ \textrm{coordinate} \\
					\hA, \hB, \cdots & (\ha, \halpha) & 11|32~ \textrm{tangent} \\
					\ul{\alpha} & (\alpha, \dalpha) & \textrm{compound notation à la \cite{Wess:1992cp}} \\
					A & (a, \ul{\alpha}) & 4|4~ \textrm{tangent} \\
					\hline
				\end{array}
			}
		\end{align*}
		\caption{\scriptsize Legend of indices. Indices of various $\bm7$-dimensional representations ($GL(7)$ coordinate, $SO(7)$ tangent, $G_2$ representation, and label for seven extra gravitini) have all been identified to avoid proliferation of notation.
		}
		\label{T:Legend}
	} 
\end{table}

In a series of recent papers \cite{Becker:2016xgv, Becker:2016rku, Becker:2016edk, Becker:2017zwe, Becker:2018phr}, a formulation of linearized 11D supergravity  with four off-shell supersymmetries was presented. In this approach, fields are linear fluctuations around a rigid Minkowski background $\mathbb R^{11|32}$. On the bosonic manifold, a global, constant 3-form $\varphi$ is chosen which, in some bosonic Cartesian coordinates $(x^m, y^\um)$, takes the form
\begin{align}
\label{E:varphi-def}
	\varphi = - (e^{123} + e^{145} + e^{167} + e^{246} - e^{257} - e^{347} - e^{356}), 
\end{align}
where $e^{\ul{mnp}} := dy^\um \wedge dy^\un \wedge dy^\up$.  The submanifold obtained by setting $x^m =0$ then define the ``internal" 7-manifold $Y = \mathbb R^7$ with coordinates $(y^\um)$, and the remaining 4 dimensions are external. $\varphi$ satisfies the properties of being a $G_2$ structure, and hence reduces the structure group of Y from $GL(7)$ to $G_2$. Using the real commuting spinor \eqref{E:eta<=>varphi} associated with the $G_2$ structure (more details on how this works later), one can naturally identify 4 special Grassmann coordinates out of the 32 in $\mathbb R^{11|32}$, which combine with the external bosonic coordinates to give a 4D, $N=1$ superspace. Component fields of 11D supergravity are then embedded in ``prepotential" superfield representations of the 4D, $N=1$ superconformal algebra. In this section we sketch a lightning review of these prepotentials and their transformations.

\subsection{Prepotentials and their transformations}
\label{S:Sect2.1-prepotentials}

Under the decomposition of 11D spacetime coordinates into four external coordinates $x^m$, and seven internal coordinates $y^\um$, the components of the 3-form break up into an abelian tensor hierarchy of forms  in external spacetime of degrees 0 through 3:
\begin{align}
	C_{3} ~~ \rightarrow  ~~ C_{\underline{mnp}}, ~~ C_{m \underline{np}}, ~~ C_{mn \up}, ~~ C_{mnp}
\end{align}
It is known \cite{Becker:2016xgv} how to embed these bosonic $p$-forms into 4D $N=1$ superfields. One needs a chiral superfield $\Phi_{\underline{mnp}}$, a real vector superfield $V_{\um \un}$, a chiral spinor superfield $\Sigma_{\alpha \um}$, and a real superfield $X$. Their abelian gauge transformation is derived by decomposing the 2-form abelian gauge transformations of $C_{3}$ in 11D:
\begin{subequations}
	\label{E:ATHtransf}
	\begin{align}
		\delta \Phi_{\underline{mnp}} &= 3 \partial_{[\um} \Lambda_{\underline{np}]} ~,\\
		\delta V_{\um \un} &= \tfrac 1{2i} \left( \Lambda_{\um \un} - \bar \Lambda_{\um \un} \right) - 2 \partial_{[\um} U_{\un ]}  ~,\\
		\delta \Sigma_{\alpha \um} &= - \tfrac 14 \bar D^2 D_{\alpha} U_{\um} + \partial_{\um} \Upsilon_\alpha ~,\\
		\delta X &= \tfrac 1{2i} \left( D^{\alpha} \Upsilon_{\alpha} - \bar D_{\dalpha} \bar \Upsilon^\dalpha  \right)~,
	\end{align}
\end{subequations}
where the gauge parameters are a chiral superfield $\Lambda_{\um \un}$, a real superfield $U_\um$, and a chiral spinor superfield $\Upsilon_\alpha$. Moreover, internal diffeomorphisms appear as a non-abelian gauge symmetry from a 4D perspective, the gauge field being the Kaluza-Klein vector of the 11D vielbein. The abelian tensor hierarchy, when gauged by internal diffeomorphisms in this way, is called a non-abelian tensor hierarchy. The Kaluza-Klein vector is embedded in the prepotential $\cV^\um$, a real vector  superfield. The only prepotentials that transform under linearized non-abelian gauge transformations are $V_{\underline{mn}}$ and $\cV^\um$:
\begin{align}
\label{E:NATHtransf-lin}
	\delta \cV^\um = \lambda^\um + \bar \lambda^{\um}~, \qquad
	\delta V_{\um \un} = -i \varphi_{\ul{mnp}} \left( \lambda^\up - \bar \lambda^\up  \right)~,
\end{align}
where the non-abelian gauge parameter $\lambda^\um$ is chiral, $\bar \lambda^\um$ is antichiral, and $\varphi_{\ul{mnp}}$ is the $G_2$ structure on the background Minkowski superspace
$\mathbb R^{4|4} \times \mathbb R^7$.
One can construct field strength superfields invariant under transformations \footnote{One can fully non-linearize the non-abelian transformations by modifying the 4D $N=1$ derivatives to include a Kaluza-Klein connection, both when the 4D superspace is curved (but $y$-independent)
\cite{Becker:2018phr} and in the fully general case \cite{Becker:2020hym}. Field strengths and Chern-Simons action can be constructed in both scenarios.} \eqref{E:ATHtransf} and \eqref{E:NATHtransf-lin}, and from them construct the Chern-Simons action. The Chern-Simons action, at the component level, contains kinetic terms for the 4D vector fields, but not those of the scalars and 2-forms. For that, one needs to add a K\"{a}hler-type term.

This embedding of spin $\le 1$ component fields of 11D supergravity has a surfeit of component fields. In addition to the 11D components with spin $\le 1$, there are $16$ extra scalars, and $15$ extra fermions. Moreover, the purely external graviton and spin $\tfrac 32$ part of the gravitino are still missing. The 4D $N=1$ gravitino and graviton belong together in a real superfield $H_{\alpha \dalpha} = (\sigma^a)_{\alpha \dalpha} H_a$. This is the linearized prepotential of 4D $N=1$ conformal supergravity, transforming under linearized local superconformal transformation as
\begin{align}
	\delta H_{\alpha \dalpha} &= D_\alpha \bar L_\dalpha - \bar D_{\dalpha} L_\alpha~,
\end{align}
where the gauge parameter $L_\alpha$ is unconstrained. The original 11D $N=1$ gravitino has $32$ components, and $H_{\alpha \dalpha}$ encodes four of them. For the remaining 28, we have seven additional gravitini which are embedded in spinor superfields $\Psi_{\um \alpha}$. These are called ``matter" gravitini. The basic matter gravitino model subjects the prepotentials to gauge transformations \cite{Gates:1979gv}
\begin{align}
	\delta \Psi_{\um \alpha} &= \Xi_{\um \alpha} + D_\alpha \Omega_\um
\end{align}
where $\Xi_{\um \alpha}$ is chiral, and $\Omega_\um$ is an unconstrained complex superfield. These transformations contain the non-manifest supersymmetry. With the introduction of $H_{\alpha \dalpha}$ and $\Psi_{\um \alpha}$, all 11D supergravity component fields have been embedded into 4D $N=1$ prepotentials. The only problem remaining is that these prepotentials have more components in them than just the physical fields in 11D supergravity plus auxiliary fields necessary for four off-shell supersymmetries. These extra fields are accompanied by additional local gauge symmetries (lying within $\Xi_{\um \alpha}$ and $\Omega_\um$) that can remove them. The prepotentials of the tensor hierarchy turn out to transform under these additional transformations as
\begin{subequations}
	\begin{align}
		\delta \Phi_{\underline{mnp}} &= - \tfrac i2  \psi_{\ul{mnpq}} \bar D^2 \bar \Omega^\uq \\
		\delta V_{\um \un} &= \tfrac 1{2i} \varphi_{\ul{mnp}} \left( \Omega^\up - \bar \Omega^\up \right) \\
		\delta \Sigma_{\alpha \um} &= - \Xi_{\um \alpha} \\
		\delta X &= D^\alpha L_\alpha + \bar D_\dalpha \bar L^\dalpha \\
		\delta \cV_\um &= - \tfrac 12 \left( \Omega_\um + \bar \Omega_\um  \right)
	\end{align}
\end{subequations}
and the matter gravitini are given local superconformal transformations with parameters $L_\alpha$: \begin{align}
	\delta \Psi_{\um \alpha} &= 2i \partial_{\um} L_\alpha~.
\end{align}
For instance, the superfield $G = - \tfrac 14 \bar D^2 X$ plays the role of chiral compensator in 4D $N=1$ conformal supergravity.

The prepotentials introduced so far can be used to construct superfields which have as the leading terms in their $\theta$-expansion  the physical 11D gauge connections ({\it i.e.}\ the frame, 3-form, and gravitini), and auxiliary fields added to the spectrum to achieve 4D $N=1$ off-shell supersymmetry. This explicit construction was performed in \cite{Becker:2021oiz}, and will also be reviewed in sections \ref{S:Sect3.4-4d-supervielbein} to \ref{S:Sect3.7}. Counting the degrees of freedom encoded in the prepotentials, one finds that  the real superdimension of the auxiliary field space adds up to $201|56$, rendering the total spectrum $376|376$ dimensional.

\subsection{Partial invariants}
\label{S:Sect2.2}
The explicit construction of \cite{Becker:2021oiz} was given at the component level. Our goal in this article is to describe the supergeometry of the $4|4+7$ superspace, whose underlying prepotential structure and component field content is precisely that of \cite{Becker:2021oiz}. An obvious problem to overcome is that the local gauge symmetries underlying our prepotentials are significantly larger than those of 11D supergravity. As a prelude towards building that supergeometry, we will introduce a number of building blocks by sequentially eliminating gauge symmetries by trading prepotentials for composite curvatures. 
Once these basic ingredients are determined we will turn to the construction of the supergeometry in section \ref{S:Sect3}.

\subsubsection*{Invariants of the abelian tensor hierarchy}
\label{S:Sect2.2.1-ATH-invariants}
We start by trading prepotentials of the abelian tensor hierarchy for their curvatures:
\begin{subequations}
	\label{E:ATH-field-strengths}
	\begin{align}
		E_{\underline{mnpq}} &:= 4 \pa_{[\um} \Phi_{\underline{npq]}} \\
		F_{\underline{mnp}} &:= \tfrac 1{2i} \left( \Phi_{\underline{mnp}} - \bar \Phi_{\underline{mnp}} \right) - 3 \pa_{[\um} V_{\underline{np} ]}  \\
		W_{ \um \un \alpha} &:= - \tfrac 14 \bar D^2 D_\alpha V_{\um \un} + 2 \pa_{[\um} \Sigma_{| \alpha | \un ]} \\
		H_\um &:= \tfrac 1{2i} \left( D^\alpha \Sigma_{\alpha \um} - \bar D_\dalpha \bar \Sigma^\dalpha {}_\um  \right) - \pa_\um X \\
		G&:= - \tfrac 14 \bar D^2 X
	\end{align}
\end{subequations}
These superfields are invariant under the abelian tensor hierarchy transformations \eqref{E:ATHtransf}.

\subsubsection*{$ L_{\alpha} $ invariants}
\label{S:Sect2.2.2-L-alpha-invariants}
Now let us introduce superfields that are $L_\alpha$ invariant. Since only $H_{\alpha \dalpha}$, $\Psi_{\um \alpha}$, and $X$ (and thus $G$ and $H_\um$) transform at the linearized level, we trade these for the following curvatures:
\begin{subequations}
	\begin{align}
		W_{\gamma \beta \alpha} &:= \tfrac i{16} \bar D^2 \pa_{(\gamma} {}^\dgamma D_\beta H_{\alpha) \dgamma} \\
		R &:= - \tfrac 1{24} \bar D^2 \bar G + \tfrac i{24} \bar D^2 \pa_{\alpha \dalpha} H^{\dalpha \alpha} \\
		G_{\alpha \dalpha} &:= - \tfrac i6 \pa_{\alpha \dalpha} \left( G- \bar G \right) + \left[ \tfrac 12 \square - \tfrac 1{32} \left\{ D^2, \bar D^2 \right\} \right] H_{\alpha \dalpha} + \left[ \tfrac 14 \pa_{\alpha \dalpha} \pa^{\beta \dbeta}  + \tfrac 1{12} \Delta_{\alpha \dalpha} \Delta^{\beta \dbeta} \right] H_{\dbeta \beta} \\
		X_{\um \alpha \dalpha} &:= \tfrac 1{2i} \left( \bar D_\dalpha \Psi_{\um \alpha} + D_\alpha \bar \Psi_{\um \dalpha}  \right) + \pa_\um H_{\alpha \dalpha} \\
		\Psi_{\um \un \alpha} &:= 2 \pa_{[\um} \Psi_{\un] \alpha} \\
		\hH_\um &:= H_\um + \tfrac 1{2i} \left( D^\alpha \Psi_{\um \alpha} - \bar D_\dalpha \bar \Psi_{\um } {}^\dalpha  \right)
	\end{align}
\end{subequations}
where $\Delta_{\alpha \dalpha} := - \tfrac 12 [D_\alpha , \bar D_\dalpha]$. The first three above are familiar from conventional $N=1$ superspace \cite{Wess:1992cp}, while the last three covariantize $H_\um$ and $\Psi_{\um \alpha}$.

\subsubsection*{$ \Xi_{\um \alpha} $ invariants}
\label{S:Sect2.2.3-Xi-invariants}
The only prepotentials suffering $\Xi_{\um \alpha}$ transformations are $\Psi_{\um \alpha}$ and $\Sigma_{\alpha \um}$. So we investigate $W_{\um \un \alpha }, X_{\um \alpha\dalpha}, \Psi_{\um \un \alpha}, \textrm{and} ~ \hH_\um$. We observe that $\hH_\um, X_{\um \alpha \dalpha}$ are already $\Xi_{\um \alpha}$-invariant, and we can easily trade $W_{\um \un \alpha}$ and $\Psi_{\um \un\alpha}$ for their invariant sum,
\begin{align}
	\hat W_{\um \un \alpha} &:= W_{\um \un \alpha} + \Psi_{\um \un \alpha}~.
\end{align}

\subsubsection*{$ \lambda_{\um} $ invariants}
\label{S:Sect2.2.4-nath-invaraints}
Prepotentials transforming under internal diffeomorphisms are $\cV_\um$ and  $V_{\um \un}$.
The Kaluza-Klein field strength
\begin{align}
	\cW_{\alpha \um} &:= - \tfrac 14 \bar D^2 D_\alpha \cV_\um
\end{align}
is invariant under $\lambda_\um$ (as well as $L_\alpha$ and $\Xi_{\um \alpha}$).
Similarly, $E_{\underline{mnp}}$ and $\hat W_{\um \un \alpha}$ are $\lambda_{\um}$-invariant. $F_{\underline{mnp}}$ transforms, but its spinor derivative can be made $\lambda_{\um}$ invariant by defining
\begin{align}
\label{E:F_alpha-mnp--lambda-inv}
	F_{\alpha \underline{mnp}} &:= D_\alpha F_{\underline{mnp}} - 3i \varphi_{\uq [ \um \un} \pa_{\up]} D_\alpha \cV^\uq
\end{align}
Its complex conjugate is denoted by $F_{\dalpha \ul{mnp}}$.

\subsection{$\Omega_\um$ transformations}
\label{S:Sect2.3}
The only transformation left is $\Omega_\um$, which contains the extended supersymmetry. The partial invariants that we have defined each transform under this symmetry as
\begin{subequations}
    \label{E:OmegaTrafos}
	\begin{align}
		\delta_\Omega X_{\um \alpha \dalpha} &= \tfrac 1{2i} \left( \bar D_\dalpha D_\alpha \Omega_\um + D_\alpha \bar D_\dalpha \bar \Omega_\um  \right) \\
		\delta_\Omega \hH_\um  &= \tfrac 1{2i} \left( D^2 \Omega_\um - \bar D^2 \bar \Omega_\um \right)  \\
		\delta_\Omega \cW_{\alpha \um} &= \tfrac 18 \bar D^2 D_\alpha \left( \Omega_\um + \bar \Omega_\um \right) \\
		\delta_\Omega \hat W_{\ul{mn} \alpha} &= \tfrac i8 \varphi_{\ul{mnp}} \bar D^2 D_\alpha \left( \Omega^\up - \bar \Omega^\up \right) + 2 \pa_{[\um} D_{|\alpha|} \Omega_{\un]} \\
		\delta_\Omega F_{\alpha \ul{mnp}} &= - \tfrac 14  \psi_{\ul{mnpq}} D_\alpha \bar D^2 \bar \Omega^\uq + 3i \varphi_{\uq [ \ul{mn}} \pa_{\up]} D_\alpha \Omega^\uq \\
		\delta_\Omega E_{\ul{mnpq}} &= 2i  \psi_{[\ul{mnp} |\ur|} \pa_{\uq]} \bar D^2 \bar \Omega^\ur
 	\end{align}
\end{subequations}
The conventional $N=1$ superfields $W_{\gamma \beta \alpha}$, $R$, and $G_{\alpha \dalpha}$ are $\Omega_\um$ invariant. These will serve as the building blocks in subsequent sections.

As a starting point, we note that the first $\Omega_\um$-invariant that we can build out of the fields in \eqref{E:OmegaTrafos} lies at dimension $\tfrac{1}{2}$:
\begin{align}
	\lambda_{\um \alpha} &= \varphi_\um {}^{\ul{np}} \hat W_{\alpha \ul{np}} - \tfrac i6  \psi_\um {}^{\ul{npq}} F_{\alpha \ul{npq}} + D_\alpha \hH_\um + 2i \cW_{\alpha \um} - 2 \bar D^\dalpha X_{\um \alpha \dalpha}  ~.
\end{align}
This is proportional to the equation of motion of the matter gravitino superfield $\Psi_{\um \alpha}$ and matches the dimension $\tfrac{1}{2}$ auxiliary field identified in \cite{Becker:2021oiz}.

\section{Linearized supergeometry of $4|4+7$ superspace}
\label{S:Sect3}
In this section, we will describe how a linearized $4|4+7$ superspace is constructed out of the prepotential constituents given in the previous section.

The fermionic coordinates of 11D $N=1$ superspace are a $32$-component Majorana fermion. The $11|32$ superspace coordinates are combined in $z^\hM = (x^\hm, \theta^{\hat \mu})$, where $x^\hm$ are $11$ bosonic coordinates, and $\theta^{\hat \mu}$ are $32$ fermionic ones. We introduce a vielbein $\hat E_\hM{}^\hA$, the corresponding frame $\hat E^\hA=dz^\hM \hat E_\hM{}^\hA $, and a spin connection one form $\hOmega$ valued in $SO(10,1)$. The torsion is the covariant derivative of the vielbein
\begin{equation}
	\hat T^{\hA} = {\cal D} \hat E^\hA =  \rd \hat E^{\hA}+\hat E^{\hB} \wedge \hat \Omega\indices{_\hB^\hA}=\frac{1}{2} \hat E^{\hC} \wedge \hat E^{\hB} \, \hat T\indices{_\hB _\hC ^\hA},
\end{equation}
and the curvature is defined in terms of the spin connection $\hat \Omega$
\begin{equation}
	\hat R\indices{_\hA ^\hB} = \rd \hat \Omega\indices{_\hA^\hB}+ \hat\Omega\indices{_\hA^\hC} \wedge \hat \Omega\indices{_\hC ^\hB}~.
\end{equation}
The torsion and the Riemann tensors satisfy the Bianchi identities
	\begin{align}
		{\cal D}\hat T^\hA = \hat E^\hB \wedge \hat R\indices{_\hB ^\hA} ~, \qquad
		{\cal D} \hat R_\hB {}^\hA = 0~.
	\end{align}
For convenience, we quote the components of these tensors
\begin{subequations}
	\begin{align}
		\hT_{\hN\hM}{}^\hA &= 2 \,\hat\cD_{[\hN} \hE_{\hM]}{}^\hA = 2 \,\pa_{[\hN} \hE_{\hM]}{}^\hA - 2\, \hE_{[\hN}{}^{\hB}\hOmega_{\hM] \,\hB}{}^{\hA} (-)^{m b}~, \\
		\hR_{\hN\hM \, \hA}{}^{\hB} &= 2\, \pa_{[\hN} \hOmega_{\hM]}{}_\hA{}^{\hB} - 2 \,\hOmega_{[\hN}{}_{|\hA|} {}^\hC \,\hOmega_{\hM]}{}_\hC{}^\hB~,
	\end{align}
\end{subequations}
and their Bianchi identities
\begin{subequations}
\label{E:torsion-curvature-BI}
	\begin{align}
		{\cal D}_{[ \hD} \hat T_{\hC \hB]} {}^\hA + \hat T_{[ \hD \hC} {}^\hF \hat T_{|\hF| \hB]} {}^\hA &= \hat R_{[\hD \hC \hB]} {}^\hA~, \\
		{\cal D}_{[\hE} \hat R_{\hD \hC \hB]} {}^\hA + \hat T_{[\hE \hD} {}^\hF \hat R_{|\hF| \hC \hB]} {}^\hA &= 0~.
	\end{align}
\end{subequations}
We suppress gradings above, and in all following component expressions. The spin connection and Riemann tensor are both $SO(10,1)$ valued, so that they obey
\begin{align}
\label{E:spin-connection-lorentz-valued}
	\hOmega_{\ha \hb} = - \hOmega_{\hb\ha}~, \qquad
	\hOmega_\halpha{}^\hbeta = \frac{1}{4} \hOmega_{\ha \hb} (\hat \Gamma^{\ha \hb})_\halpha{}^\hbeta~,
\end{align}
with other components vanishing (similarly for $\hat R_\hA{}^\hB$).
The spectrum of 11D supergravity also contains a 3-form, so we introduce one in superspace,
\begin{equation}
	\hC_3 = \frac{1}{3!} \rd z^\hM \wedge \rd z^\hN \wedge \rd z^\hP \, \hC_{\hP \hN \hM}.
\end{equation}
The associated 4-form field strength $\hat G_4 = \rd \hat C_3$ satisfies the Bianchi identity
\begin{equation}
\label{E:4form-curvature-BI}
	\rd \hat G_4 = 0 = \tfrac 1{4!} {\cal D}_{[\hE} \hat G_{\hD \hC \hB \hA]} + \tfrac 1{3! 2!} \hat T_{[\hE \hD} {}^\hF \hat G_{|\hF| \hC \hB \hA]}~.
\end{equation}

\subsection{Review: On-shell 11D supergravity in superspace}
\label{S:Sect3.1}

The Lagrangian of 11D supergravity is given by 
\begin{align}
\hat e^{-1} \cL &= -\frac{1}{2} \hat {\mathcal R}
        + \frac{1}{2} \hat \psi_\hm{}^\halpha (\Gamma^{\hm\hn\hp})_{\halpha \hbeta}
            \cD_\hn \hat \psi_\hp{}^\hbeta
        - \frac{1}{4\cdot 4!} \hat G_{\hm\hn\hp\hq} \hat G^{\hm\hn\hp\hq}
        \eol & \qquad
        - \frac{1}{12} \veps^{\hm_1 \cdots \hm_{11}}
            \hat C_{\hm_1 \hm_2 \hm_3} \hat G_{\hm_4 \cdots \hm_7} \hat G_{\hm_8 \cdots \hm_{11}}
        + \cdots
\end{align}
where we have omitted higher order fermionic terms. An 11D superspace is said to be on-shell if its component projection satisfies equations of motion derived from the above Lagrangian. It is called off-shell otherwise.

We give a quick review of the on-shell $11|32$ superspace, initially constructed in \cite{Brink:1980az, Cremmer:1980ru}, in our notations and conventions. In the process, we point out why this superspace is necessarily on-shell, and motivate our construction of a partially off-shell, albeit linearized around a background, $4|4+7$ superspace. 

Suppose we augment the superspace data (following \cite{Brink:1980az, Cremmer:1980ru}) by the constraints for $\hat G_4$ 
\begin{subequations}
\label{E:B-H-G_4-constraints}
	\begin{align}
	\hat G_{\halpha \hbeta \hgamma \hdelta} &= 0 = \hat G_{\ha \hbeta \hgamma \hdelta} = \hat G_{\ha \hb \hc \hdelta}~, \\
	\hat G_{\ha \hb \hgamma \hdelta} &= 2 (\hat \Gamma_{\ha \hb})_{\hgamma \hdelta}~,
	\end{align}
\end{subequations}
and the constraints for the torsion
\begin{subequations}
\label{E:B-H-torsion-constraints}
	\begin{align}
	\hat T_{\hgamma \hbeta} {}^\ha &= 2 (\hat \Gamma^\ha)_{\hgamma \hbeta} ~, ~~~ \hat T_{\hgamma \hbeta} {}^\ha = 0 = \hat T_{\hgamma \hbeta} {}^\halpha~, \\
	\hat T_{\hc \hb} {}^\ha &= 0~.
	\end{align}
\end{subequations}
The $\hat G_4$ Bianchi identities then determine the following non-zero components of the torsion:
\begin{subequations}
\label{E:non-zero-torsion}
	\begin{align}
	\hT_{\ha \hbeta}{}^\hgamma &= - \tfrac{1}{36} \hG_{\ha\hb\hc\hd} (\hat \Gamma^{\hb\hc\hd})_{\hbeta}{}^\hgamma - \tfrac{1}{288} (\hat \Gamma_{\ha\hb\hc\hd\he})_{\hbeta}{}^\hgamma \hG^{\hb\hc\hd\he}~, \\
	\hT_{\ha \hb}{}^\halpha &= -\tfrac{1}{84} (\hat \Gamma^{\hc\hd})^\halpha{}^\hbeta \cD_\hbeta \hG_{\ha\hb\hc\hd}~.
\end{align}
\end{subequations}
From (\ref{E:non-zero-torsion}b), it follows that the gravitino field strength $\hat T_{\ha \hb} {}^\halpha$ satisfies the Rarita-Schwinger equation of motion
\begin{align}
	(\hat \Gamma^{\ha \hb \hc})_\halpha {}^\hbeta \hat T_{\hb \hc, \hbeta} &= 0~.
\end{align}
One then imposes the torsion Bianchi identities to find that the equations of motion for the vielbein, and the 4-form field strength are also satisfied. This establishes that the superspace of refs. \cite{Brink:1980az, Cremmer:1980ru} is on-shell.

At the component level, this has the following consequence. If we normalize the 11D gravitino conventionally as $\hat \psi_{\hm}{}^{\halpha} = 2\hE_\hm{}^\halpha \vert_{\theta=0}$, then the 11D SUSY transformations of the component fields can be written
\begin{subequations}
\begin{align}
	\delta \hat e_\hm{}^\ha &= -\veps^\halpha (\hat \Gamma^\ha)_{\halpha \hbeta} \Psi_\hM{}^\hbeta~, \\
	\delta \hat\psi_{\hm}{}^\halpha &= 2 \hat\cD_\hm \veps^\halpha
	+ 2 \,\hat e_\hm{}^\ha\, \veps^\hbeta
		\Big(
		\frac{1}{36} \hG_{\ha \hb \hc \hd} (\hat \Gamma^{\hb \hc \hd})_{\hbeta}{}^\halpha
		+ \frac{1}{288} (\hat \Gamma_{\ha \hb \hc \hd \he})_{\hbeta}{}^\halpha \hat G^{\hb \hc \hd \he}
		\Big)~,\\
	\delta \hC_{\hm \hn \hp} &=
	- 3 \,\veps^\halpha
        (\hat \Gamma_{[\hm \hn})_{|\halpha \hbeta|} \hat\Psi_{\hp]}{}^\hbeta~,
\end{align}
\end{subequations}
and these supersymmetry transformations close only up to the equations of motion of 11D supergravity. 

The root cause of the on-shell nature of the superspace lies in the choice of constraints \eqref{E:B-H-G_4-constraints} and \eqref{E:B-H-torsion-constraints}. One may be able to throw the superspace off-shell by cleverly relaxing these. In this scenario, auxiliary fields would be present in the spectrum, playing their usual important role in off-shell closure of the SUSY algebra, and superspace Bianchi identities would not imply field equations. However, the new set of constraints must also have the property that we get back the on-shell theory upon imposing field equations. So far it has not been possible to achieve this in the fully non-linear setting. We present a partial solution to the problem by constructing a linearized superspace which is also truncated, resulting in partially off-shell SUSY. In the next two sections, we sequentially describe these two steps -- linearization around a background and restriction to a subspace that keeps only 4 fermionic coordinates and throws away the other 28.

\subsection{Linearizing around a background}

We expand to linear order about a background superspace which satisfies the superspace Bianchi identities. All background quantities are denoted by placing a circle $\mathring{}$ on top, and linear fluctuations are denoted by bold letters. We will compute components of the torsion and curvature (and associated Bianchi identities), and four-form (and associated Bianchi identity) in terms of the linear fluctuations. Our choice of background is the flat $11|32$ superspace. In Minkowski coordinate system, we can choose the background superframe (up to rigid Lorentz transformations)
\begin{align}
\label{E:11DBackground.E}
	\mathring E^\ha = \rd x^\ha - \theta^\halpha (\Gamma^\ha)_{\halpha \hbeta} \rd \theta^\hbeta~, \qquad
	\mathring E^\halpha = \rd \theta^\halpha~.
\end{align}
The linearized supervielbein is given by
\begin{align}
	\hat E_\hM{}^\hA &= \mathring E_\hM{}^\hA + \mathring E_\hM{}^\hB \, {\bm H}_\hB{}^\hA~,
\end{align}
where ${\bm H}_\hB{}^\hA$ is the linearized fluctuation. The spin connection has no background value, so $\hOmega_\hM {}_\hA {}^\hB = {\bm \Omega}_\hM{}_\hA{}^\hB$.

We will shortly give an explicit construction of the linear fluctuations of the supervielbein (restricted to $4|4+7$ superspace)  in terms of the prepotential constituents introduced in section \ref{S:Sect2}. A guiding principle in this construction is the transformation properties of these objects under diffeomorphisms and local Lorentz transformations. Linearizing the diffeomorphism parameter, we take $\hat\xi^\hM = \mathring \xi^\hM + {\bm \xi}^\hM$. The indices can be flattened with the background vielbein. We ignore background diffeomorphisms $\mathring \xi^\hM$, as these will reduce to the isometries that preserve \eqref{E:11DBackground.E}. Consequently, the linearized transformation rules of $\bm H$ and $\bm \Omega$ are
\begin{align}\label{E:11DLinearizedTrafo}
	\delta {\bm H}_\hB{}^\hA &= - {\bm L}_\hB{}^\hA + D_\hB{\bm \xi}^\hA + {\bm \xi}^\hC \mathring T_{\hC\hB}{}^\hA~, \qquad
	\delta {\bm \Omega}_{\hM\,\hB}{}^\hA = \pa_\hM {\bm L}_\hB{}^\hA~.
\end{align}
Here, ${\bm L}_\hB{}^\hA$ are Lorentz parameters.

Next, we denote by ${\bm T}_{\hC \hB}{}^\hA$ the linearized fluctuations of the tangent space components of torsion, i.e.
\begin{align}
\hat T_{\hC \hB}{}^\hA = \mathring T_{\hC \hB}{}^\hA  + {\bm T}_{\hC \hB}{}^\hA~.
\end{align}
From the definition of torsion, it follows that
\begin{align}
\label{E:linearized-torsion-defn}
{\bm T}_{\hC\hB}{}^\hA  &= 2\, \hat D_{[\hC} {\bm H}_{\hB]}{}^\hA
    + 2 \,{\bm \Omega}_{[\hC \hB]}{}^\hA  + \mathring T_{\hC\hB}{}^\hD {\bm H}_\hD{}^\hA
    - 2\,{\bm  H}_{[\hC}{}^\hD \mathring T_{|\hD | \hB]}{}^\hA~,
\end{align}
which is invariant under the linearized transformations \eqref{E:11DLinearizedTrafo}. The linearized curvature tensor is
\begin{equation}
{\bm R}_\hB{}^\hA = \rd {\bm \Omega}_\hB{}^\hA
\end{equation}
which in components becomes
\begin{equation}
{\bm R}_{\hD \hC}{}_\hB{}^\hA = 2 \,D_{[\hD} {\bm \Omega}_{\hC] \hB}{}^\hA
    + \mathring T_{\hD \hC}{}^\hF {\bm \Omega}_\hF{}_\hB{}^\hA~.
\end{equation}
The linearized torsion and curvature Bianchi identities read
\begin{align}
\hat D_{[\hD} {\bm T}_{\hC \hB]}{}^\hA + \mathring T_{[\hD \hC}{}^\hF {\bm T}_{|\hF| \hB]}{}^{\hA}
    + {\bm T}_{[\hD \hC}{}^\hF \mathring T_{|\hF |\hB]}{}^{\hA}
    = {\bm R}_{[\hD \hC \hB]}{}^\hA~, \qquad
\hat D_{[\hE} {\bm R}_{\hD \hC] \hB}{}^\hA = 0~.
\end{align}

We will also need similar results for the linearized 3-form and its 4-form field strength. For the background 4-form, we take
\begin{equation}
	\mathring G_{\halpha \hbeta \hc \hd} = 2 (\Gamma_{\hc \hd})_{\halpha \hbeta}~,
\end{equation}
and all other components of $\mathring G$ are set to zero.
We denote by ${\bm C}$ the linearized fluctuation of cotangent space components of the 3-form, i.e. $\hat C = \mathring C + \bm{C}$. We expand this fluctuation in the background frames,
\begin{align}
	{\bm C} &= \tfrac 1{3!} \mathring E^\hA \wedge \mathring E^\hB \wedge \mathring E^\hC
        \,\,{\bm C}_{\hC \hB \hA} 
\end{align}
The linearized fluctuation to the cotangent space components of $G_4$ is denoted by ${\bm G}_{\hD \hC \hB \hA}$, so that
\begin{align}
{\bm G}_{\hD\hC\hB\hA} = 4 D_{[\hD} {\bm C}_{\hC \hB \hA]} + 6 \,\mathring T_{[\hD \hC}{}^\hF {\bm C}_{|\hF| \hB \hA]}
    - 4 {\bm H}_{[\hD|}{}^\hF \mathring G_{\hF| \hC \hB \hA]}~.
\end{align}
This is invariant under the linearized diffeomorphisms and gauge transformations
\begin{align}
\delta {\bm C}_{\hM \hN \hP}
    &= 3 \pa_{[\hM} {\bm \Lambda}_{\hN \hP]}
    + {\bm \xi}^R \mathring G_{\hR \hM \hN \hP}~.
\end{align}
The Bianchi identity that $\bm G$ obeys can be shown to be
\begin{align}
\frac{1}{4!} D_{[\hE} {\bm G}_{\hD \hC \hB \hA]}
    + \frac{1}{2!} \frac{1}{3!} \mathring T_{[\hE \hD}{}^{\hF} {\bm G}_{|\hF |\hC \hB \hA]}
    + \frac{1}{2!} \frac{1}{3!} {\bm T}_{[\hE \hD}{}^{\hF} \mathring G_{|\hF |\hC \hB \hA]}
    = 0.
\end{align}
The fact that the background solves the superspace Bianchi identities can be explicitly checked. The only non-trivial check is the $G_4$ Bianchi identity involving $\mathring G_{\halpha \hbeta \hc \hd}$, equivalent to
\begin{align}
	(\hat \Gamma_\ha)_{(\halpha \hbeta} (\hat \Gamma^{\ha \hb})_{\hgamma \hdelta)} &= 0~,
\end{align}
which is the fundamental 11D gamma matrix identity.

\subsection{Reduction to $4|4 + 7$ superspace}
Having described linearization around a background, we now move on to the second key step -- reduction to a $4|4+7$ superspace. Choosing a global 3-form \eqref{E:varphi-def} splits the bosonic background into a 4 dimensional ``external" space $\mathbb R^4$ and a 7 dimensional ``internal" space $\mathbb R^7$. We can use this structure to pick out four special fermionic directions in the following manner. The background now has a reduced structure group $SO(3,1) \times G_2$. 
A spinor $\psi^\halpha$ of $SO(10,1)$ decomposes under a reduction to $SO(3,1) \times SO(7)$ as 
\begin{align}
	\psi^\halpha &= (\psi^{\alpha \, I}, \bar \psi_{\dalpha \, I})~,
\end{align}
where $\alpha, \dalpha$ are indices of Weyl spinors of $SO(3,1)$, and $I$ is the index of an 8-component spinor of $SO(7)$. Breaking $SO(7)$ further to its $G_2$ subgroup results in $\boldsymbol 8_{SO(7)} = \boldsymbol 1_{G_2} + \boldsymbol 7_{G_2}$ in the following way:
\begin{subequations}
	\begin{align}
		\psi^{\alpha I} &= \eta^I \psi^{\alpha} + i (\Gamma_\um \eta)^I \psi^{\um \alpha} \\
		\psi_{\dalpha I} &= \eta_I \psi_{\dalpha} + i (\Gamma^\um \eta)_I \psi_{\um \dalpha}~.
	\end{align}
\end{subequations}
The constant real spinor $\eta^I$ of unit norm defines the embedding of $G_2$ into $SO(7)$ and $\Gamma^\um$ are the $SO(7)$ gamma matrices. Our conventions for these are discussed in appendix \ref{A:Gamma-matrices} and the $G_2$ embedding is elaborated a bit in appendix \ref{A:AppendixC}. The real $G_2$ structure $\varphi_{\ul{mnp}}$ is given in terms of $\eta$ by
\begin{align}
	\varphi_{\ul{mnp}} &= i \eta^\rT \Gamma_{\ul{mnp}} \eta~, \qquad \psi_{\ul{mnpq}} = \eta^\rT \Gamma_{\ul{mnpq}} \eta~.
\end{align}
Applying this decomposition to the fermionic coordinates $\theta^{\hat \mu}$ of 11D $N=1$ superspace gives
\begin{align}
	\theta^{\hat \mu} = (\theta^\mu, \theta^{\um \mu}, \bar \theta_{\dot \mu}, \bar \theta_{\um \dot \mu})~.
\end{align}
Now we are in a position to define our $4|4+7$ superspace. It is the hypersurface
\begin{align}
	\theta^{\um \mu} = 0 = \bar \theta_{\um \dot \mu}~.
\end{align}
For consistency, the following one forms also get set to zero:
\begin{align}
	\rd \theta^{\um \mu} = 0 = \rd \bar \theta_{\um \dot \mu}~.
\end{align}
The bosonic coordinates undergo the simple split $x^\hm = (x^m, y^\um)$. Together with $\theta^\mu, ~ \bar \theta_{\dot \mu}$, they form the coordinates of our $4|4+7$ superspace: $z^\hM \longrightarrow z^\hM |_{4|4+7} = (z^M, y^\um)$, where $z^M = (x^m, \theta^\mu, \bar \theta_{\dot \mu})$ are the usual 4D $N=1$ superspace coordinates. The background superframe becomes
\begin{subequations}
\begin{align}
\mathring E^a &= \rd x^a
    + \theta^\alpha (\gamma^a)_{\alpha \dbeta} \rd \bar \theta^\dbeta
    + \bar\theta_\dalpha (\gamma^a)^{\dalpha \beta} \rd \theta_\beta~, \\
\mathring E^\um &= \rd y^\um~, \\
\mathring E^\alpha &= \rd \theta^\alpha~, \\
\mathring E^{\um\alpha} &= 0~.
\end{align}
\end{subequations}
We emphasize that the projection to $4|4+7$ superspace means that form indices of all geometric objects will run through the range of $4|4+7$. In particular, forms will not have legs along the ``extra" fermionic directions. However, we still keep the full (background + linear fluctuation) ``superframe" $\hat E^\hA = (\hat E^\ha, \hat E^\halpha)$ as a collection of $11+32$ one forms:
\begin{align}
	\hat E^\hA = (\hat E^a, \hat E^{\underline a}, \hat E^\alpha, \hat E^{\underline a \alpha}, \hat E_\dalpha, \hat E_{\underline a \dalpha})~.
\end{align} 
Here $\hat E^A = (\hat E^a, \hat E^\alpha, \hat E_\dalpha)$ is the $4|4$ supervielbein, $\hat E^{\underline a}$ is a Kaluza-Klein photon, and $\hat E^{\underline a \alpha}, ~ \hat E_{\underline a \dalpha}$ are seven additional gravitini. All of these have form legs only along the $4|4+7$ superspace. This is not a vielbein in the traditional sense since it is rectangular. 


An important comment about the internal bosonic indices is in order. There are three different indices that take seven values in the non-linear theory -- the internal coordinate index $\um$, the $SO(7)$ vector index $\underline a$, and a $G_2$ index, say $i$. We have already implicitly set $\underline a = i$ by convention. The remaining two indices $\um$ and $\underline a$ are related by the vielbein $\hat E_\um {}^{\underline a}$. In the linearized theory, the background value for this internal component being $\delta_\um {}^{\underline a}$, we can identify $\underline a$ with $\um$.

\subsection{Components of the 4D supervielbein and spin connection}
\label{S:Sect3.4-4d-supervielbein}
In the remainder of section \ref{S:Sect3}, we will give explicit expressions for the linear fluctuations of the vielbein, spin connection, 3- and 4-form etc. in terms of the off-shell prepotentials described in section \ref{S:Sect2}. These definitions will be guided by how these linear fluctuations transform under different symmetries, and by torsion constraints. 

In this subsection, we consider the purely 4D part ${\bm H}_B{}^A$. This should be built out of $H_{\alpha \dalpha}$ and $G$ and $\bar G$. There are actually a number of ways of doing this, and different choices are related to how one defines the various transformation parameters. A simple choice is \cite{Binetruy:2000zx}
\begin{align}
{\bm H}_{\alpha}{}^{\dbeta \beta}
    := i D_\alpha H^{\dbeta \beta}~, \qquad
{\bm H}_\alpha{}^\beta := \delta_\alpha{}^\beta H~, \qquad
{\bm H}_{\alpha \, \dbeta} := 0~,
\end{align}
where
\begin{align}
H = \frac{1}{12} D_\alpha \bar D_\dalpha H^{\dalpha \alpha}
    - \frac{i}{6} \pa_{\alpha \dalpha} H^{\dalpha \alpha}
    - \frac{1}{6} G
    + \frac{1}{3} \bar G~.
\end{align}
These transform consistently as
\begin{subequations}
\begin{align}
\delta {\bm H}_\alpha{}^{\dbeta \beta}
    &= D_\alpha {\bm \xi}^{\dbeta \beta} + 4i \, \delta_\alpha{}^\beta \, \bar {\bm\veps}^{\dbeta}~, \\
\delta {\bm H}_\alpha{}^\beta
    &= D_\alpha {\bm \veps}^\beta -{\bm  L}_\alpha{}^\beta
    = \frac{1}{2} \delta_\alpha{}^\beta D_\gamma {\bm \veps}^\gamma~, \\
\delta {\bm H}_\alpha{}_\dbeta
    &= D_\alpha \bar{\bm \veps}_{\dbeta} = 0~,
\end{align}
\end{subequations}
where we identify the linearized parameters
\begin{subequations}
\begin{align}
{\bm \xi}_{\alpha \dalpha} &:= -i (D_\alpha \bar L_\dalpha + \bar D_\dalpha L_\alpha)~, \\
{\bm \veps}_{\alpha} &:= -\frac{1}{4} \bar D^2 L_\alpha~, \\
{\bm L}_{\alpha\beta} &:= D_{(\alpha} {\bm \veps}_{\beta)}~.
\end{align}
\end{subequations}
The other components of ${\bm H}_B{}^A$, and the linearized spin connection, are obtained by solving the expression for the purely 4D part of the linearized torsion,
\begin{align}
	{\bm T}_{CB}{}^A  &= 2\, D_{[C} {\bm H}_{B]}{}^A + 2 \bm \Omega_{[C B]}{}^A  + \mathring T_{CB}{}^D {\bm H}_D{}^A - 2\, {\bm H}_{[C}{}^D \mathring T_{| D | B]}{}^A~,
\end{align}
subject to the constraints
\begin{equation}
\begin{split}
\textrm{dimension~} 0 : ~~~~ &{\bm T}_{\underline{\alpha \beta}} {}^c =0~, \\
\textrm{dimension~} \tfrac 12 : ~~~~&{\bm T}_{\underline{\alpha \beta}} {}^{\underline \gamma}=0~,  \quad {\bm T}\indices{_{\underline \alpha}_b^c}={\bm T}\indices{_a_{\underline \beta}^c}=0~, \\
\textrm{dimension~} 1 : ~~~~&{\bm T}\indices{_a_b^c}=0~,
\end{split}
\end{equation}
which correspond to the linearized version of eq.~(14.25) of ref.~\cite{Wess:1992cp}.

First, let us consider the ${\bm T}^a$ conditions. From ${\bm T}_{\beta \dbeta}{}^a = 0$, we obtain
\begin{align}
{\bm H}_{\beta \dbeta}{}^{\dalpha \alpha}
    &= \frac{1}{2} [D_\beta, \bar D_\dbeta] H^{\dalpha \alpha}
    - \delta_\beta{}^\alpha \delta_\dbeta{}^\dalpha \Big[
    \frac{1}{3} (G + \bar G)
    + \frac{1}{6} [D_\gamma, \bar D_\dgamma] H^{\dgamma \gamma}
    \Big]~.
\end{align}
This transforms consistently as
\begin{align}
	\delta {\bm H}_b{}^a &= \pa_b {\bm \xi}^a - {\bm L}_b{}^a~,
\end{align}
with $\bm L_{ba} = - (\sigma_{ba})^{\alpha \beta} \bm L_{\alpha \beta} + (\bar \sigma_{ba})^{\dalpha \dbeta} \bar{\bm L}_{\dalpha \dbeta}$. The condition ${\bm T}_{\gamma \beta}{}^a = 0$ is already satisfied. Alternatively, this condition would have told us $\bm H_\alpha{}_\dbeta=0$ if we hadn't already fixed it. From ${\bm T}_{\dgamma b} {}^a = 0 = \bm T_{\gamma b} {}^a$, one finds that
\begin{subequations}
\label{E:Omega4d-fermion-leg}
	\begin{align}
	{\bm \Omega}_{\dgamma b}{}^a &= -\bar D_\dgamma {\bm H}_b {}^a + \pa_b {\bm H}_\dgamma {}^a - 2i {\bm H}_b {}^\beta (\sigma^a)_{\beta \dgamma}~. \\
	\bm \Omega_{\gamma b} {}^a &= - D_\gamma {\bm H}_b{}^a + \pa_b {\bm H}_\gamma {}^a + 2i {\bm H}_b {}^\dbeta (\sigma^a)_{\gamma \dbeta}~.
	\end{align}
\end{subequations}
The components $\bm H_b {}^{\underline \beta}$ on the right hand side of the above equations have not been defined yet. We note that, by virtue of \eqref{E:spin-connection-lorentz-valued}, $\bm \Omega_{\underline \gamma, \, \beta \alpha}$ and $\bm \Omega_{\underline \gamma, \, \dbeta \dalpha}$ are derived from $\bm \Omega_{\underline \gamma, \, ba}$. From ${\bm T}_{c b}{}^a = 0$, one determines ${\bm \Omega}_{cb}{}^a$ in the usual manner.

Next, we consider the ${\bm T}^\alpha$ conditions. ${\bm T}_{\dgamma \dbeta}{}^\alpha = 0$ holds identically. $\bm T_{\dgamma \beta}{}^\alpha = 0$ leads to
\begin{align}
	{\bm \Omega}_{\dgamma \beta}{}^\alpha = - \bar D_\dgamma \bm H_\beta {}^\alpha - 2 i {\bm H}_{\beta \dgamma} {}^\alpha~.
\end{align}
The two equations for this component of the spin connection imply
\begin{subequations}
\label{E:H_b,alpha--4DOmega-ferm-leg}
\begin{align}
	{\bm H}_{\beta \dbeta}{}^\alpha &= \frac{i}{8} \bar D^2 D_\beta H_\dbeta{}^\alpha - i \delta_\beta{}^\alpha \bar D_\dbeta \bar H~, \\
	{\bm \Omega}_{\dgamma, \, \beta \alpha} &= \frac{1}{4} \bar D^2 D_{(\beta} H_{\alpha) \dgamma}~, \\
	{\bm \Omega}_{\dgamma, \, \dbeta \dalpha} &= -2 \eps_{\dgamma (\dbeta} \bar D_{\dalpha)} \bar H~,
\end{align}
\end{subequations}
and these transform as
\begin{align}
\delta \bm H_{\beta \dbeta} {}^\alpha &= \pa_{\beta \dbeta} \bm \veps^\alpha~, \qquad
\delta {\bm \Omega}_{\gamma, \, \beta \alpha} = D_\gamma {\bm L}_{\beta \alpha}~, \qquad
\delta {\bm \Omega}_{\dgamma, \, \beta \alpha} = \bar D_\dgamma {\bm L}_{\beta \alpha}~.
\end{align}
The remaining conditions on the torsion tensor are solved very similarly to chapter XV of ref.~\cite{Wess:1992cp}, and using the definitions for $R$, $G_{\alpha \dalpha}$, and $W_{\gamma\beta\alpha}$ in section \ref{S:Sect2.3} all conditions are satisfied in a lengthy computation we do not repeat here.

\subsection{Extending the supervielbein}
\label{S:Sect3.5-extending-supervielbein}
Let us now compute the following components of the fluctuations of the supervielbein
\begin{align}
	({\bm H}_{\hB}{}^{\hA}) {\big|}_{4|4+7} = \begin{pmatrix}
									{\bm H}_B{}^A & {\bm H}_B{}^\um \\
									{\bm H}_\un{}^A & {\bm H}_\un{}^\um
								  \end{pmatrix}
							\equiv
								  \begin{pmatrix}
									{\bm H}_B{}^A & {\bm \cA}_B{}^\um \\
									{\bm \chi}_\un{}^A & {\bm H}_\un{}^\um
								   \end{pmatrix}~.
\end{align}
Typically, in a Kaluza-Klein setting, the bosonic component ${\bm \chi}_\un{}^a$ in the lower left block would vanish. This would involve the gauge-fixing of $SO(10,1)$ to its $SO(3,1) \times SO(7)$ subgroup. It turns out that, in the 4D $N=1$ superfield setting that we are employing, this gauge choice is extremely inconvenient while maintaining the irreducible superfield content we have identified. That is, we find
\begin{align}
\label{aii}
	{\bm H}_\um{}^{\dalpha \alpha} = {\bm \chi}_\um{}^{\dalpha \alpha} = - \frac{1}{2} ( \bar D^\dalpha \Psi_\um{}^\alpha - D^\alpha \bar\Psi_\um{}^\dalpha)
\end{align}
and setting this to vanish would impose an awkward constraint on $\Psi_\um^\alpha$. This transforms as
\begin{align}\label{ai}
	\delta {\bm \chi}_\um{}^{\dalpha \alpha} = \pa_\um{\bm \xi}^{\dalpha \alpha} - {\bm L}_\um{}^{\dalpha \alpha}~,
\end{align}
where
\begin{align}\label{aiii}
	{\bm L}_\um{}^{\dalpha \alpha} &= \frac{1}{2} \bar D^\dalpha D^\alpha \Omega_\um - \frac{1}{2} D^\alpha \bar D^\dalpha \bar \Omega_\um
\end{align}
has the obvious interpretation of being the higher-dimensional $SO(10,1)$ transformation that has not been gauge-fixed. We will sometimes refer to this as the mixed Lorentz parameter. Next, we define
\begin{align}
	2 {\bm \chi}_{\um, \alpha} =  \psi_{\um, \alpha} = \tfrac{i}{4} \Big[ \bar D^2 \Psi_{\um \alpha} + \tfrac{2i}{3} \Big(D_\alpha \hat H_\um + 2 D^\dalpha X_{\um \alpha \dalpha} \Big) - \tfrac{8}{3} \cW_\alpha{}_\um \Big] + d_1\, \lambda_{\um \alpha} , 
\end{align}
where $d_1$ is a constant. Since $\lambda_{\um \alpha}$ has the same engineering dimension and index structure as the other terms, its addition corresponds merely to a field redefinition of the gravitino by a covariant piece. We will leverage this arbitrariness of $d_1$ to simplify certain things later in this section. $\bm \chi_\um {}^\alpha$ transforms simply as
\begin{align}
	\delta{\bm \chi}_\um{}^\alpha = \pa_\um {\bm \veps}^\alpha
\end{align}
This completes the construction of ${\bm H}_B{}^A$ and $\bm H_\um {}^A$, leaving ${\bm H}_B{}^\um$ and $\bm H_\un {}^\um$. A natural choice for ${\bm H}_\alpha{}^\um$ would seem to be
\begin{align}
	{\bm H}_\alpha{}^\um = i D_\alpha \cV^\um~.
\end{align}
 This transforms as
\begin{align}
\delta {\bm H}_\alpha{}^\um
    &= i D_\alpha \lambda^\um - \frac{i}{2} D_\alpha (\Omega^\um + \bar \Omega^\um) 
    = D_\alpha \bm \xi^\um + 2 i \bm  \veps_{\um \alpha}
\end{align}
where we define
\begin{align}
	\bm \xi^\um := \tfrac i2 (\Omega^\um - \bar \Omega^\um) + i (\lambda^\um - \bar \lambda^\um)~, \qquad
	\bm \veps_{\um \alpha} := - \tfrac 12 D_\alpha \Omega_\um
\end{align}
The parameter $\bm \veps_{\um \alpha}$ describes the additional 28 supersymmetries present in 11D supergravity. We also choose
\begin{align}
	\bm H_{\alpha \dalpha}{}^\um &= \frac{1}{2} [D_\alpha, \bar D_\dalpha] \cV^\um = -\Delta_{\alpha \dalpha} \cV^\um
\end{align}
which transforms as
\begin{align}
\delta \bm H_{\alpha \dalpha}{}^\um
    = i \pa_{\alpha \dalpha} (\lambda^\um - \bar \lambda^\um)
    + \frac{1}{2} \Delta_{\alpha \dalpha} (\Omega^\um + \bar \Omega^\um) 
    = \pa_{\alpha \dalpha} \bm \xi^\um - \bm L_{\alpha \dalpha}{}^\um
\end{align}
where $\bm L_{a}{}^\um = - \bm L^{\um}{}_a$ correctly accounts for the Lorentz transformation of this component with a mixed Lorentz parameter.

Finally, we consider $\bm H_\un{}^\um$. We choose a gauge where this is symmetric, implying that it can be identified with the internal metric.
\begin{align}
\bm H_\un{}^\um = \frac{1}{2} g_\un{}^\um = \tfrac 14 \varphi_{ \ul{rs} ( \un}  F^{\um) \ul{rs}} - \tfrac 1{36} \delta_\un^\um \varphi^{\ul{pqr}} F_{\ul{pqr}} 
\end{align}
This transforms as
\begin{align}
\label{E:delta-g_nm}
\delta \bm H_\un{}^\um = \pa_\un \bm \xi^\um - \bm L_\un{}^\um~, \qquad
\bm L_{\un \um} := \pa_{[\un} \bm \xi_{\um]}~.
\end{align}
This completes the construction of the linearized supervielbein with all indices restricted to $4|4+7$. The extra gravitini in $\bm H^{\um \underline \alpha}$ will be constructed in the next subsection. Symmetrization of the bosonic components of the vielbein defines the linearized 11D metric. We don't give a detailed account of this here, except to introduce
\begin{align}
	g_{\alpha \ul{mn}} &:= D_\alpha \left[ g_{\ul{mn}} - 2i \pa_{(\um} \cV_{\un )} \right]
\end{align}
which is a $\lambda^\um$ -invariant quantity constructed from the spinor derivative of $g_{\um \un}$. In fact, it is the $\bm 1 + \bm{27}$ $G_2$ projection of $F_{\alpha \ul{mnp}}$ defined in \eqref{E:F_alpha-mnp--lambda-inv}. Its complex conjugate is denoted by $g_{\dalpha \um \un}$. The $\bm 7$ piece $F_{\alpha \um}$ of $F_{\alpha \ul{mnp}}$ descends from the $\bm 7$ piece $F_\um$ of $F_{\ul{mnp}}$:
\begin{align}
	F_\um := - \tfrac 1{12}  \psi_{\ul{mnpq}} F^{\ul{npq}}~, \qquad
	F_{\alpha \um} := - \tfrac 1{12}  \psi_{\ul{mnpq}} F_\alpha {}^{\ul{npq}}	
\end{align}
The complex conjugate of $F_{\alpha \um}$ is denoted by $F_{\dalpha \um}$. The quantities $g_{\alpha \um \un}$ and $F_\um$ will be useful in the construction of components of the extra gravitini below.

\subsection{Extra gravitini}
\label{S:Sect3.6-extra-gravitini}
Now we focus on the so far undefined seven extra gravitini $\psi^{\um \alpha}$. We define components of $\psi^{\um \alpha}$ with bosonic form legs,
\begin{subequations}
	\begin{align}
		\bm H_b {}^{\un \alpha} &:= \frac{1}{2} \psi_b {}^{\un \alpha},  \\
		\psi_{\beta \dbeta, \um \alpha} &:= D_{(\alpha} X_{|\um| \beta) \dbeta} + \epsilon_{\beta \alpha} \left(  \tfrac 13 \bar D_\dbeta \hat H_\um + \tfrac {2i}3 \bar \cW_{\dbeta \um} + \tfrac 16 D^\gamma X_{\um \gamma \dbeta}  \right) + d_2 \, \epsilon_{\beta \alpha} \bar \lambda_{\um \dbeta} 
	\end{align}
\end{subequations}
and
\begin{subequations}
	\begin{align}
		\bm H_\um{}^{\un \alpha} &:= \frac{1}{2} \psi_\um{}^{\un \alpha}, \\
		\psi_{\um, \un \alpha} &:= \tfrac i2 g_{\alpha \um \un} - \tfrac 12 \left( \hat W_{\alpha \um \un} - \tfrac 16 \varphi_{\ul{mnp}} \varphi^{\ul{prs}} \hat W_{\alpha \ul{rs}}  \right) - \tfrac i{72} \varphi_{\ul{mnp}}  \psi^{\ul{pqrs}} F_{\alpha \ul{qrs}} \nonumber \\
		~~ & + \tfrac 16 \varphi_{\ul{mnp}} D_\alpha \hat H^\up + d_3 \, \varphi_{\ul{mnp}} \lambda^\up {}_\alpha~.
	\end{align}
\end{subequations}
Again, the $\lambda_{\um \alpha}$ terms come with (as of now) arbitrary constants $d_2$ and $d_3$. These components of the gravitini transform as
	\begin{align}
		\delta \psi_{\beta \dbeta, \um \alpha} = 2 \, \pa_{\beta \dbeta} \bm \veps_{\um \alpha}~, \qquad
		\delta \psi_{\um, \un \alpha} = 2 \, \pa_\um \bm \veps_{\un \alpha}
	\end{align}
The remaining components are $\psi_\alpha {}^{\un \beta}$ and $\psi_\dalpha {}^{\un \beta}$.
These are given by
\begin{subequations}
\begin{alignat}{2}
\bm H_\alpha{}^{\un \beta} &:= \frac 12 \psi_\alpha {}^{\un \beta} 
    :=  - \tfrac{1}{4} \delta_\alpha {}^\beta \left( F^\un - i \hat H^\un \right)~, 
    &\qquad
\bm H^{\dalpha , \un \beta} &:= \frac 12 \psi^{\dalpha, \un \beta} :=  -\frac{i}{2} X^{\un \dalpha \beta}~, \\
\delta \bm H_\alpha{}^{\un \beta}
    &= D_\alpha \bm \veps^{\un \beta} - \tfrac{1}{4} \delta_\alpha{}^\beta \varphi^{\ul{npq}} \bm L_{\ul{pq}}~, &\qquad
\delta \bm H^{\dalpha , \un \beta} &=  \bar D^\dalpha \bm \veps^{\un \beta} + \frac{1}{2} \bm L_\un {}^{\beta \dalpha}~.
\end{alignat}
\end{subequations}

In addition to linearized SUSY transformations, these components also undergo internal Lorentz transformations. These higher Lorentz transformations naturally descend from 11D superspace in the following manner. Recall that the linearized torsion on the extended space is of the form \eqref{E:linearized-torsion-defn}.
The linearized spin connection $\bm \Omega$ is valued over all of $SO(10,1)$. Reducing this to $4|4+7$, the form indices $\hC$ and $\hB$ will run over $4|4+7$ superspace, while $\hA$ will include the extra gravitini as well. Let us look at the cases when $\hA$ is a spinor index. Looking only at the spin connection contributions, we find
\begin{align}
	\bm T_{\hC \hB}{}^\alpha &\sim \tfrac{1}{4} \delta_\hB{}^\beta \, \bm \Omega_{\hC de} (\gamma^{de})_\beta{}^\alpha - \hC \leftrightarrow \hB~, \\
	\bm T_{\hC \hB}{}^{\um\alpha} &\sim + \tfrac{1}{2} \delta_{\hB \dbeta} \, \bm \Omega_{\hC \, d} {}^\um (\sigma_{d})^{\dbeta \alpha} + \tfrac{1}{4} \delta_\hB{}^\beta \, \varphi^{\um \un \up} \bm \Omega_{\hC \un \up} - \hC \leftrightarrow \hB~.
\end{align}
Here, we have used 11D gamma matrices outlined in Appendix \ref{A:Gamma-matrices}. The first equation with $\hA = \alpha$ just means that the torsion tensor $\bm T^\alpha$ gets no contribution from spin connections with either one or both Lie algebra indices along the internal manifold. This is consistent with the fact that the vielbeins involved in this torsion do not suffer Lorentz transformations with $\bm L_{a \um}$ and $\bm L_{\um \un}$. However, $\bm T^{\um \alpha}$ gets spin connection contributions of these kinds to negate the $\bm L_{a \um}$ and $\bm L_{\um \un}$ transformations of $\bm H_\hB{}^{\um \alpha}$
\begin{align}
	\delta \bm H_{\hB}{}^{\um\alpha} |_{\textrm{Lorentz}} &= - \tfrac{1}{2} \delta_{\hB \dbeta} \, \bm L_{d} {}^\um (\sigma_{d})^{\dbeta \alpha} - \tfrac{1}{4} \delta_\hB{}^\beta \, \varphi^{\ul{mnp}} \bm L_{\ul{np}}~.
\end{align}
This concludes the construction of the entire supervielbein.


\subsection{Extending the linearized spin connection}

The purely 4D components of $\bm \Omega$ were determined in section \ref{S:Sect3.4-4d-supervielbein}. The remaining spin connection components will be given now. Being valued in $SO(10,1)$, the only independent spin connections are $\bm \Omega^{\hb \ha}$, and the rest are determined in terms of them: $\bm \Omega^{\hbeta \halpha} = \tfrac 14 (\Gamma_{\hb \ha})^{\hbeta \halpha} \bm \Omega^{\hb \ha}$, ~ $\bm \Omega^{\hb \halpha} = 0$. One guiding principle in their definitions can be their transformation rules under local $SO(10,1)$: $\delta \bm \Omega_{\hC} {}^{\hb \ha} = D_{\hC} \bm L^{\hb \ha}$, with the form index $\hC$ restricted to $4|4+7$. Another approach is to give definitions of torsion components, which are covariant quantities, and derive from them the spin connections. We will take the second approach.

As in ordinary general relativity, setting the bosonic components $\bm T_{\hc \hb}{}^\ha = 0$ lets one determine the purely bosonic part of the spin connection as
\begin{align}
	\bm \Omega_{\hc \hb \ha} &= -\frac{1}{2} \Big( K_{\hc \hb \ha} + K_{\ha \hb \hc} + K_{\ha \hc \hb} \Big)~,
\end{align}
where $K_{\ha \hb \hc} = \pa_\ha \bm H_{\hb \hc} -\pa_\hb \bm H_{\ha \hc}$ are the anholonomy coefficients. 

The rest of $\bm T^\ha$ is going to involve parts with fermionic legs. The purely 4D components $\bm \Omega_{\underline \gamma, ~ \underline{\alpha \beta}}$ are in \eqref{E:H_b,alpha--4DOmega-ferm-leg}. Now we look at the cases where at least one Lie algebra index is internal. Consider $\bm T^a$, with one fermionic form leg and one internal leg. This is
\begin{align}
	\bm T_{\beta \um}{}^a &= D_\beta \bm H_\um{}^a - \pa_\um \bm H_\beta{}^a  + \bm \Omega_\beta{}_\um{}^a + 2i \bm H_{\um, \dgamma} (\sigma^a)_\beta {}^\dgamma~.
\end{align}
This component of the torsion tensor is dimension $\tfrac 12$ and can include $\lambda_\um{}^\alpha$. We rewrite
\begin{align}
\label{E:Omega_beta-m^a-I}
	 \bm \Omega_\beta{}_\um{}^a &= - \bm T_{\um \beta}{}^a + \pa_\um \bm H_\beta{}^a - D_\beta \bm H_\um{}^a - 2i \bm H_{\um, \dgamma} (\sigma^a)_\beta {}^\dgamma~.    
\end{align}
We also notice that, from the definition of $\bm T_{\beta a} {}^\um$,
\begin{align}
\label{E:E:Omega_beta-m^a-II}
	\bm \Omega_\beta{}_a {}^\um  = - \bm T_{a \beta}{}^\um + \pa_a \bm H_\beta{}^\um - D_\beta \bm H_a{}^\um - 2i \delta^{\um \un} \bm H_{a, \un \beta}~.
\end{align}
This must be opposite in sign from the previous expression \eqref{E:Omega_beta-m^a-I}, so we get
\begin{align}
	\bm T_{\um \beta, a} + \bm T_{a \beta, \um} &= \pa_a \bm H_{\beta, \um} - D_\beta \bm H_{a, \um} + \pa_\um \bm H_{\beta, a} - D_\beta \bm H_{\um, a} - 2i \bm H_{\um \dgamma} (\sigma_a)_\beta{}^\dgamma - 2i \bm H_{a, \um \beta}
    ~,
\end{align}
which is independent of the spin connection. We can write out the right hand side explicitly from the given definitions of the supervielbein,
\begin{align}
	(\sigma^a)_{\alpha \dalpha} (\bm T_{\um \beta, a} + \bm T_{a \beta, \um} ) &= i\, \eps_{\beta \alpha} (2d_1 + d_2) \bar\lambda_{\um \dalpha}~.
\end{align}
We choose to populate $\bm T_{a \beta} {}^\um$ by $\lambda_{\um \alpha}$, with an undetermined proportionality factor $d_4$, and consequently
\begin{align}
	\bm T_{a \beta}{}^\um &= i \,d_4\, (\sigma_a)_{\beta \dbeta} \bar\lambda^{\dbeta \um}~, \\
	\bm T_{\um \beta}{}^a &= i \,(d_1 + \tfrac{1}{2} d_2 - d_4) (\sigma^a)_{\beta \dbeta} \bar\lambda^{\dbeta} {}_\um ~, \\
	\bm \Omega_{\beta a}{}^\um &= (\sigma_a)_{\beta \dbeta} (\bar \cW^{\dbeta \um} + i \,d_4 \bar \lambda^{\dbeta \um}) - 2 i \, \bm H_{a, \beta}{}^\um~.
\end{align}
The definition of $\bm T_{\un \beta}{}^\um$ gives
\begin{align}
	\bm T_{\un \beta}{}^\um &= \pa_\un \bm H_\beta{}^\um - D_\beta \bm H_\un{}^\um - \bm \Omega_{\beta}{}_\un{}^\um - 2i\, \delta^{\ul{mp}} \bm H_{\un, \up \beta}~.
\end{align}
The symmetric part of this in $\un \um$ gives the SUSY transformation of $g_{\un \um}$ with extra SUSY parameters $\bm \veps_{\um \alpha}$. Because there are no such terms in \eqref{E:delta-g_nm}, this symmetric part is zero. The antisymmetric part amounts to a choice of $\bm \Omega_\beta{}_\un{}^\um$, and its $\rep{7}$ part can be chosen to have a $\lambda$ piece. We find
\begin{align}
	\bm T_{\beta \un}{}^\um &= - \bm T_{\un \beta }{}^\um = i \,d_5 \, \varphi_{\un}{}^{\um \up} \lambda_{\up \beta}~, \\
	\bm \Omega_{\beta \um \un} &= i (d_5 - d_3) \varphi_{\um \un}{}^\up \lambda_{\up \beta} + i D_\beta \pa_{[\um} \cV_{\un]}+ \tfrac{i}{2} (\hat W_{\beta \um \un} - \tfrac{1}{6} \varphi_{\ul{mnp}} \varphi^{\ul{pqr}} \hat W_{\beta \ul{qr}} )
    \eol & \quad
    - \tfrac{1}{72} \varphi_{\ul{mnp}} \psi^{\ul{pqrs}} F_{\beta \ul{qrs}} - \tfrac{i}{6} \varphi_{\ul{mnp}} D_\beta \hat H_\up~.
\end{align}
Again, $d_5$ is undetermined. For the dimension zero components, we find as expected
\begin{align}
	0 &= \bm T_{\alpha \beta}{}^\um = 2 D_{(\alpha} \bm H_{\beta)}{}^\um~, \\
	0 &= \bm T_{\alpha \dbeta}{}^\um = D_\alpha \bm H_\dbeta{}^\um  + \bar D_\dbeta \bm H_\alpha{}^\um + 2i  \delta^{\um \un} \bm H_{\alpha, \un \dbeta}~.
\end{align}
At this point, we have chosen all components of the spin connection.


\subsection{Evaluating the remaining torsion components}
Torsion components for which explicit prepotential expressions have not yet been given can now be computed. The purely external parts of $\bm T^{\alpha}$ work out in the standard way. The parts with at least one internal leg, such as
\begin{align}
	\bm T_{\um \beta}{}^\alpha = \pa_\um \bm H_\beta{}^\alpha - D_\beta \bm H_\um{}^\alpha + \bm \Omega_\um{}_\beta{}^\alpha~,
\end{align}
are more complicated, and presumably do not vanish because there are invariants with that dimension and representation. We decompose this into irreps of $SL(2, \mathbb C)$:
\begin{align}
	 \bm T_{\um \beta}{}^\alpha &=: i \,\delta_\beta{}^\alpha S_\um + S_\um{}_{\beta}{}^{\alpha}
\end{align}
where the second term is traceless. Explicitly,
\begin{align}
	i \,S_\um &= \pa_\um H + \tfrac{1}{2} D^\gamma H_{\um \gamma} \eol
	&= -\tfrac{i}{6} \pa_{\alpha \dalpha} X_\um{}^{\dalpha \alpha} - \tfrac{1}{24} \bar D^2 H_\um + \tfrac{1}{24} D^2 H_\um - \tfrac{i}{6} D^\alpha \cW_\alpha{}_\um + \tfrac{1}{4} d_1 D^\alpha \lambda_{\um \alpha} \eol
	&= \tfrac{i}{6} J(\cV)_\um - \tfrac{i}{36} \bm G_{\ul{mnpq}} \varphi^{\ul{npq}} + \tfrac{1}{4} d_1 D^\alpha \lambda_{\um \alpha}
\end{align}
and
\begin{align}
	S_\um{}_{\beta\alpha} &= - D_{(\beta} \bm H_{|\um| \alpha)} + \bm \Omega_\um{}_{\beta \alpha} \eol
	&= -\tfrac{1}{12} D_{(\beta} \bar D^\dgamma X_{|\um| \alpha) \dgamma} - \tfrac{i}{4} \pa_{(\beta}{}^\dgamma X_{\um \alpha) \dgamma} + \tfrac{i}{12} D_{(\beta} \cW_{\alpha)}{}_\um - \tfrac{1}{2} d_1 D_{(\beta} \lambda_{\um \alpha)} \eol
	&= -\tfrac{i}{12} \varphi_{\um} {}^{\ul{np}} \,\bm G_{\alpha \beta \,\ul{np}} + (\tfrac{1}{24} - \tfrac{d_1}{2}) D_{(\beta} \lambda_{|\um| \alpha)}
\end{align}
Above, $J(\cV)_\um$ denotes the equation of motion of $\cV_\um$.
Next,
\begin{align}
	\bm T_{\um}{}^{\dbeta}{}^\alpha \equiv -i S_{\um c} (\bar \sigma^c)^{\dbeta \alpha} &= - \bar D^\dbeta \bm H_\um{}^\alpha \eol
	&= -\tfrac{1}{2} d_1 \bar D^\dbeta \lambda_\um{}^\alpha + \tfrac{1}{12} \bar D^\dbeta D^\alpha \hat H_\um + \tfrac{1}{12} \bar D^2 X_\um{}^{\dbeta \alpha} \eol
	&= -\tfrac{1}{2} d_1 \bar D^\dbeta \lambda_\um{}^\alpha + \tfrac{1}{12} \Big[ - \tfrac{1}{2} \bar D^\dbeta \lambda_\um{}^\alpha - \tfrac{1}{2} D^\alpha \bar \lambda_\um{}^\dbeta + 2 \tilde G^{\dbeta \alpha}{}_\um + \tfrac{i}{6} \psi_\um {}^{\ul{npq}} \bm G^{\dbeta \alpha} {}_{\ul{npq}} \Big]
\end{align}
The lowest dimension components of $\bm T^{\um \alpha}$ are with dimension=1/2:
\begin{align}
	\bm T_{\gamma \beta} {}^{\um \alpha} &= 2 D_{(\gamma} \bm H_{\beta)} {}^{\um \alpha} + \tfrac{1}{2} \varphi^{\ul{mnp}} \delta_{(\beta}{}^{\alpha} \bm \Omega_{\gamma) \ul{np}}~, \\
	\bm T_{\gamma \dbeta}{}^{\um \alpha} &= D_{\gamma} \bm H_{\dbeta}{}^{\um \alpha} + \bar D_{\dbeta} \bm H_{\gamma} {}^{\um \alpha} + 2 i \bm H_{\gamma \dbeta} {}^{\um \alpha} - \tfrac{1}{2} \bm \Omega_{\gamma} {}^{d \um} (\sigma_d)^\alpha {}_\dbeta + \tfrac{1}{4} \varphi^{\ul{mnp}} \delta_{\gamma} {}^{\alpha} \bm \Omega_{\dbeta \, \ul{np}}~, \\
	\bm T_{\dgamma \dbeta}{}^{\um \alpha} &= 2 \bar D_{(\dgamma} \bm H_{\dbeta)} {}^{\um \alpha} -  \bm \Omega_{(\dgamma} {}^{d \um} (\sigma_{|d|})^\alpha {}_{\dbeta)} ~.
\end{align}
These must be invariants, and may involve the gravitino equation of motion. The last one then must vanish. The other two can be chosen to vanish by choosing the spinor spin connections appropriately, but this means there may be
tension with other components of the torsion. We stick with our previous choices of spin connections instead. Explicitly, we find
\begin{align}
	\bm T_{\gamma \beta} {}^{\um \alpha} &= 3i\, (d_5 - d_3) \, \delta_{(\gamma}{}^\alpha  \lambda^\um {}_{\beta)}~, \\
	\bm T_{\gamma \dbeta}{}^{\um \alpha} &= i \delta_\gamma{}^\alpha \bar \lambda^\um {}_{\dbeta} \Big( -\tfrac{3}{2} (d_2 - d_3 + d_5) - d_4 \Big)~, \\
	\bm T_{\dgamma \dbeta}{}^{\um \alpha} &= 0
\end{align}
Remarkably, we can turn off all dimension $\tfrac 12$ torsion components by setting all of the $d_i$ to zero. We make this convenient choice. Explicit expressions for the remaining dimension 1 components can be similarly obtained by expanding the right hand sides of
\begin{subequations}
	\begin{align}
	\bm T_{b \gamma} {}^{\um \alpha} &= \pa_b \bm H_{\gamma} {}^{\um \alpha} - D_\gamma \bm H_b {}^{ \um\alpha} + \bm \Omega_{b,  \gamma} {}^{\um \alpha}~, \\
	\bm T_{b \dgamma} {}^{\um \alpha} &= \pa_b \bm H_{\dgamma} {}^{\um \alpha} - \bar D_\dgamma \bm H_b {}^{ \um\alpha} + \bm \Omega_{b,  \dgamma} {}^{\um \alpha}~,	
	\end{align}
\end{subequations}
and their complex conjugate equations where $\um \alpha \longrightarrow \um \dalpha$. These components of the torsion can be decomposed into $SL(2, \mathbb{C})$ irreducible pieces. It turn out that, as a consequence of Bianchi identities, some of these irreducible pieces are related to the $S$ superfields introduced above, and spinor derivatives of $\lambda_{\um \alpha}$. The explicit prepotential expressions of the torsion components, although useful in their own right, do not immediately make it transparent  that such relationships exist. Without the knowledge of Bianchi identities, one could still discover these relationships by computing various derivatives of the torsion components explicitly, and linearly combining them (respecting dimension and representation theory, of course). Obviously, a more systematic approach would be to use the Bianchi identities to extract the exhaustive list of relations, identify which pieces of the torsion components do not participate in Bianchi identities and, hence, are new/independent superfields (as opposed to being derivatives of superfields that are in lower dimensional torsion components). This approach will be presented in section \ref{S:Sect4}.

The dimension $\tfrac 32$ components are gravitino curls $\bm T_{\hb \hc} {}^\halpha$, and the same comment applies to them. Instead of giving explicit prepotential expressions for these, we will use Bianchi identities in section \ref{S:Sect4} to find which parts of these are independent, and which parts get determined in terms of lower dimensional stuff.


\subsection{Components of the 3-form gauge field and its 4-form curvature}
\label{S:Sect3.7}
Before going to the Bianchi identities, let us not forget the super 3-form gauge field and its field strength. The highest dimensional components of $G_4$ are fully bosonic components with dimension 1. We start with its purely internal part, $\bm G_{\ul{mnpq}}$. Under abelian tensor hierarchy transformations alone, this is encoded by $E_{\ul{mnpq}}$, which no longer is invariant under $\Omega$ transformations. So we covariantize it:
\begin{align}
	\bm G_{\ul{mnpq}} &= \textrm{Re} E_{\ul{mnpq}} + 2  \psi_{\ur [\ul{mnp}} \pa_{\uq]} \hH^{\ur}
\end{align}
which is an invariant quantity, a dimension 1 curvature. Similar exercise of covariantization gives the rest of the components as follows: 
\begin{subequations}
	\begin{align}
		(\sigma^a)^{\alpha \dalpha} \bm G_{a \ul{mnp}} &= \tfrac 12 \left[  D_\alpha F_{\dalpha \ul{mnp}} - \bar D_\dalpha F_{\alpha \ul{mnp}} - 6 \varphi_{\uq [\ul{mn}} \pa_{\up]} X^\uq {}_{\alpha \dalpha} +  \psi_{\ul{mnpq}} \pa_{\alpha \dalpha} \hH^\uq  \right] \\
		\bm G_{ab \um \un} &= - (\sigma_{ab})^{\alpha \beta} G_{\alpha \beta \um \un} + \textrm{h.c.} \nonumber \\
		&= \tfrac i2 (\sigma_{ab})^{\alpha \beta}  \left[ D_{(\alpha} \hat W_{\beta) \um \un} + i \varphi_{\ul{mn}} {}^\up \, \pa_{(\alpha} {}^\dgamma X_{|\up| \beta) \dgamma}   \right]  + \textrm{h.c.} \\
		\bm G_{abc \um} &= \epsilon_{abcd} \tilde G^d {}_\um \nonumber \\
		&=  \tfrac 18 \epsilon_{abcd} (\bar \sigma^d)^{\dalpha \alpha} \left( [D_\alpha, \bar D_\dalpha] \hat H_\um - D^2 X_{\um \alpha \dalpha} - \bar D^2 X_{\un \alpha \dalpha}   \right) \\
		\bm G_{abcd} &= 3i \epsilon_{abcd} (R - \bar R)~.
	\end{align}
\end{subequations}
It is going to be useful to decompose $\bm G_{a \ul{mnp}}$ into its $\rep{1} + \rep{27}$ and $\rep{7}$ bits in the canonical way. We find
\begin{subequations}
	\begin{align}
		(\sigma^a)_{\alpha \dalpha} G_{a \um \un} &= \tfrac{1}{2} D_\alpha g_{\dalpha \um \un} - \tfrac{1}{2} \bar D_\dalpha g_{\alpha \um \un} - 2 \,\pa_{(\um} X_{ \un) \alpha \dalpha} \\    
		(\sigma^a)_{\alpha \dalpha} G_{a \um} &= \tfrac{1}{2} D_\alpha  F_{\dalpha \um} - \tfrac{1}{2} \bar D_\dalpha F_{\alpha \um} - \varphi_{\um}{}^{\un \up} \pa_\un X_{\up \alpha \dalpha} + \pa_{\alpha \dalpha} \hat H_\um~.
	\end{align}
\end{subequations}
We also choose to define the dimension $\tfrac 12$ component
\begin{align}
	\bm G_{\alpha b \um \un} &= - \tfrac 16 (\sigma_b)_{\alpha \dalpha} \varphi_{\ul{mnp}} \bar \lambda^{\up \dalpha}~,
\end{align}
and set all remaining components of $\bm G$ to zero.
From the above expressions for the curvature components, we can derive components of the three-form gauge field up to exact pieces. These are:
\begin{subequations}
	\begin{align}
		\bm C_{\ul{mnp}} &= \tfrac 12 \left( \Phi_{\ul{mnp}} + \bar \Phi_{\ul{mnp}} \right) + \tfrac 12  \psi_{\ul{mnpq}} \hat H^\uq \\
		(\sigma^a)_{\alpha \dalpha} \bm C_{a \ul{mn}} &= \tfrac 12 [D_\alpha, \bar D_\dalpha] V_{\um \un} - \varphi_{\ul{mnp}} \pa_{\alpha \dalpha} \cV^\up + \varphi_{\ul{mnp}} X^\up {}_{\alpha \dalpha} \\
		\bm C_{ab \um} & = - (\sigma_{ab})^{\alpha \beta} C_{\alpha \beta \um} + \textrm{h.c.},\nonumber \\
		&= - (\sigma_{ab})^{\alpha \beta} \Big[- \tfrac i2 \left( D_{(\alpha} \Sigma_{\beta) \um} + D_{(\alpha} \Psi_{|\um| \beta)}  \right) \Big] + \textrm{h.c.} \\
		\bm C_{abc} &= \epsilon_{abcd} \tilde C^d  \nonumber \\
		&= - \tfrac 12 \epsilon_{abcd} (\bar \sigma^d)^{\dalpha \alpha} \Big[ - \tfrac 14 \Big(  [D_\alpha, \bar D_\dalpha] X + D^2 H_{\alpha \dalpha} + \bar D^2 H_{\alpha \dalpha} \Big)  \Big]~.
	\end{align}
\end{subequations}


\section{Solving the Bianchi identities}
\label{S:Sect4}
In the previous sections, we have constructed a $4|4+7$ superspace (with extra gravitini and gauge fields) from prepotential ingredients. Now we change gears to take a ``first principles" supergeometric approach in which we deal with torsions, curvatures, and their Bianchi identities. The linearized Bianchi identities in $11|32$ superspace are \eqref{E:torsion-curvature-BI}, \eqref{E:4form-curvature-BI}. We will project these to $4|4+7$ by restricting all form indices and solve them in order of their engineering dimensions starting from the lowest dimension, which is $\tfrac{1}{2}$. When we say we solve Bianchi identities, we mean that we write down an exhaustive list of derivative relations satisfied by various torsion and curvature components, and identify components that are not constrained by Bianchi identities at all. We turn a blind eye to all the information presented in section \ref{S:Sect3} except, crucially, to pose a set of torsion constraints that are inspired by the explicit construction. These constraints will be the analogue of eq.~(14.25) in \cite{Wess:1992cp}. Together with the Bianchi identities, they define the supergeometry of a linearized $4|4+7$ superspace, a solution to which (in terms of unconstrained prepotentials) is the one we presented in section \ref{S:Sect3}.


\subsection{Restriction of linearized Bianchi identities to $4|4+7$}

Recall that the spinorial superframes $\hE^\halpha$ decompose as 
\begin{subequations}
	\begin{align}
		\hE^{\alpha I} &= \eta^I \hE^{\alpha} + i (\Gamma_\um \eta)^I \hE^{\um \alpha} \\
		\hE_{\dalpha I} &= \eta_I \hE_{\dalpha} + i (\Gamma^\um \eta)_I \hE_{\um \dalpha}~. 	
	\end{align}
\end{subequations}
The dual superspace derivatives become
\begin{subequations}
	\begin{align}
		D_{\alpha I} &= \eta_I D_{\alpha} + i (\Gamma^\um \eta)_I D_{\um \alpha} \\
		\bar D^{\dalpha I} &= \eta^I \bar D^{\dalpha} + i (\Gamma_\um \eta)^I \bar D^{\um \dalpha}~.
	\end{align}
\end{subequations}
The derivatives $D_{\um \alpha}$, $\bar D_{\um \dalpha}$ do not appear in the Bianchi identities restricted to $4|4+7$.

Bianchi identities satisfied by $\bm T^A$ in $11|32$ are:
\begin{equation}
	D_{[\hD} \bm T_{\hC \hB ]} {}^A + \mathring T_{[\hD \hC} {}^{\hat f} \bm T_{|\hat f | \hB ]} {}^A + \bm T_{[\hD \hC} {}^\hgamma \mathring T_{|\hgamma| \hB ]} {}^A = \bm R_{[\hD \hC \hB ]} {}^A~.
\end{equation}
Restricted to $4|4+7$, this gives rise to the following:
\begin{subequations}
\label{E:T^A-BI-4|4+7}
	\begin{align}
	D_{[ D} \bm T_{CB ]} {}^A + \mathring T_{[DC} {}^f \bm T_{| f | B ]} {}^A + \bm T_{[ DC} {}^F \mathring T_{|F| B ]} {}^A &= \bm R_{[DCB ]} {}^A	\\
	\pa_\um \bm T_{CB} {}^A + 2 D_{[C} \bm T_{B ] \um} {}^A + \mathring T_{CB} {}^f \bm T_{f \um} {}^A + 2 \bm T_{\um [C} {}^F \mathring T_{|F| B ]} {}^A &= 2 \bm R_{\um [C B ] } {}^A + \bm R_{CB \um} {}^A \\
	2 \pa_{[ \un} \bm T_{\um ] B} {}^A + D_B \bm T_{\ul{nm}} {}^A + \bm T_{\ul{nm}} {}^F \mathring T_{FB} {}^A &= \bm R_{\underline{nm} B} {}^A + 2 \bm R_{B [\underline{nm}]} {}^A \\
	\pa_{[ \un} \bm T_{\ul{mp}]} {}^A &= \bm R_{[\ul{nmp}]} {}^A~.
	\end{align}
\end{subequations}
For $\bm T^{\ul a}$, we have
\begin{equation}
	D_{[\hD} \bm T_{\hC \hB ]} {}^{\ul a} + \mathring T_{[\hD \hC} {}^{\hat f} \bm T_{|\hat f| \hB ]} {}^{\ul a} + \bm T_{[\hD \hC} {}^{\hgamma} \mathring T_{|\hgamma| \hB ]} {}^{\ul a} = \bm R_{[\hD \hC \hB ]} {}^{\ul a}~.
\end{equation}
Restricted to $4|4+7$, this gives rise to
\begin{subequations}
\label{E:T^i-BI-4|4+7}
	\begin{align}
\label{E:T^i-BI-4|4+7.a}
		D_{[ D} \bm T_{CB ]} {}^{\ul a} + \mathring T_{[DC} {}^f \bm T_{| f| B ]} {}^{\ul a} + \bm T_{[ DC} {}^\hgamma \mathring T_{|\hgamma| B ]} {}^{\ul a} &= \bm R_{[DCB ]} {}^{\ul a}	\\
\label{E:T^i-BI-4|4+7.b}
		\pa_\um \bm T_{CB} {}^{\ul a} + 2 D_{[C} \bm T_{B ] \um} {}^{\ul a} + \mathring T_{CB} {}^f \bm T_{f \um} {}^{\ul a} + 2 \bm T_{\um [C} {}^\hgamma \mathring T_{|\hgamma| B ]} {}^{\ul a} &= 2 \bm R_{\um [C B ] } {}^A + \bm R_{CB \um} {}^{\ul a} \\
\label{E:T^i-BI-4|4+7.c}
		2 \pa_{[ \un} \bm T_{\um ] B} {}^{\ul a} + D_B \bm T_{\ul{nm}} {}^{\ul a} + \bm T_{\ul{nm}} {}^\hgamma \mathring T_{\hgamma B} {}^{\ul a} &= \bm R_{\ul{nm} B} {}^{\ul a} + 2 \bm R_{B [\ul{nm}]} {}^{\ul a} \\
\label{E:T^i-BI-4|4+7.d}
		\pa_{[ \un} \bm T_{\ul{mp}]} {}^{\ul a} &= \bm R_{[\ul{nmp}]} {}^{\ul a}~.
	\end{align}
\end{subequations}
For $\bm T^{\um \ul \alpha}$, we have
\begin{equation}
	D_{[\hD} \bm T_{\hC \hB ]} {}^{\ul{m \alpha}} + \mathring T_{[\hD \hC} {}^{\hat f} \bm T_{|\hat f| \hB ]} {}^{\ul{m \alpha}}  = \bm R_{[\hD \hC \hB \}} {}^{\ul{m \alpha}}~.
\end{equation}
Restricted to $4|4+7$,
\begin{subequations}
\label{E:T^i.alpha-BI-4|4+7}
	\begin{align}
		D_{[D} \bm T_{CB ]} {}^{\ul{m \alpha}} + \mathring T_{[ DC } {}^f \bm T_{|f| B ] } {}^{\ul{m \alpha}} &= \bm R_{[ DCB ]} {}^{\ul{m \alpha}} \\
		\pa_\un \bm T_{CB} {}^{\ul{m \alpha}} + 2 D_{[C} \bm T_{B ] \un} {}^{\ul{m \alpha}} + \mathring T_{CB} {}^f \bm T_{f \un} {}^{\ul{m \alpha}} &= 2 \bm R_{\un [CB ]} {}^{\ul{m \alpha}} + \bm R_{CB \un} {}^{\ul{m \alpha}} \\
		2 \pa_{[ \un} \bm T_{\up] B} {}^{\ul{m \alpha}} + D_B \bm T_{\ul{np}} {}^{\ul{m \alpha}} &= \bm R_{\ul{np} B} {}^{\ul{m \alpha}} + 2 \bm R_{B \ul{np}} {}^{\ul{m \alpha}} \\
		\pa_{[ \un} \bm T_{\ul{pq} ]} {}^{\ul{m \alpha}} &= \bm R_{[\ul{npq}]} {}^{\ul{m \alpha}}~.
	\end{align}
\end{subequations}

The linearized $G_4$ Bianchi identities in $11|32$ superspace are
\begin{equation}
	D_{[\hE} \bm G_{\hD \hC \hB \hA ]} + 2 \mathring T_{[\hE \hD} {}^\hF \bm G_{|\hF| \hC \hB \hA ]} + 2 \bm T_{[\hE \hD} {}^\hF \mathring G_{|\hF| \hC \hB \hA ]} = 0
\end{equation}
Restricted to $4|4+7$, 
\begin{subequations}
\label{E:G4-BI-4|4+7}
	\begin{align}
\label{E:G4-BI-4|4+7.a}
		D_{[E} \bm G_{DCBA ]} + 2 \mathring T_{[ED} {}^f \bm G_{|f|CBA ]} + 2 \bm T_{[ED} {}^\hF \mathring G_{|\hF| CBA ]} &= 0 \\
\label{E:G4-BI-4|4+7.b}
		D_{[\um} \bm G_{DCBA ]} + 2 \mathring T_{[ \um D} {}^f \bm G_{|f|CBA ]} + 2 \bm T_{[\um D} {}^\hF \mathring G_{|\hF| CBA ]} &= 0 \\
\label{E:G4-BI-4|4+7.c}
		D_{[\um} \bm G_{\un CBA ]} + 2 \mathring T_{[ \ul{mn}} {}^f \bm G_{|f|CBA ]} + 2 \bm T_{[\ul{mn}} {}^\hF \mathring G_{|\hF| CBA ]} &= 0 \\
\label{E:G4-BI-4|4+7.d}
		D_{[\um} \bm G_{ \ul{np} BA ]} + 2 \mathring T_{[ \ul{mn}} {}^f \bm G_{|f| \up BA ]} + 2 \bm T_{[ \ul{mn}} {}^\hF \mathring G_{|\hF| \up BA ]} &= 0 \\
\label{E:G4-BI-4|4+7.e}
		D_{[\um} \bm G_{ \ul{npq} A ]} + 2 \bm T_{[ \ul{mn}} {}^\hF \mathring G_{|\hF| \ul{pq} A ]} &= 0 \\
\label{E:G4-BI-4|4+7.f}
		\pa_{[\um} \bm G_{ \ul{npqr} ]} &= 0
	\end{align}
\end{subequations}

One can further decompose the $4|4$ indices $A, B, C$ into bosonic and fermionic indices, but we don't show it explicitly here. The curvature two forms $\bm R_{\ul \alpha} {}^{\ul{m \beta}}$ appear in the torsion Bianchi identities. From $\hat R_\halpha{}^\hbeta = \frac{1}{4} \hat R_{\ha \hb} (\hat \Gamma^{\ha \hb})_\halpha{}^\hbeta$, we deduce that the only non-zero $\bm R_{\hbeta} {}^\halpha$'s are
\begin{subequations}
\label{E:R-(hbeta, halpha)}
	\begin{align}
		\bm R_\beta {}^\alpha &= \tfrac 14 \bm R_{ab} (\gamma^{ab})_\beta {}^\alpha~, \\
		\bm R_{\um \beta} {}^\alpha &= - \tfrac 14 \delta_\beta {}^\alpha \varphi_{\ul{mnp}} \bm R^{\ul{np}}~, \\
		\bm R_\beta {}^{\um \alpha} &= \tfrac 14 \delta_\beta {}^\alpha \varphi^{\ul{mnp}} \bm R_{\ul{np}}~, \\
		\bm R_{\un \beta} {}^{\um \alpha} &= \tfrac 14 \delta_\un {}^\um \bm R_{cd} (\gamma^{cd})_\beta {}^\alpha + \tfrac 14 \delta_\beta {}^\alpha \left[ \psi_\un {}^{\ul{mpq}} \bm R_{\ul{pq}} + \bm R_\un {}^\um - \bm R^\um {}_\un \right]~, \\
		\bm R^{\un \dbeta , \alpha} &= - \tfrac 12 (\bar \sigma^a)^{\dbeta \alpha } \bm R_a {}^\un~, \\
		\bm R^{\dbeta , \um \alpha} &= \tfrac 12 (\bar \sigma^b)^{\dbeta \alpha} \bm R_b {}^\um~, \\
		\bm R^{\un \dbeta , \um \alpha} &= - \tfrac 12 (\bar \sigma^d)^{\dbeta \alpha} \varphi^{\ul{mnp}} \bm R_{d \up}~,
	\end{align}
\end{subequations}
and their complex conjugates. Of these, the ones with the first index along the extra $\theta$ directions do not appear in the restricted Bianchi identities.


\subsection{Constraints}
\label{S:Sect4.2:Torsion-constraints}

We postulate the following conventional constraints for the torsion and the 4-form.
All dimension 0 components of the torsion vanish:
\begin{align}
	\bm T_{\ul{\alpha \beta}} {}^c = \bm T_{\ul{\alpha \beta}} {}^\um = 0~.
\end{align}
All dimension $\tfrac 12$ components of $\bm T$ vanish:
\begin{align}
	\bm T_{a \ul{\beta}}{}^c &= \bm T_{\um {\ul\beta}}{}^c =
	\bm T_{a \ul{\beta}} {}^\um = \bm T_{\un \ul{\beta}} {}^\um = 0~, \qquad
	\bm T_{\ul{\alpha \beta}}{}^{\ul\gamma} = 
	\bm T_{\ul{\alpha \beta}} {}^{\um \gamma} = \bm T_{\ul{\alpha \beta} , \um \dgamma} = 0~.
\end{align}
All dimension 1 components of $\bm T$ determining bosonic spin connections vanish:
\begin{align}
	\bm T_{ab} {}^c &= \bm T_{a \um} {}^c = \bm T_{\ul{mn}} {}^c
	= \bm T_{ab} {}^\um  = \bm T_{a \un} {}^\um = \bm T_{\ul{np}} {}^\um = 0~.
\end{align}
All components of $\bm G$ with dimensions $-1$, $- \tfrac 12$, and 0 vanish:
\begin{align}
	\bm G_{\ul{\alpha \beta \gamma \delta}} = 0~, \qquad
	\bm G_{a \ul{\beta \gamma \delta}} = \bm G_{\um\, \ul{\beta \gamma \delta}} = 0~, \qquad
	\bm G_{ab \,\ul{\gamma \delta}} = \bm G_{a \um\, \ul{\gamma \delta}} = \bm G_{\ul{mn}\, \ul{\gamma \delta}}= 0~.
\end{align}
The 4-form components at dimension $\tfrac 12$ are $\bm G_{\ul{\alpha mnp}}$, $\bm G_{\ul{\alpha} abc}$, $\bm G_{\ul{\alpha} ab \um}$, and $\bm G_{\ul{\alpha} b \ul{mn}}$. The first one is set to zero by fiat:
\begin{align}
	G_{\ul{\alpha mnp}} &= 0~.
\end{align}
The next two will be found to vanish as consequence of dimension $\tfrac 12$ Bianchi identities. The last one is non-zero, and, in general, a 2-form of $G_2$. We impose the constraint that it belongs in the $\boldsymbol{7}$ of $G_2$, and is proportional to $\lambda_{\um \alpha}$. Choosing a normalization:
\begin{align}
	\label{E:G_4-dim-1/2}
	\bm G_{\alpha b \ul{mn}} &= - \tfrac 16 (\sigma_b)_{\alpha \dalpha} \varphi_{\ul{mnp}} \bar \lambda^{\up \dalpha}~.
\end{align}

The above constraints are informed by the explicit construction of section \ref{S:Sect3}. In the next subsection, we state the consequences of the Bianchi identities subject to these constraints.


\subsection{Results}
We present our results organized by dimension. At each dimension, we list all components of $\bm T$, $\bm R$, and $\bm G$, and decompose them into their $SL(2, \mathbb{C})$ and $G_2$ irreducible pieces. Then we state which of the irreducible pieces are determined by Bianchi identities to be derivatives of lower dimensional components, and which ones are new/independent superfields. We end the discussion at each dimension by reiterating important relations/properties (such as reality or chirality of various pieces of the components) implied by the Bianchi identities. 

The purely 4D part of the analysis follows in the standard way \cite{Wess:1992cp}. We do not explain their derivations but merely quote the results.


\subsubsection{Dimension $\leq \tfrac 12$}
All dimension $-1$ components (namely $\bm G_{\underline{\alpha \beta \gamma \delta}}$), all dimension $- \tfrac 12$ components (namely $\bm G_{\hat a \ul{\beta \gamma \delta}}$), all dimension 0 components (namely $\bm G_{\ha \hb \ul{\gamma \delta}}$ and $\bm T_{\ul{\alpha \beta}, \hc}$), and all dimension $\tfrac 12$ torsion components (namely $\bm T_{\ha \ul \beta, \hc}$ and $\bm T_{\ul{\alpha \beta}, \hgamma}$) vanish by fiat. The dimension $\tfrac 12$ components $\bm G_{\ul{\alpha mnp}}$ are chosen to vanish by fiat.
The lowest dimensional Bianchi identities are a set of $G_4$ equations at dimension $\tfrac 12$:
\begin{subequations}
	\label{E:G_4-dim1/2BI}
	\begin{align}
		(\sigma^c)_{(\delta | \dbeta |} \bm G_{\gamma ) cba} &= 0 = (\sigma^c)_{\delta (\dbeta} \bm G_{\dgamma ) cba} \\
		(\sigma^b)_{(\delta |\dbeta|} \bm G_{\gamma ) ab \um} &= 0 = (\sigma^b)_{\delta (\dbeta} \bm G_{\dgamma ) ab \um} \\
		(\sigma^b)_{(\delta |\dbeta|} \bm G_{\gamma ) b \ul{mn}} &= 0 = (\sigma^b)_{\delta (\dbeta} \bm G_{\dgamma ) b \ul{mn}}~.
	\end{align}
\end{subequations}
Decomposing all external indices under $SL(2, \mathbb{C})$, these imply
\begin{align}
	\bm G_{\ul\delta\, abc } = \bm G_{\ul\delta \,ab \um } = 0~,\qquad
	\bm G_{\alpha \,b \ul{mn}} = - \tfrac 16 (\sigma_b)_{\alpha \dalpha} \bar \lambda^\dalpha {}_{\ul{mn}}~,
\end{align}
where $\lambda_{\alpha \ul{mn}}$ is an arbitrary internal 2-form. It decomposes under $G_2$ as $\bm{21}_{\textrm{SO(7)}} = \bm 7_{\textrm G_2} + \bm{14}_{\textrm{G}_2}$. Our constraint \eqref{E:G_4-dim-1/2} then restricts it to its $\bm 7$ piece 
\begin{align}
	\bm G_{\alpha \,b \ul{mn}} = - \tfrac 16 \varphi_{\ul{mnp}} (\sigma_b)_{\alpha \dalpha} \bar \lambda^{\up \dalpha}~.
\end{align}
This is the only non-zero component field up to dimension $\tfrac 12$.


\subsubsection{Dimension 1}
\paragraph{Torsion.}
Torsion components at dimension 1 are
\begin{itemize}
	\item $\bm T_{\hb \hc, \ha}$: these vanish by constraint via a choice of spin connection.
	\item $\bm T_{\hb \hgamma, \halpha} \xrightarrow{4|4+7}  \bm T_{\hb \ul \gamma, \halpha}$: we decompose these into irreps of $SL(2, \mathbb{C})$ and $G_2$.
\end{itemize}
The purely external components, namely $\bm T_{b \ul \gamma, \ul \alpha}$, follow from \cite{Wess:1992cp}:

\begin{subequations}
	\begin{align}
		\bm T_{b \gamma, \alpha} &= - \tfrac 12 (\bar \sigma_b)^{\dbeta \beta} \bm T_{\beta \dbeta, \gamma, \alpha} \nonumber \\
		& \bm T_{\beta \dbeta, \gamma, \alpha} = - \tfrac i4 (\epsilon_{ \beta  \alpha} G_{\gamma \dbeta} - 3 \epsilon_{\gamma \beta} G_{\alpha \dbeta} - 3 \epsilon_{\gamma \alpha} G_{\beta \dbeta}) \\
		\bm T_{b \dgamma, \dalpha} &= -\tfrac 12 (\bar \sigma_b)^{\dbeta \beta} \bm T_{\beta \dbeta, \dgamma, \dalpha} \nonumber \\
		& \bm T_{\beta \dbeta, \dgamma, \dalpha} = - \tfrac i4 (\epsilon_{ \dbeta  \dalpha} G_{\beta \dgamma} - 3 \epsilon_{\dgamma \dbeta} G_{\beta \dalpha } - 3 \epsilon_{\dgamma \dalpha} G_{\beta \dbeta}) \\
		\bm T_{b \gamma, \dalpha} &= - \tfrac 12 (\bar \sigma_b)^{\dbeta \beta} \bm T_{\beta \dbeta, \gamma, \dalpha} \nonumber \\
		 & \bm T_{\beta \dbeta, \gamma, \dalpha} = 2i \epsilon_{\gamma \beta} \epsilon_{\dbeta \dalpha } R^{\dagger}~, ~~ ~~ ~~ R^{\dagger} = \bar R \\
		 \bm T_{b \dgamma, \alpha} &= - \tfrac 12 (\bar \sigma_b)^{\dbeta \beta} \bm T_{\beta \dbeta, \dgamma, \alpha} \nonumber \\
		 & \bm T_{\beta \dbeta, \dgamma, \alpha} = 2i \epsilon_{\dgamma \dbeta} \epsilon_{\beta \alpha } R~.
	\end{align}
\end{subequations}
The independent superfields here are $G_{\alpha \dalpha}$ and $R$. It is implied by a Bianchi identity that $G_{\alpha \dbeta}$ is real. 

The internal pieces of $\bm T_{\ul\alpha}$ are decomposed as
\begin{subequations}
\label{E:defn-S's}
	\begin{align}
		\bm T_{\um \beta, \alpha} &= -i \epsilon_{\beta \alpha} S_{\um} + S_{\um \beta \alpha} \\
		\bm T_{\um \dbeta, \alpha} &= -i (\sigma^c)_{\alpha \dbeta} S_{\um c} = -i S_{\um \alpha \dbeta } \\
		\bm T_{\um \dbeta, \dalpha} &= -i \epsilon_{ \dbeta  \dalpha} S_{\um} + \bar S_{\um \dbeta \dalpha}~.
	\end{align}
\end{subequations}
Here, $S_{\um \alpha \beta}$ and $\bar S_{\um \dalpha \dbeta}$ are symmetric in the two spinor indices. From $\bm T^\dalpha = (\bm T^\alpha)^*$ and the above definitions, the following complex conjugation relations follow
\begin{align}
	(S_{\um \alpha \beta})^* = - \bar S_{\um \dalpha \dbeta}~, \qquad
	\bar S_{\um c} := (S_{\um c})^* ~.
\end{align}
A dimension 1 Bianchi identity \eqref{E:G4-BI-4|4+7.b} for $G_4$ implies that $S_\um$ is real \begin{align}
	\bar S_\um &:= (S_\um)^* = S_\um~,
\end{align}
which we have already taken into account in \eqref{E:defn-S's}.
The independent superfields here are $S_{\um}$, $S_{\um \alpha \beta} = - (\bar S_{\um \dalpha \dbeta})^*$, and $S_{\um c} = (\bar S_{\um c})^*$. 

Next we have the pieces of $\bm T^{\ul{m \alpha}}$. These are $\bm T_{\un, \ul \beta}{}^{\ul{m \alpha}}$ and $\bm T_{b \ul\beta}{}^{\ul{m\alpha}}$. We start with the first of these. When the two spinor indices are both dotted (or both undotted), a dimension 1 torsion Bianchi identity \eqref{E:T^i-BI-4|4+7.b} implies that the $\epsilon$-traceless (spin 1) piece is (proportional to) a curvature component. The trace piece (spin 0) does not participate in \eqref{E:T^i-BI-4|4+7.b}, and hence it is proportional to an independent superfield $Z_{\um, \un}$. The comma between indices denotes that it belongs to $\bm 7 \times \bm 7$ of $G_2$ and should be decomposed into irreps of $G_2$. When one spinor index is undotted, and the other is dotted, we split the corresponding torsion component into its real and imaginary parts. The real part is set equal to a curvature component (times $i$, since the curvature component in question is imaginary), and the imaginary part is new, denoted by $X_{\alpha \dbeta \um, \un}$. Explicitly, we have
\begin{subequations}
\label{E:4.25}
	\begin{align}
		\bm T_{\um ~ \gamma,  ~ \un \beta } &= - \tfrac i4 \bm R_{\gamma \beta \ul{mn}} + \tfrac 12 \epsilon_{\gamma \beta} Z_{\um, \un} \\
		\bm T_{\um ~ \dgamma,  ~ \un \dbeta } &=  \tfrac i4 \bm R_{\dgamma \dbeta \ul{mn}} - \tfrac 12 \epsilon_{\dgamma \dbeta} \bar Z_{\um, \un} \\
		\bm T_{\um ~ \gamma, ~ \un \dbeta } &=  \tfrac i4 \bm R_{\gamma \dbeta \ul{mn}} - i X_{\gamma \dbeta \um, \un} \\
		\bm T_{\um ~ \dgamma, ~ \un \beta } &=  - \tfrac i4 \bm R_{\beta \dgamma \ul{mn}} - i X_{\beta \dgamma \um, \un}~,
	\end{align}
\end{subequations}
where
\begin{align}
	\bar Z_{\um, \un} &:= (Z_{\um, \un})^* ~, \qquad 
	X_{\beta \dgamma ~ \um, \un} = (X_{\gamma \dbeta ~ \um, \un })^* \\
	\bm R_{\dgamma \dbeta, \ul{mn}} &= - (\bm R_{\gamma \beta \ul{mn}})^*~, ~~~ \bm R_{\dgamma \beta, \ul{mn}} = - (\bm R_{\gamma \dbeta \ul{mn}})^*
\end{align}
More information about $X_{a \um , \un} = - \tfrac 12 (\bar \sigma_a)^{\dalpha \alpha} X_{\alpha \dalpha \um , \un}$ is hidden in the dimension 1 $G_4$ Bianchi identities 
\eqref{E:G4-BI-4|4+7.c} and \eqref{E:G4-BI-4|4+7.d}.
The first implies that $X_{a [\um, \un]}$ lies only in $\bm 7$ of $G_2$, which means that the full $X_{a \um, \un}$ is in $\bm{35}$ of $SO(7)$, an internal 3-form. The second implies that this internal 3-form is nothing but $\bm G_{a \ul{mnp}}$.  We $G_2$-decompose $X_{a \um, \un}$ (alternatively $\bm G_{a \ul{mnp}}$). Its $G_2$ singlet is denoted $X_a$, its traceless symmetric piece ($\bm{27}$) is denoted by $\tilde X_{a \um \un}$, and its antisymmetric piece ($\bm 7$) is encoded in $X_{a \um}$. Their complex conjugates are $\bar X_a$, $\tilde{\bar X}_{a \um \un}$, and $\bar X_{a \um}$ respectively:
\begin{subequations}
	\begin{align}
		X_{a \um, \un} &=:  \tilde X_{a \ul{mn}} + \tfrac 17 \delta_{\um \un} X_a + \tfrac 13 \varphi_{\ul{mn}} {}^{\up} X_{a \up} \\
		\delta^{\um \un} X_{a \um , \un} &= X_a~, ~~~~ \tilde X_{a \um \un} = \tilde X_{a \un \um}~, ~~~~ X_{a [\um , \un]} = \tfrac 13 \varphi_{\ul{mnp}} X_a {}^\up ~.
	\end{align}
\end{subequations}
The identification of this superfield with $\bm G_{a \ul{mnp}}$ is through the following 
\begin{align}
	\bm G_{a \ul{mnp}} &= -6 X_{a[\um ,} {}^{\uq} \varphi_{\ul{np}] \uq} ~,
\end{align}
which implies
\begin{subequations}
\label{E:Xamn=Gamnp}
	\begin{align}
		X_a &= - \tfrac 1{36} \varphi^{\ul{mnp}} \bm G_{a \ul{mnp}} \\
		 \tilde X_{a \ul{mn}} &= - \tfrac 18 \varphi_{(\um} {}^{\ul{pq}} \bm G_{|a| \un ) \ul{pq}} + \tfrac 1{56} \delta_{\ul{mn}} \phi^{\ul{pqr}} \bm G_{a \ul{pqr}} \\
		X_{a \um} &= \tfrac 1{48} \psi_\um {}^{\ul{npq}} \bm G_{a \ul{npq}} ~.
	\end{align}
\end{subequations}
Equations \eqref{E:Xamn=Gamnp} can be easily inverted to express irreducible pieces of $\bm G_{a \ul{mnp}}$ in terms of those of $X_{a \um, \un}$. This is not all. Bianchi identity \eqref{E:G4-BI-4|4+7.c} determines completely the $\bm 7$ piece $X_{a \um}$, and the curvature component $\bm R_{\alpha \dbeta \um \un}$ in terms of previously introduced independent superfields: 
\begin{subequations}
	\begin{align}
		i X_{\alpha \dbeta \um} &= \tfrac 1{16} (D_{\alpha} \bar \lambda_{\um \dbeta} + \bar D_{\dbeta} \lambda_{\um \alpha}) - \tfrac {3i}4 (S_{\um \alpha \dbeta} + \bar S_{\um \alpha \dbeta}) \\
		\bm R_{\alpha \dbeta \ul{mn}} &= \varphi_{\ul{mn}} {}^\up \left[ \tfrac i{12} (D_{\alpha} \bar \lambda_{\up \dbeta} - \bar D_\dbeta \lambda_{\up \alpha}) - (S_{\up \alpha \dbeta} - \bar S_{\up \alpha \dbeta}) \right]
	\end{align}
\end{subequations}
Notice that $\bm R_{\alpha \dbeta \um \un}$ is forced to lie in the $\bm 7$. The components $\bm R_{\alpha \beta \ul{mn}}$ and $\bm R_{\dalpha \dbeta \ul{mn}}$ are not fully determined by dimension 1 Bianchi identities. Their $\bm 7$ pieces are forced (by (\ref{E:T^i.alpha-BI-4|4+7}a)) to vanish
\begin{align}
	\varphi^{\ul{pmn}} \bm R_{\alpha \beta \ul{mn}} &= 0 = \varphi^{\ul{pmn}} \bm R_{\dalpha \dbeta \ul{mn}}~,
\end{align}
while their $\bm{14}$ pieces are left unconstrained.

We will now $G_2$ decompose $Z_{\um, \un}$. For its symmetric part, we denote the $\bm 1$ piece by $\tilde{\bar R}$, and the $\bm{27}$ piece by $\tilde{\bar R}_{\um \un}$. For the anti-symmetric part, we denote the $\bm 7$ piece by $\bar R_\um$, and the $\bm{14}$ piece by $L_{[\um \un]_{14}}$. Their complex conjugates are $\tilde R$, $\tilde R_{\um \un}$, $R_\um$, and $\bar L_{[\um \un]_{14}}$ respectively:
\begin{subequations}
	\begin{align}
		Z_{\um, \un} &=: \tfrac 12 \tilde{\bar R}_{\ul{mn}} + \tfrac 1{14} \delta_{\um \un} \tilde{\bar R} + \tfrac 16 \varphi_{\ul{mnp}} \bar R^{\up} + L_{[\ul{mn}]_{\boldsymbol{14}}} \\
		\delta^{\um \un} Z_{\um,\un} &= \tfrac 12 \tilde{\bar R}~, ~~~~ \tilde{\bar R}_{\um \un} = \tilde{\bar R}_{\un \um}~, ~~~~ Z_{[\um, \un]} =  \tfrac 16 \varphi_{\ul{mnp}} \bar R^{\up} + L_{[\ul{mn}]_{\boldsymbol{14}}}~.
	\end{align}
\end{subequations}
The dimension 1 Bianchi identity \eqref{E:G4-BI-4|4+7.c} for $G_4$ implies
\begin{subequations}
	\begin{align}
		L_{[\ul{mn}]_{\boldsymbol{14}}} &= \bar L_{[\ul{mn}]_{\boldsymbol{14}}}  \\
		R_{\um} - \bar R_{\um} &=  6i S_{\um} + \tfrac 14 D^{\alpha} \lambda_{\alpha \um} - \tfrac 14 \bar D_{\dalpha} \bar \lambda^{\dalpha} {}_{\um}
	\end{align}
\end{subequations}
Thus, new/independent superfields in the components \eqref{E:4.25} are $\bm R_{\alpha \beta \ul{mn}} = - (\bm R_{\dalpha \dbeta \ul{mn}})^*$, $\tilde X_{a \ul{mn}}$, $X_a$, $L_{[\ul{mn}]_{14}}$ (which is real), $\operatorname{Re}(R_{\um})$, $\tilde{\bar R}_{\ul{mn}} $, and $\tilde{\bar R}$.

The remaining torsion components at this dimension are fully determined by Bianchi identities in the following way (we skip the details of the derivation):
\begin{subequations}
	\begin{align}
		\bm T_{\gamma b, \um \alpha} &= - \tfrac 12 (\bar \sigma_b)^{\dbeta \beta} \bm T_{\gamma, \beta \dbeta, \um \alpha} \nonumber \\
		 &=  - \tfrac 12 (\bar \sigma_b)^{\dbeta \beta}  \Big[ 2i \epsilon_{\gamma \beta}\bar S_{\um \alpha \dbeta} + \epsilon_{\gamma \alpha} \left[ \tfrac i4  \bar S_{\um \beta \dbeta} + \tfrac {3i}4  S_{\um \beta \dbeta} + \tfrac 1{16}  (D_{\beta} \bar \lambda_{\dbeta \um} - \bar D_{\dbeta} \lambda_{\beta \um}) \right] \Big] \\
		 \bm T_{\dgamma b, \um \dalpha} &= - \tfrac 12 (\bar \sigma_b)^{\dbeta \beta} \bm T_{\dgamma, \beta \dbeta, \um \dalpha} \nonumber \\
		 &=  - \tfrac 12 (\bar \sigma_b)^{\dbeta \beta}  \Big[ 2i \epsilon_{\dgamma \dbeta} S_{\um \beta \dalpha} + \epsilon_{\dgamma \dalpha} \left[ \tfrac i4  S_{\um \beta \dbeta} + \tfrac {3i}4  \bar S_{\um \beta \dbeta} + \tfrac 1{16}  ( \bar D_{\dbeta} \lambda_{\beta \um} - D_{\beta} \bar \lambda_{\dbeta \um}) \right]  \Big] \\
		 \bm T_{ \gamma b, \um \dalpha} &= - \tfrac 12 (\bar \sigma_b)^{\dbeta \beta} \bm T_{\gamma, \beta \dbeta, \um \dalpha} \nonumber \\
		 &=  - \tfrac 12 (\bar \sigma_b)^{\dbeta \beta}  \Big[ i \epsilon_{\gamma \beta} \epsilon_{\dbeta \dalpha} S_{\um} + 3 \epsilon_{\gamma \beta} \bar S_{\um \dbeta \dalpha} - \epsilon_{\dbeta \dalpha} S_{\um \gamma \beta} \Big] \\
		 \bm T_{\dgamma b, \um  \alpha} &= - \tfrac 12 (\bar \sigma_b)^{\dbeta \beta} \bm T_{\dgamma, \beta \dbeta, \um \alpha} \nonumber \\
		 &=  - \tfrac 12 (\bar \sigma_b)^{\dbeta \beta} \Big[ i \epsilon_{\dgamma \dbeta} \epsilon_{\beta \alpha} S_{\um} + 3 \epsilon_{\dgamma \dbeta}  S_{\um \beta  \alpha} - \epsilon_{ \beta \alpha} \bar S_{\um \dgamma \dbeta} \Big]
	\end{align}
\end{subequations}
There are no new superfields here.

\paragraph{Curvature.}
The curvature components at dimension 1 are
\begin{itemize}
	\item $\bm R_{\hdelta \hgamma \hat b \hat a}  \xrightarrow{4|4+7}  \bm R_{\ul{\delta \gamma} \hb \ha}$ 
\end{itemize}
Remember we do not need to separately consider $\bm R_{\hdelta \hgamma \hbeta}{}^{\halpha}$
since these get determined through \eqref{E:R-(hbeta, halpha)}.
Purely external components follow in the standard way as in \cite{Wess:1992cp}. The Lie algebra indices on $\bm R_{\ddelta \dgamma~ ba}$ are decomposed into self-dual and anti-self-dual pieces,
\begin{align}
	\bm R_{\ddelta \dgamma ~ ba} &= - (\sigma_{ba})^{\beta \alpha} \bm R_{\ddelta \dgamma ~ \beta \alpha} + (\bar \sigma_{ba})^{\dbeta \dalpha} \bm R_{\ddelta \dgamma ~ \dbeta \dalpha}~,
\end{align}
and similarly for $\bm R_{\delta \gamma ~ ba}$, $\bm R_{\ddelta \gamma ~ ba}$, and $\bm R_{\delta \dgamma ~ ba}$. A Bianchi identity forces
\begin{align}
	\bm R_{\ddelta \dgamma ~ \beta \alpha} &= 0 = \bm R_{\delta \gamma ~ \dbeta \dalpha}~,
\end{align}
and the rest of the components are determined in terms of the superfields $G_{\alpha \dalpha}$ and $R$:
\begin{subequations}
	\begin{align}
		\bm R_{\ddelta \dgamma \dbeta \dalpha} &= 4 \left( \epsilon_{\ddelta \dalpha} \epsilon_{\dgamma \dbeta} +  \epsilon_{\dgamma \dalpha} \epsilon_{\ddelta \dbeta} \right) R~, \qquad
		\bm R_{\delta \gamma \beta \alpha} = 4 \left( \epsilon_{ \delta  \alpha} \epsilon_{ \gamma \beta } +  \epsilon_{\gamma  \alpha} \epsilon_{ \delta \beta} \right) R^\dagger \\
		\bm R_{\delta \dgamma \dbeta \dalpha} &= - \left(  \epsilon_{\dgamma \dbeta} G_{\delta \dalpha} + \epsilon_{\dgamma \dalpha } G_{\delta \dbeta}  \right)~, \qquad
		\bm R_{\delta \dgamma \beta \alpha} = - \left(  \epsilon_{\delta \beta} G_{\alpha \dgamma} + \epsilon_{\delta \alpha } G_{\beta \dgamma}  \right)~.
	\end{align}
\end{subequations}
Similarly, components with internal indices are given to be
\begin{subequations}
	\begin{align}
		\bm R_{\gamma \beta, \um a} &= -8 (\sigma_{ac})_{\beta \gamma} \bar S_{\um} {}^c~, \qquad
		\bm R_{\dgamma \dbeta, \um a} = -8 (\bar \sigma_{ac})_{\dbeta \dgamma}  S_{\um} {}^c \\
		\bm R_{\gamma \dbeta, \um a} &= 2i \left[ (\sigma_a)_{\beta \dbeta} S_{\um \gamma} {}^{\beta} - (\sigma_a)_{\gamma \dgamma} \bar S_{\um \dbeta} {}^{\dgamma}  \right] \\
		\bm R_{\alpha \dbeta \ul{mn}} &= \varphi_{\ul{mn}} {}^{\up} \left[  \tfrac i{12} (D_{\alpha} \bar \lambda_{\dbeta \up} - \bar D_{\dbeta} \lambda_{\alpha \up} ) - (S_{\up \alpha \dbeta} - \bar S_{\up \alpha \dbeta})  \right] ~.
	\end{align}
\end{subequations}
The components $\bm R_{\alpha \beta \ul{mn}}$ and $\bm R_{\dalpha \dbeta \ul{mn}}$ appear in the irrep. decomposition of torsion components $\bm T_{\um,\alpha}  {}^{\un \beta} $ (and c.c.). They lie in the $\bm{14}$. We have already catalogued these as new/independent superfields in the above paragraph for torsion.

\paragraph{4-form.}
4-form components at dimension 1 are
\begin{itemize}
	\item $ \bm G_{\hat a \hat b \hat c \hat d} \xrightarrow{4|4+7} \bm G_{\hat a \hat b \hat c \hat d}   $
\end{itemize}
Most of these are fully determined by Bianchi identities:
\begin{subequations}
	\begin{align}
		\bm G_{abcd} &= 3i \epsilon_{abcd} \left( R - \bar R \right) \\
		\bm G_{abc \um} &= 3i \epsilon_{abcd} \left( \bar S_{\um} {}^d -  S_{\um} {}^d \right) \\
		\bm G_{ab \ul{mn}} &= - (\sigma_{ab})^{\alpha \beta} \left[  \varphi_{\ul{mn}} {}^\up \left( - \tfrac i{12} D_\alpha \lambda_{\beta \up} +2i S_{\up \alpha \beta} \right) - \tfrac 12 \bm R_{\alpha \beta \ul{mn}}  \right] \nonumber \\
		& ~~~ + (\bar \sigma_{ab})^{\dalpha \dbeta} \left[  \varphi_{\ul{mn}} {}^\up \left( - \tfrac i{12} \bar D_\dalpha \bar \lambda_{\dbeta \up} + 2i \bar S_{\up \dalpha \dbeta} \right) + \tfrac 12 \bm R_{\dalpha \dbeta \ul{mn}}  \right] \\
		\bm G_{a \ul{mnp}} &= -6 X_{a[\um ,} {}^\uq \varphi_{\ul{np}] \uq}~.
	\end{align}
\end{subequations}
The only 4-form component at dimension 1 completely undetermined/unconstrained by dimension 1 Bianchi identities is $\bm G_{\ul{mnpq}}$, which we $G_2$ decompose:
\begin{subequations}
	\begin{align}
		\bm G_{\ul{mnpq}} &= \tfrac 1{24 \times 7} \psi_{\ul{mnpq}} \cG + \tfrac 1{42} \varphi_{[\ul{mnp}} \cG_{\uq]} + \psi_{[\ul{mnp}} {}^{\ur} \cG_{\uq ] \ur} \\
		\cG_{\ul{mn}} &= \cG_{\ul{nm}}; ~~ \delta^{\ul{mn}} \cG_{\ul{mn}} = 0 ~.
	\end{align}
\end{subequations}
The above is $\bm{35}_{SO(7)} = \bm 1_{G_2} + \bm 7_{G_2} + \bm{27}_{G_2} $.

\paragraph{Relations.}
We reiterate some of the properties of the independent superfields implied by Bianchi identities:
\begin{subequations}
	\begin{align}
		(G_{\alpha \dbeta})^* &= G_{\beta \dalpha} \\
		\bar S_{\um} &= (S_{\um})^* = S_{\um} \\
		\varphi^{\ul{mnp}} \bm R_{\alpha \beta \ul{np}} &= 0 = \varphi^{\ul{mnp}} \bm R_{\dalpha \dbeta \ul{np}} \\
		\bm R_{\alpha \dbeta \ul{mn}} &= \varphi_{\ul{mn}} {}^{\up} \left[  \tfrac i{12} (D_{\alpha} \bar \lambda_{\dbeta \up} - \bar D_{\dbeta} \lambda_{\alpha \up} ) - (S_{\up \alpha \dbeta} - \bar S_{\up \alpha \dbeta})  \right]  \\
		i X_{\alpha \dbeta \um} &= \tfrac 1{16} (D_{\alpha} \bar \lambda_{\dbeta \um} + \bar D_{\dbeta} \lambda_{\alpha \um}) - \tfrac {3i}4 (S_{\um \alpha \dbeta} + \bar S_{\um \alpha \dbeta}) \\
		R_{\um} - \bar R_{\um} &=  6i S_{\um} + \tfrac 14 D^{\alpha} \lambda_{\alpha \um} - \tfrac 14 \bar D_{\dalpha} \bar \lambda^{\dalpha} {}_{\um}
	\end{align}
\end{subequations}

\subsubsection{Dimension $\tfrac 32$ }

\paragraph{Torsion.}
Torsion components at this dimension are 
\begin{itemize}
	\item $\bm T_{\hat b \hat c , \halpha} \xrightarrow{4|4+7} \bm T_{\hb \hc, \ul{\alpha}}, \bm T_{\hb \hc, \ul{m \alpha}}$.
\end{itemize}
Purely external components follow from \cite{Wess:1992cp}. We have $\bm T_{cb, \ul \alpha} = \tfrac 14 (\bar \sigma_c)^{\dgamma \gamma} (\bar \sigma_b)^{\dbeta \beta} \bm T_{\gamma \dgamma, \beta \dbeta, \ul \alpha}$
\begin{subequations}, and
	\begin{align}
		\bm T_{\gamma \dgamma, \beta \dbeta, \alpha} &= -2 \epsilon_{\dgamma \dbeta} W_{\gamma \beta \alpha} - \tfrac 12 \epsilon_{\dgamma \dbeta} \left( \epsilon_{\gamma \alpha} \bar D_\ddelta G_\beta {}^\ddelta + \epsilon_{\beta \alpha} \bar D_\ddelta G_\gamma {}^\ddelta  \right) \nonumber \\
		& ~~~~ + \tfrac 12 \epsilon_{\gamma \beta} \left(  \bar D_\dgamma G_{\alpha \dbeta} + \bar D_\dbeta G_{\alpha \dgamma}  \right) \\
		\bm T_{\gamma \dgamma, \beta \dbeta, \dalpha} &= -2 \epsilon_{ \gamma \beta} \bar W_{\dgamma \dbeta \dalpha} - \tfrac 12 \epsilon_{\gamma \beta} \left( \epsilon_{\dgamma \dalpha}  D_\delta G^\delta {}_\dbeta + \epsilon_{\dbeta \dalpha} D_\delta G^\delta {}_\dgamma  \right) \nonumber \\
		& ~~~~ + \tfrac 12 \epsilon_{\dgamma \dbeta} \left(  D_\gamma G_{\beta \dalpha} +  D_\beta G_{\gamma \dalpha}  \right)~.
	\end{align}
\end{subequations}
Now we deal with components with internal indices one by one. Decomposing $\bm T_{a \um, \beta}$ with respect to $SL(2, \mathbb{C})$, we have
\begin{align}
	\bm T_{a \um, \beta} & = -i (\bar \sigma_a)^{\dalpha \alpha} X_{\um \alpha \beta \dalpha} + i (\sigma_a)_{\beta \dalpha} X_{\um} {}^{\dalpha}~,
\end{align}
where $X_{\um \alpha \beta \dalpha}$ is symmetric in $\alpha \beta$, and $X_{\um \dalpha}$ is the $\epsilon$-trace piece. We use the following notation for their complex conjugates
\begin{align}
	(X_{\um \alpha \beta \dgamma})^*&= - \bar X_{\um \dalpha \dbeta \gamma}~,  \qquad (X_{\um \dalpha})^* = \bar X_{\um  \alpha}~,
\end{align}
in terms of which
\begin{align}
	\bm T_{a \um, \dbeta} &= (\bm T_{a \um, \beta})^* = -i (\bar \sigma_a)^{\dalpha \alpha} \bar X_{\um \dalpha \dbeta \alpha} - i (\sigma_a)_{\alpha \dbeta} \bar X_\um {}^\alpha ~.
\end{align}
Bianchi identities of dimension $\tfrac 32$ fully determine these superfields in terms of the lower dimensional ones:
\begin{subequations}
	\begin{align}
		X_{\um \alpha \beta \dgamma} &= \tfrac i4 D_{(\alpha}  S_{| \um | \beta ) \dgamma} - \bar D_{\dgamma} S_{\um \alpha \beta}~, \qquad
		\bar X_{\um \dalpha \dbeta \gamma} =  \tfrac i4 \bar D_{(\dalpha}  \bar S_{|\um \gamma | \dbeta ) } - D_{\gamma} \bar S_{\um \dalpha \dbeta} \\
		X_{\um \dalpha} &= - \tfrac i4 \left[  \bar D_{\dalpha} S_{\um} + \tfrac 12 D^{\beta} S_{\um \beta \dalpha}  \right]~, \qquad
		\bar X_{\um  \alpha} =   \tfrac i4 \left[  D_{ \alpha} S_{\um} - \tfrac 12 \bar D_{\dbeta} \bar S_{\um \alpha} {}^{\dbeta}  \right] ~.
	\end{align}
\end{subequations}
Here we mention that, as a consequence of the Bianchi identity (\ref{E:T^A-BI-4|4+7}b), the $\alpha \beta$ symmetric part of the derivative $D_\alpha \bar S_{\um \beta \dgamma}$ vanishes. We give a name to the remaining anti-symmetric piece:
\begin{align}
	D_\alpha \bar S_{\um b} &= \tfrac i4 (\sigma_b)_{\alpha \dbeta} \bar \rho_{\um} {}^{\dbeta}~, \qquad 
	\bar \rho_{\um \dalpha} = (\rho_{\um \alpha})^*~.
\end{align}
Remaining components of $\bm T^{\ul \alpha}$ at dimension $\tfrac 32$ are $\bm T_{\um \un} {}^{\ul \alpha}$. This is an internal 2-form, and decomposes under $G_2$ as $\bm 7 + \bm{14}$.
\begin{subequations}
	\begin{align}
		\bm T_{\ul{mn}} {}^\alpha &= \tfrac 16 \varphi_{\ul{mnp}} U^{\up \alpha} + T_{[\ul{mn}]_{14}} {}^\alpha \\
		\bm T_{\ul{mn}} {}^\dalpha &= (\bm T_{\ul{mn}} {}^\alpha)^* = \tfrac 16 \varphi_{\ul{mnp}} \bar U^{\up \dalpha} + \bar T_{[\ul{mn}]_{14}} {}^\dalpha~.
	\end{align}
\end{subequations}
The complex conjugate notations are, obviously, 
\begin{align}
	(U_{\um \alpha})^* &= \bar U_{\um \dalpha}~, \qquad
	(T_{[\ul{mn}]_{14}} {}^\alpha)^* = \bar T_{[\ul{mn}]_{14}} {}^\dalpha  ~.
\end{align}
The superfield $U_{\um \alpha}$ is fully determined by Bianchi identities,
\begin{align}
	U_{\um \alpha} &= - \tfrac {3i}2 \rho_{\um \alpha} - D_\alpha S_\um + \bar D^\dbeta \bar S_{\um \alpha \dbeta} - \tfrac i{48} [ D^2 - 2 \bar D^2 ] \lambda_{\um \alpha} - \tfrac i{24} [ D_\alpha \bar D_\dbeta + 2 \bar D_\dbeta D_\alpha ] \bar \lambda_\um {}^\dbeta~,
\end{align}
while $T_{[\um \un]_{14} \alpha}$ is left unconstrained. It is an independent superfield.

Next, we consider
\begin{subequations}
\label{E:T-(ab, m.gamma)}
	\begin{align}
		\bm T_{ab, \um \gamma} &= - (\sigma_{ab})^{\alpha \beta} \left[  T_{\um \alpha \beta \gamma} + \epsilon_{\gamma \alpha} T_{\um \beta} \right] - (\bar \sigma_{ab})^{\dalpha \dbeta} T_{\dalpha \dbeta \gamma \um} \\
		\bm T_{ab, \um \dgamma} &= (	\bm T_{ab, \um  \gamma})^* = (\bar \sigma_{ab})^{\dalpha \dbeta} \left[ - \bar T_{\um \dalpha \dbeta \dgamma} + \epsilon_{\dgamma \dalpha} \bar T_{\um \dbeta}   \right] - (\sigma_{ab})^{ \alpha \beta} \bar T_{\alpha  \beta \dgamma \um}~,
	\end{align}
\end{subequations}
where notations adopted for complex conjugation are
\begin{align}
	 (T_{\um \alpha \beta \gamma})^* &= -  \bar T_{\um \dalpha \dbeta \dgamma}~, \qquad 
	 (T_{\um \alpha})^* = \bar T_{\um \dalpha}~, \qquad
	 ( T_{\dalpha \dbeta \gamma \um} )^* = - \bar T_{ \alpha \beta \dgamma \um}~.
\end{align}
These superfields are all fully determined by Bianchi identities, 
\begin{subequations}
	\begin{align}
		T_{\um \alpha \beta \gamma} &= i D_{(\alpha} S_{| \um | \beta \gamma )}~, \qquad  \qquad \qquad ~~~~
		\bar T_{\um \dalpha \dbeta \dgamma}  = - i \bar D_{(\dalpha} \bar S_{ | \um | \dbeta \dgamma )}~. \\
		T_{\um \alpha} &= \tfrac 12 \left[ D_{\alpha} S_{\um} - \tfrac i2 \rho_{\um \alpha}  \right]~, \qquad \qquad ~
		 \bar T_{\um \dalpha} =  \tfrac 12 \left[ \bar D_{\dalpha} S_{\um} + \tfrac i2 \bar \rho_{\um \dalpha}  \right] ~.  \\
		T_{\dalpha \dbeta \gamma \um} &= \tfrac 12 \bar D_{(\dalpha} \bar S_{| \um \gamma | \dbeta )} - i D_{\gamma} \bar S_{\um \dalpha \dbeta}~, \qquad
		\bar T_{ \alpha \beta \dgamma \um} =  - \tfrac 12  D_{( \alpha} \bar S_{| \um | \beta) \dgamma } + i \bar D_\dgamma  S_{\um  \alpha  \beta}~.
	\end{align}
\end{subequations}

Next, we move on to components with two internal indices, belonging to $\bm 7 \times \bm 7 = (\bm 1 + \bm{27}) + (\bm 7 + \bm{14})$ of $G_2$. These are $\bm T_{a \um} {}^{\un \beta}$ and $\bm T_{a \um} {}^{\un \dbeta} = (\bm T_{a \um} {}^{\un \beta})^*$. We decompose these components with respect to $SL(2,\mathbb{C})$ first, and then with respect to $G_2$:
\begin{subequations}
	\begin{align}
		\bm T_{a \um, \un \beta} &= -i (\bar \sigma_a)^{\dgamma \alpha} Y_{\um , \un \alpha \beta \dgamma} + i (\sigma_a)_{\beta \dalpha} Y_{\um , \un} {}^\dalpha \\
		&= -i (\bar \sigma_a)^{\dgamma \alpha} \left[ \tfrac 12 \tilde Q_{\um \un \alpha \beta \dgamma} + \tfrac 1{14} \delta_{\um \un} \tilde Q_{\alpha \beta \dgamma} + 
	\tfrac 16 \varphi_{\ul{mnp}} Q^{\up} {}_{\alpha \beta \dgamma} + K_{[\ul{mn}]_{14} \alpha \beta \dgamma} \right]   \nonumber \\
		& ~~~~ + i (\sigma_a)_{\beta \dalpha} \left[  \tfrac 12 \tilde P_{\ul{mn}} {}^\dalpha + \tfrac 1{14} \delta_{\ul{mn}} \tilde P^\dalpha + \tfrac 16 \varphi_{\ul{mnp}} P^{\up \dalpha} + M_{[\ul{mn}]_{14}} {}^\dalpha  \right]~.
\end{align}
\end{subequations}
Here, $Y_{\um, \un \alpha \beta \dgamma}$ is symmetric in $\alpha \beta$, and belongs in $\bm 7 \times \bm 7$ of $G_2$. Its symmetric part decomposes into a $G_2$ singlet $\tilde Q_{\alpha \beta \dgamma}$ and a $\bm{27}$ (traceless, symmetric) of $G_2$ denoted $\tilde Q_{\um \un \alpha \beta \dgamma}$, while its anti-symmetric part decomposes into a $\bm 7$ of $G_2$ denoted $Q_{\up \alpha \beta \dgamma}$, and a $\bm{27}$ of $G_2$ denoted $K_{[\um \un]_{14} \alpha \beta \dgamma}$. The $\alpha \beta$ anti-symmetric part of $\bm T_{\alpha \dalpha ~ \um, ~\un \beta}$ is encoded in $Y_{\um, \un} {}^\dalpha$, which is in $\bm 7 \times \bm 7$ of $G_2$ and gets similarly decomposed. $\tilde P_{\um \un \dalpha}$ is in $\bm{27}$, $\tilde P_\dalpha$ in $\bm 1$, $P_{\up \dalpha}$ in $\bm 7$, and $M_{[\um \un]_{14} \dalpha}$ in $\bm{14}$.
All these irreducible parts are fully determined (by Bianchi identities) in terms of previously introduced superfields:
\begin{subequations}
	\begin{align}
		\tilde Q_{\um \un \alpha \beta \dgamma} &= i D_{(\alpha} \tilde X_{\beta) \dgamma \um \un}  ~, \qquad \qquad \qquad \qquad \qquad
		\tilde Q_{\alpha \beta \dgamma} = i D_{(\alpha} X_{\beta) \dgamma}~. \\
		Q_{\um \alpha \beta \dgamma} &= \tfrac 1{16} D_{(\alpha} \bar D_{|\dgamma} \lambda_{\um | \beta)} - \tfrac {3i}4 D_{(\alpha} S_{|\um | \beta ) \dgamma}~, \qquad 
		K_{[\ul{mn}]_{14} \alpha \beta \dgamma} = \tfrac i{16} \bar D_{\dgamma} \bm R_{\alpha \beta \ul{mn}}~. \\
		\tilde P_{\ul{mn} \dalpha} &= \tfrac 1{12} \bar D_{\dalpha}  \left[ \tilde{\bar R}_{\ul{mn}} -\tilde R_{\ul{mn}}  \right] - \tfrac i6 D^{\beta} \tilde X_{\beta \dalpha \ul{mn}}~, ~~ 
		\tilde P_{\dalpha} = \tfrac 1{12} \bar D_{\dalpha} \left[  \tilde{\bar R} - \tilde R \right] -i D^{\beta} X_{\beta \dalpha}~. \\
		P_{\um \dalpha} &= \tfrac 38 \bar \rho_{\um \dalpha} - \tfrac i2 \bar D_\dalpha S_\um - \tfrac i8 D^\beta S_{\um \beta \dalpha} - \tfrac 1{96} [ D^2 + \bar D^2 ] \bar \lambda_{\um \dalpha} - \tfrac 1{96} [ D^\beta \bar D_\dalpha +2 \bar D_\dalpha D^\beta ] \lambda_{\um \beta}~, \nonumber \\
		& \qquad \qquad \qquad \qquad \qquad \qquad \qquad  M_{[\ul{mn}]_{14} \dalpha} = - \tfrac i{48} \bar D^{\dbeta} \bm R_{\dbeta \dalpha \ul{mn}} - \tfrac i4 T_{[\ul{mn}]_{14}  \dalpha}~.
	\end{align}
\end{subequations}

Finally, we have the torsion components $\bm T_{\um \un, \ul{p \alpha}}$ with three internal indices. The internal indices we decompose step by step, first under $SL(7)$, then under $SO(7)$, and finally under $G_2$. The $SL(7)$ decomposition results in a totally anti-symmetric piece $V$ and a mixed symmetric piece $W$
\begin{align}
	\bm T_{\ul{mn}, \up \alpha} & \overset{SL_7}{=\joinrel =} V_{[\ul{mnp}], \alpha} 
        + W_{\um \un | \up \,\alpha} 
\end{align}
We use $\um \un | \up$ to denote the tableaux 
${\tiny\begin{ytableau}
	\um & \up \\
		\un
\end{ytableau}}$.
This is $\bm{21} \times \bm 7 = \bm{35} + \bm{112}$ for $SL(7)$. Under $G_2$, the $\bm{35}$ decomposes into $\bm 1 + \bm 7 + \bm{27}$ in the following manner:
\begin{align}
	V_{[\ul{mnp}], \alpha} &= \tfrac 1{42} \varphi_{\ul{mnp}} V_{\alpha} + \tfrac 1{24} \psi_{\ul{mnpq}} V^{\uq} {}_{\alpha} + \tfrac 34 \varphi^{\uq} {}_{[\ul{mn}} V_{\up] \uq, \alpha} ~,
\end{align}
where $V_\alpha$ is in $\bm 1$, $V_{\um \alpha}$ is in $\bm 7$, and $V_{\um \un \alpha}$ is a traceless symmetric $\bm{27}$:
\begin{align}
	V_{\um \un \alpha} &= V_{\un \um \alpha}~, \qquad ~ \delta^{\um \un} V_{\um \un \alpha} = 0~.
\end{align}
Under $SO(7)$, $W$ decomposes into $\bm 7 + \bm{105}$:
\begin{subequations}
	\begin{align}
		W_{\um \un |\up \,\alpha} &= J_{\ul{mn}| \up, \alpha} + \delta_{\ul{pm}} \Upsilon_{\un \alpha} - 					\delta_{\ul{pn}} \Upsilon_{\um \alpha} \\
		J_{\um \un| \up \alpha} &= - J_{\un \um| \up \alpha}~, \qquad J_{[\um \un| \up] \alpha} = 0~, \qquad \delta^{\ul{np}} J_{\um \un| \up \alpha} = 0~.
	\end{align}
\end{subequations}
Under $G_2$, the $\bm{105}$ decomposes further into $\bm{14} + \bm{27} + \bm{64}$:
\begin{subequations}
	\begin{align}
		J_{\ul{mn}| \up, \alpha} &= J^{14}_{\ul{mn}| \up, \alpha} + J^{27}_{\ul{mn}| \up, \alpha} + J^{64}_{\ul{mn}| \up, \alpha}
	\end{align}
\end{subequations}
where $\um \un|\up$ now denotes the irreducible hook representation of $SO(7)$.
Let us parameterize $J^{14}$ by a 2-form $J_{\ul{mn}, \alpha}$ in the $\bm{14}$ and $J^{27}$ by a rank 2 symmetric traceless tensor $I_{\ul{mn}, \alpha}$ as
\begin{subequations}
	\begin{align}
		J^{14}_{\ul{mn}| \up, \alpha} &= \varphi_{\ul{mn}} {}^\uq J_{\ul{pq}, \alpha} - \tfrac 12 \varphi_{\ul{np}} {}^\uq J_{\ul{qm}, \alpha} + \tfrac 12 \varphi_{\ul{mp}} {}^\uq J_{\ul{qn}, \alpha} \\
		J^{27}_{\ul{mn}| \up, \alpha} &= \varphi_{\ul{mn}} {}^\uq I_{\ul{qp}, \alpha} - \tfrac 12 \varphi_{\ul{np}} {}^\uq I_{\ul{qm}, \alpha} + \tfrac 12 \varphi_{\ul{mp}} {}^\uq I_{\ul{qn}, \alpha} ~,
	\end{align}
\end{subequations}
which can be inverted as
\begin{subequations}
	\begin{align}
		\varphi_\um {}^{\ul{pq}} J^{14}_{\ul{pq}| \un, \alpha} &= 9 J_{\ul{mn}, \alpha}~, ~~ \qquad ~~  \varphi_\um {}^{\ul{pq}} J^{14}_{\ul{np}| \uq, \alpha} = - \tfrac 92 J_{\ul{mn}, \alpha} \\
		\varphi_\um {}^{\ul{pq}} J^{27}_{\ul{pq}| \un, \alpha} &= 7 I_{\ul{mn}, \alpha}, ~~ \qquad ~~  \varphi_\um {}^{\ul{pq}} J^{27}_{\ul{np}| \uq, \alpha} = - \tfrac 72 I_{\ul{mn}, \alpha}~.
	\end{align}
\end{subequations}
We do not give an explicit parameterization of $J^{64}$, since it is just the remaining piece of $J_{\um \un| \up \alpha}$, and denote it instead by $Z_{\um \un| \up \alpha}$. From the fact that $J$, $J^{14}$, $J^{27}$ have the same mixed-symmetry,  $Z_{\um \un| \up \alpha}$ must satisfy
\begin{align}
	Z_{\ul{mn}| \up, \alpha} &= - Z_{\ul{nm}| \up, \alpha}~, \qquad 
	\delta^{\ul{np}} Z_{\ul{mn}| \up, \alpha} = 0~, \qquad
	Z_{[\ul{mn}| \up], \alpha} = 0~,
\end{align}
in addition to the irreducibility conditions\footnote{Both contractions belong to the $\bm 7 \times \bm 7$ of $G_2$ which decomposes as $\bm 1 + \bm 7 + \bm{14} + \bm{27}$, with no $\bm{64}$.}
\begin{align}
	\varphi_\uq {}^{\ul{mn}} Z_{\ul{mn}| \up, \alpha} &= 0~, \qquad 
	\varphi_\uq {}^{\ul{np}} Z_{\ul{mn}| \up, \alpha} = 0~.
\end{align}
Combining everything, we have the following equations:
\begin{subequations}
	\begin{align}
		\varphi_\um {}^{\ul{pq}} \bm T_{\ul{np}, \uq, \alpha} &= \tfrac 17 \delta_{\ul{mn}} V_\alpha + (V_{\ul{mn}, \alpha} - \tfrac 72 I_{\ul{mn} ,\alpha}) - \tfrac 92 J_{\ul{mn}, \alpha} + \varphi_{\ul{mn}} {}^\up (\tfrac 16 V_{\up \alpha} - \Upsilon_{\up \alpha}) \\
		\varphi_\um {}^{\ul{pq}} \bm T_{\ul{pq}, \un, \alpha} &= \tfrac 17 \delta_{\ul{mn}} V_\alpha + (V_{\ul{mn}, \alpha} + 7 I_{\ul{mn} ,\alpha}) +  9 J_{\ul{mn}, \alpha} + \varphi_{\ul{mn}} {}^\up (-\tfrac 16 V_{\up \alpha} + 2 \Upsilon_{\up \alpha})~.
	\end{align}
\end{subequations}
Plugging these irrep decompositions into the Bianchi identities, one finds
\begin{subequations}
	\begin{align}
		V_\alpha &= i D_\alpha \tilde{\bar R} \\
		V_{\um \alpha} &= - \tfrac {3i}2 \rho_{\um \alpha} - \bar D^\dbeta [ \tfrac 32 S_{\um \alpha \dbeta} + \bar S_{\um \alpha \dbeta}] + 2 D_\alpha S_\um \nonumber \\
		&~~~~ + \tfrac i{24} [ D^2 + 4 \bar D^2] \lambda_{\um \alpha} + \tfrac 1{12} [D_\alpha \bar D_\dbeta + 2 \bar D_\dbeta D_\alpha ] \bar \lambda_\um {}^\dbeta + \tfrac 13 \varphi_\um {}^{\ul{np}} \pa_\un \lambda_{\up \alpha}  \\
		V_{\ul{mn} \alpha} &= \tfrac i{18} D_{\alpha} (\tilde{\bar R}_{\ul{mn}} - \tilde R_{\ul{mn}}) + \tfrac 49 \bar D^{\dbeta} \tilde X_{\alpha \dbeta \ul{mn}} + \tfrac 19 \partial_{(\um} \lambda_{\un)_{\textrm{traceless}}, \alpha} \\
		\Upsilon_{\um \alpha} &= \tfrac i{12} D_\alpha \bar R_\um - \tfrac {3i}{16} \rho_{\um \alpha} - \tfrac 18 \bar D^\dbeta [ \tfrac 32 S_{\um \alpha \dbeta} + \bar S_{\um \alpha \dbeta} ] + \tfrac 14 D_\alpha S_\um \nonumber \\
		& ~~~~ + \tfrac i{192} [D^2 + 4 \bar D^2] \lambda_{\um \alpha} + \tfrac i{96} [ D_\alpha \bar D_\dbeta + 2 \bar D_\dbeta D_\alpha ] \bar \lambda_\um {}^\dbeta + \tfrac 1{24} \varphi_\um {}^{\ul{np}} \pa_\un \lambda_{\up \alpha} \\
		J_{\ul{mn} \alpha} &= \tfrac i9 D_{\alpha} L_{[\ul{mn}]_{\boldsymbol{14}}} + \tfrac 1{54} D^{\beta} R_{\beta \alpha \ul{mn}} \\
		I_{\ul{mn} \alpha} &= - \tfrac i{14} D_{\alpha} [\tfrac {17}{18} \tilde{\bar R}_{\ul{mn}} + \tfrac 1{18} \tilde R_{\ul{mn}} ] + \tfrac 2{63} \bar D^{\dbeta} \tilde X_{\alpha \dbeta \ul{mn}} + \tfrac 1{126} \partial_{(\um} \lambda_{\un)_{\textrm{traceless}} ,\alpha}~.
	\end{align}
\end{subequations}
$Z_{\ul{mn}| \up \alpha}$ is unconstrained by Bianchi identities at this dimension.
Therefore, the new/independent superfields in the torsion components at dimension $\tfrac 32$ are $T_{[\ul{mn}]_{14} \alpha}$ and $Z_{\ul{mn}| \up \alpha}$.

\paragraph{Curvature.}
Curvature components at dimension $\tfrac 32$ are
\begin{itemize}
	\item $\bm R_{\hd \hgamma ~ \hb \ha}  \xrightarrow{4|4+7}  \bm R_{\hd \ul \gamma ~ \hb \ha}$ 
\end{itemize}
Bianchi identities at this dimension fully determine these curvature components. The purely 4D components are as in \cite{Wess:1992cp},
\begin{subequations}
	\begin{align}
		\bm R_{\alpha b ~ cd} &= i \left[ (\sigma_b)_{\alpha \dbeta} \bm T_{cd} {}^\dbeta - (\sigma_c)_{\alpha \dbeta} \bm T_{db} {}^\dbeta - (\sigma_d)_{\alpha \dbeta} \bm T_{bc} {}^\dbeta  \right] \\
		\bm R_{\dalpha b ~ cd} &= - i \left[ (\sigma_b)_{\beta \dalpha} \bm T_{cd} {}^\beta - (\sigma_c)_{\beta \dalpha} \bm T_{db} {}^\beta - (\sigma_d)_{\beta \dalpha} \bm T_{bc} {}^\beta  \right]~.
	\end{align}
\end{subequations}
Similarly, components with at least one internal index are as follows:
\begin{subequations}
	\begin{align}
		\bm R_{\alpha \um, ab} &= -i \bm T_{ab, \um \alpha} - 2i  (\sigma_{[a})_{|\alpha \dbeta|} \bm T_{b] \um} {}^\dbeta~, \qquad
		\bm R_{\dalpha \um, ab} = i \bm T_{ab, \um \dalpha} + 2i  (\sigma_{[a})_{|\beta \dalpha|} \bm T_{b] \um} {}^\beta \\
		\bm R_{\alpha a, b \um} & = i \bm T_{ab, \um \alpha} + 2i (\sigma_{(a})_{|\alpha \dbeta|} \bm T_{b) \um} {}^\dbeta~, \qquad ~
		\bm R_{\dalpha a, b \um} = -i \bm T_{ab, \um \dalpha} - 2i (\sigma_{(a})_{| \beta \dalpha |} \bm T_{b) \um} {}^\beta \\
		\bm R_{\alpha a, \ul{mn}} &= i (\sigma_{a})_{\alpha \dbeta} \bm T_{\ul{mn}} {}^\dbeta + 2i \bm T_{a [\um, \un] \alpha}~, \qquad ~~
		\bm R_{\dalpha a, \ul{mn}} = -i (\sigma_{a})_{\beta \dalpha} \bm T_{\ul{mn}} {}^\beta - 2i \bm T_{a [\um, \un] \dalpha} \\
		\bm R_{\alpha \um, a \un} &= i (\sigma_a)_{\alpha \dbeta} \bm T_{\ul{mn}} {}^\dbeta - 2i \bm T_{a (\um, \un) \alpha}~, \qquad~~
		\bm R_{\dalpha \um, a \un} = - i (\sigma_a)_{\beta \dalpha} \bm T_{\ul{mn}} {}^\beta + 2i \bm T_{a (\um, \un) \dalpha} \\
		\bm R_{\alpha \um , \ul{np}} &= i [ \bm T_{\ul{mn}, \up \alpha} - \bm T_{\ul{np}, \um \alpha} + \bm T_{\ul{pm}, \un \alpha}  ], ~~~
		\bm R_{\dalpha \um , \ul{np}} = -i [ \bm T_{\ul{mn}, \up \dalpha} - \bm T_{\ul{np}, \um \dalpha} + \bm T_{\ul{pm}, \un \dalpha}  ]~.
	\end{align}
\end{subequations}

\paragraph{4-form.}
There are no 4-form components beyond dimension 1.

\paragraph{Relations.}

We repeat (and in some cases state for the first time) some of the important relations implied by the Bianchi identities at dimension $\tfrac 32$:
\begin{subequations}
	\begin{align}
		\bar D_{\dalpha} R &= 0 = D_{\alpha} \bar R \\
		D^{\beta} G_{\beta \dalpha} &= \bar D_{\dalpha} \bar R \\
		D_{\alpha} \bar S_{\um b} &= \tfrac i4 (\sigma_b)_{\alpha \dbeta} \bar \rho_{\um} {}^{\dbeta} \iff D_{\alpha} \bar S_{\um \beta \dgamma} = \tfrac i2 \epsilon_{\alpha \beta} \bar \rho_{\um \dgamma} \\
		D_\alpha \left( R_\um - \bar R_\um \right) &= 6i D_\alpha S_\um + 4i \bar D^\dbeta \bar S_{\um \alpha \dbeta} - \tfrac 18 D^2 \lambda_{\um \alpha} - \tfrac 14 D_\alpha \bar D_\dbeta \bar \lambda_\um {}^\dbeta \\
		\tfrac 1{42} D_\alpha \cG_\um &= V_{\um \alpha} = - \tfrac {3i}2 \rho_{\um \alpha} - \bar D^\dbeta [ \tfrac 32 S_{\um \alpha \dbeta} + \bar S_{\um \alpha \dbeta}] + 2 D_\alpha S_\um + \tfrac i{24} [ D^2 + 4 \bar D^2] \lambda_{\um \alpha} \nonumber \\
		&\qquad \qquad  + \tfrac 1{12} [D_\alpha \bar D_\dbeta + 2 \bar D_\dbeta D_\alpha] \bar \lambda_\um {}^\dbeta + \tfrac 13 \varphi_\um {}^{\ul{np}} \pa_\un \lambda_{\up \alpha} \\
		\bar D_\dalpha \left( \tfrac 12 \tilde R + \tfrac 16 \tilde{\bar R}  \right) &= \tfrac i3 \partial^\um \bar \lambda_{\um \dalpha} + 3i D^\beta X_{\beta \dalpha} \\
		D_\alpha \left( \tilde{\bar R} - \tfrac i{48} \cG  \right) &= 0 \\
		D_\alpha \cG_{\ul{mn}} &= - \tfrac {2i}{21} D_\alpha \left( 4 \tilde{\bar R}_{\ul{mn}} - \tilde R_{\ul{mn}}  \right) - \tfrac {16}{21} \bar D^\dbeta \tilde X_{\alpha \dbeta \ul{mn}} - \tfrac 4{21} \partial_{(\um} \lambda_{\un)_{\textrm{traceless}}, \alpha}~.
	\end{align}
\end{subequations}

\subsubsection{Dimension 2}
There are no torsion components beyond dimension $\tfrac 32$, and no 4-form components beyond dimension 1. Dimension 2 Bianchi identities either determine various components of the curvature tensor, or imply the algebraic Bianchi identities satisfied by it.

\paragraph{Curvature.}
The curvature component at dimension 2 is
\begin{itemize}
	\item $\bm R_{\hd \hc ~ \hb \ha}  $ 
\end{itemize}
Some Bianchi identities at dimension 2 imply the familiar algebraic identities satisfied by the bosonic Riemann tensor, namely
\begin{align}
	\bm R_{[\hd \hc, \hb] \ha} &= 0~,
\end{align}
which, in turn, imply the pair-exchange symmetry $\bm R_{\hd \hc ~ \hb \ha} = \bm R_{\hb \ha ~ \hd \hc}$.

The purely 4D torsion Bianchi identities of dimension 2 are 
\begin{subequations}
	\begin{align}
		\bm R_{[dc, b]a} & = 0 \\
		\bm R_{dc, \beta \alpha} &= D_\beta \bm T_{dc, \alpha} + \pa_d \bm T_{c \beta, \alpha} - \pa_c \bm T_{d \beta, \alpha}~, \qquad 
		\bm R_{dc, \dbeta \dalpha} = \bar D_\dbeta \bm T_{dc, \dalpha} + \pa_d \bm T_{c \dbeta, \dalpha} - \pa_c \bm T_{d \dbeta, \dalpha} \\
		0 &= D_\beta \bm T_{dc, \dalpha} + \pa_d \bm T_{c \beta, \dalpha} - \pa_c \bm T_{d \beta, \dalpha}~, \qquad \qquad ~
		0 = \bar D_\dbeta \bm T_{dc, \alpha} + \pa_d \bm T_{c \dbeta, \alpha} - \pa_c \bm T_{d \dbeta, \alpha}
	\end{align}
\end{subequations}
The component $\bm R_{dc, ba}$ can be decomposed into $SL(2,\mathbb{C})$ irreducible pieces as follows:
\begin{align}
	R_{\delta \ddelta, \gamma \dgamma, \beta \dbeta, \alpha \dalpha} &= (\sigma^d)_{\delta \ddelta} (\sigma^c)_{\gamma \dgamma} (\sigma^b)_{\beta \dbeta} (\sigma^a)_{\alpha \dalpha} \bm R_{dc, ba} \nonumber \\
	&= 4 \left[ \epsilon_{\ddelta \dgamma} \epsilon_{\dbeta \dalpha} X_{(\delta \gamma) (\beta \alpha)} + \epsilon_{\delta \gamma} \epsilon_{\beta \alpha} \bar X_{(\ddelta \dgamma) (\dbeta \dalpha)} - \epsilon_{\ddelta \dgamma} \epsilon_{\beta \alpha} \Psi_{(\delta \gamma) (\dbeta \dalpha)} - \epsilon_{\delta \gamma} \epsilon_{\dbeta \dalpha} \bar \Psi_{(\ddelta \dgamma) (\beta \alpha)} \right]~,
\end{align}
where $\bar \Psi$ and $\bar X$ denote complex conjugates of $\Psi$ and $X$ respectively. The algebraic Bianchi identity is satisfied if and only if the following conditions are met:
\begin{align}
	\Psi _{(\delta \gamma) (\dbeta \dalpha)} &= \bar \Psi_{(\dbeta \dalpha)(\ddelta \dgamma)} ~, \qquad 
	\epsilon^{\beta \delta} X_{(\alpha \beta)(\gamma \delta)} = \epsilon_{\alpha \gamma} \Lambda~, \qquad \bar \Lambda = \Lambda
\end{align}
The other two Bianchi identities determine completely the irreducible pieces $X$ and $\Psi$ in terms of lower dimensional superfields ($W_{\gamma \beta \alpha}$, $G_{\alpha \dbeta}$, $R$), and imply some derivative relations between the said lower dimensional superfields. We have,
\begin{subequations}
	\begin{align}
		X_{(\delta \gamma)(\beta \alpha)} &= - D_{(\alpha} W_{\beta) \delta \gamma} + \tfrac 12 \epsilon_{(\delta |(\beta} D_{\alpha)|} \bar D^\dgamma G_{\gamma)\dgamma}  + \tfrac i2 \epsilon_{(\beta | (\delta} \pa_{\gamma)|} {}^\dgamma G_{\alpha ) \dgamma} \\
		\bar \Psi_{(\ddelta \dgamma)} {}^{(\beta \alpha)} &= - \tfrac 14 \left[  D^{(\beta} \bar D_{(\ddelta} G^{\alpha)} {}_{\dgamma)} - \bar D_{(\ddelta} D^{(\beta} G^{\alpha)} {}_{\dgamma)}  \right] ~.
	\end{align}
\end{subequations}
Now we move on to the following components with one internal index.
\begin{subequations}
\label{E:R_{mc,ba}}
	\begin{align}
		\bm R_{\um c, ba} &= - (\sigma_{ba})^{\beta \alpha} \bm R_{\um c, \beta \alpha} - (\bar \sigma_{ba})^{\dbeta \dalpha} \bm R_{\um c, \dbeta \dalpha} \\[2ex]
\label{E:R_{mc,ba.b}}
		R_{\um, \gamma \dgamma, \beta \alpha} &= (\sigma^c)_{\gamma \dgamma} \bm R_{\um c, \beta \alpha}    
		= (\sigma^c)_{\gamma \dgamma} \left(  \partial_{\um} \bm T_{c \beta, \alpha} + \partial_c \bm T_{\beta \um, \alpha} - D_\beta \bm T_{c \um , \alpha}   \right) \eol
		&= - \tfrac i4 \left[ \epsilon_{\gamma \alpha} \partial_{\um} G_{\beta \dgamma} - 3 \epsilon_{\beta \gamma} \partial_\um G_{\alpha \dgamma} - 3 \epsilon_{\beta \alpha} \partial_\um G_{\gamma \dgamma}  \right] + \partial_{\gamma \dgamma} \left( i \epsilon_{\beta \alpha}  S_\um - S_{\um \beta \alpha} \right) \nonumber \\
		& ~~ + \tfrac 12 \epsilon_{\beta (\gamma} D^2 S_{| \um | \alpha) \dgamma} + 2i D_\beta \bar D_\dgamma S_{\um \gamma \alpha} + \tfrac 12 \epsilon_{\gamma \alpha} \left[ D_{\beta} \bar D_\dgamma S_\um - \tfrac 14 D^2 S_{\um \beta \dgamma} \right] \\[2ex]
\label{E:R_{mc,ba.c}}
		R_{\um, \gamma \dgamma, \dbeta \dalpha} &= (\sigma^c)_{\gamma \dgamma}  \bm R_{\um c, \dbeta \dalpha} 
		= (\sigma^c)_{\gamma \dgamma} \left(  \partial_\um \bm T_{c \dbeta, \dalpha} + \partial_c \bm T_{\dbeta \um, \dalpha} - \bar D_\dbeta \bm T_{c \um , \dalpha}   \right)  \eol
		&=  \tfrac i4 \left[ \epsilon_{\dgamma \dalpha} \partial_{\um} G_{\gamma \dbeta} - 3 \epsilon_{\dbeta \dgamma} \partial_\um G_{\gamma \dalpha} - 3 \epsilon_{\dbeta \dalpha} \partial_\um G_{\gamma \dgamma}  \right] + \partial_{\gamma \dgamma} \left( - i \epsilon_{\dbeta \dalpha}  S_\um + \bar S_{\um \dbeta \dalpha} \right) \nonumber \\
		& ~~ + \tfrac 12 \epsilon_{\dbeta (\dgamma} \bar D^2 \bar S_{| \um \gamma | \dalpha)} - 2i \bar D_\dbeta D_\gamma \bar S_{\um \dgamma \dalpha} + \tfrac 12 \epsilon_{\dgamma \dalpha} \left[ - \bar D_{\dbeta} D_\gamma S_\um - \tfrac 14 \bar D^2 \bar S_{\um \gamma \dbeta} \right] 
	\end{align}
\end{subequations}
This clearly means that the $\alpha \beta$ or $\dalpha \dbeta$ antisymmetric pieces above will vanish, yielding relations between derivatives of the superfields involved, and the expressions for $R_{\ul{mn}, \beta \alpha}$ and $R_{\ul{mn}, \dbeta \dalpha}$ will contain only the symmetric pieces of the right hand sides. Also, from pair exchange symmetry, we conclude that $\bm R_{ba, \um c} = \bm R_{\um c, ba}$. 

The rest of the components of the Riemann tensor have more than one internal index and are to be $G_2$ decomposed. First, we consider this component with two internal indices:
\begin{subequations}
	\begin{align}
		\bm R_{\ul{mn}, ba} &= - (\sigma_{ba})^{\beta \alpha} \bm R_{\ul{mn}, \beta \alpha} - (\bar \sigma_{ba})^{\dbeta \dalpha} \bm R_{\ul{mn}, \dbeta \dalpha} \\
		\bm R_{\ul{mn}, \beta \alpha} &= 2 \partial_{[\um} \bm T_{\un] \beta, \alpha} + D_\beta \bm T_{\ul{mn}, \alpha} \\
		\bm R_{\ul{mn}, \dbeta \dalpha} &= 2 \partial_{[\um} \bm T_{\un] \dbeta, \dalpha} + \bar D_\dbeta \bm T_{\ul{mn}, \dalpha}
	\end{align}
\end{subequations}
This again has consequences similar to \eqref{E:R_{mc,ba}} which we do not spell out in detail. We note that $\bm R_{\um \un, ba}$ is an internal 2-form, and hence decomposes into a $\bm 7$ and a $\bm{14}$ of $G_2$.

The component $\bm R_{dc, \ul{np}}$ is antisymmetric in the two internal indices, decomposing into a $\bm 7$ and a $\bm{14}$ of $G_2$. We have
\begin{align}
	\bm R_{dc, \ul{np}} &= \tfrac 16 \varphi_{\ul{np}} {}^\uq \mathcal R_{dc \uq} + R_{dc, [\ul{np}]_{14}}~. 
\end{align}
Each of these irreducible components get fully determined by Bianchi identities:
\begin{subequations}
	\begin{align}
		\mathcal R_{dc \um} &= 6i \pa_{[d} \left(  S_{| \um | c]} + \bar S_{| \um | c]}  \right) - \tfrac 14 (\bar \sigma_{[c})^{\dgamma \gamma} \pa_{d]} \left( D_{\gamma} \bar \lambda_{\um \dgamma} - \bar D_\dgamma \lambda_{\um \gamma}  \right) \\
		R_{dc, [\ul{mn}]_{14}} &= R_{[\ul{mn}]_{14} , dc} \qquad \textrm{(by pair exchange)} \\
		&= - (\sigma_{dc})^{\beta \alpha} \left[  2 \pa_{[\um} S_{\un]_{14} \beta, \alpha} + D_\beta T_{[\ul{mn}]_{14}, \alpha}  \right] - (\bar \sigma_{dc})^{\dbeta \dalpha} \left[  2 \pa_{[\um} \bar S_{\un]_{14} \dbeta, \dalpha} + \bar D_\dbeta \bar T_{[\ul{mn}]_{14}, \dalpha}  \right].
	\end{align}
\end{subequations}
The other component with two internal indices is $\bm R_{\un a, b \um}$ belonging in $\bm 7 \times \bm 7$ of $G_2$. First, we $G_2$ decompose it,
\begin{subequations}
	\begin{align}
		\bm R_{\un a, b \um} & = \tfrac 17 \delta_{\ul{nm}} \mathcal S_{ab} + \tilde{\mathcal S}_{ab \ul{nm}}  + \tfrac 16 \varphi_{\ul{nm}} {}^\up \mathcal S_{ab \up} + \mathcal S_{ab [\ul{nm}]_{\boldsymbol{14}}}  \\
		\tilde{\mathcal S}_{ab \ul{nm}} &= \tilde{\mathcal S}_{ab \ul{mn}}~, \qquad
		\delta^{\ul{nm}} \bm R_{\un a, b \um} = \mathcal S_{ab}~.
 	\end{align}
\end{subequations}
The two 4D vector indices $a$ and $b$ on each $G_2$-irreducible piece above can be decomposed into symmetric and antisymmetric parts. We choose not to do this explicitly. Instead, we give the full expressions for $\mathcal S_{ab}$, $\tilde{\mathcal S}_{ab \un \um}$, $\mathcal S_{ab \um}$ and $\mathcal S_{ab [\un \um]_{14}}$ as determined by Bianchi identities of dimension 2:
\begin{subequations}
	\begin{align}
		\mathcal S_{ab} &= i \eta_{ab} \pa^\un S_\un - 3 (\bar \sigma_{ab})^{\dalpha \dgamma} \pa^\un \bar S_{\un \dalpha \dgamma} + (\sigma_{ab})^{\beta \gamma} \pa^\un S_{\un \beta \gamma}  \nonumber \\
		& ~~~ -2i \pa_a X_b - \tfrac i{24} \eta_{ab} D^2 \left( \tilde R - \tilde{\bar R} \right) - \tfrac 12 \eta_{ab} D^\rho \bar D^{\dot \rho} X_{\rho \dot \rho} - (\sigma_{ab})^{\beta \gamma} D_\beta \bar D^{\dot \rho} X_{\gamma \dot \rho} \nonumber \\
		& ~~~ - \tfrac 12 (\bar \sigma_a)^{\dgamma \gamma} (\bar \sigma_b)^{\dalpha \beta} D_\beta \bar D_{( \dalpha } X_{|\gamma| \dgamma)} \\
		\tilde{\mathcal S}_{ab \ul{nm}} &= \pa_{(\un} \left[ i \eta_{|ab|}  S_{\um)} - 3 (\bar \sigma_{|ab|})^{\dalpha \dbeta} \bar S_{\um) \dalpha \dbeta} + (\sigma_{|ab|})^{\alpha \beta}  S_{\um)  \alpha \beta} \right]_{(\ul{nm})_{\textrm{traceless}}} \nonumber \\
		& ~~~ -2i \pa_a \tilde X_{b \ul{nm}} - \tfrac 12 (\bar \sigma_a)^{\dgamma \gamma} (\bar \sigma_b)^{\dalpha \beta} D_\beta \bar D_{(\dalpha} \tilde X_{|\gamma| \dgamma) \ul{nm}} - \tfrac 1{12} \eta_{ab} D^\rho \bar D^{\dot \rho} \tilde X_{\rho \dot \rho \ul{nm}}  \nonumber \\
		& ~~~ + \tfrac 16 (\sigma_{ab})^{\alpha \beta} D_\alpha \bar D^{\dot \rho} \tilde X_{\beta \dot \rho \ul{nm}} + \tfrac i{24} \eta_{ab} D^2 \left( \tilde R_{\ul{nm}} - \tilde{\bar R}_{\ul{nm}}  \right) \\
		\mathcal S_{ab \um} &= \varphi_\um {}^{\ul{np}} \pa_\un \left[ i \eta_{ab} S_\up - 3 (\bar \sigma_{ab})^{\dalpha \dbeta} \bar S_{\up \dalpha \dbeta} + (\sigma_{ab})^{\alpha \beta} S_{\up \alpha \beta}   \right] - \eta_{ab} D^2 \left[ \tfrac 12 S_\um - \tfrac i{48} \bar D^\dbeta \bar \lambda_{\um \dbeta} \right] \nonumber \\
		& ~~~ + (\bar \sigma_b)^{\dbeta \beta} \pa_a \left[  \tfrac 18 \left(  D_\beta \bar \lambda_{\um \dbeta} + \bar D_\dbeta \lambda_{\um \beta}  \right) - \tfrac {3i}2 \left(  S_{\um \beta \dbeta} + \bar S_{\um \beta \dbeta}   \right)   \right] \nonumber \\
		& ~~~ + (\bar \sigma_b)^{\dbeta \beta} (\bar \sigma_a)^{\dalpha \alpha} D_\beta \left[  \tfrac i{16} \bar D_{(\dbeta} D_{|\alpha} \bar \lambda_{\um| \dalpha)} + \tfrac 34 \bar D_{(\dbeta} \bar S_{| \um \alpha| \dalpha)}  \right]  \nonumber  \\
		& ~~~ - \tfrac 18 (\sigma_a \bar \sigma_b)^{\alpha \beta} D_\beta \left[ 3i \rho_{\um \alpha} -  \bar D^\dbeta \bar S_{\um \alpha \dbeta} - \tfrac i{12} \bar D^2 \lambda_{\um \alpha} + \tfrac i{12} \bar D^\dbeta D_\alpha \bar \lambda_{\um \dbeta}   \right] \\
		\mathcal S_{ab [\ul{nm}]_{14}} &= \pa_{[\un} \left[ i \eta_{|ab|}  S_{\um ]_{14}} + \tfrac 32 (\bar \sigma_{|b} \sigma_{a|})^{\dalpha \dgamma} \bar S_{\um]_{14} \dalpha \dgamma} - 3 (\bar \sigma_{|ab|})^{\dalpha \dgamma} \bar S_{\um]_{14} \dalpha \dgamma} + (\sigma_{|ab|})^{\beta \gamma} S_{\um]_{14} \beta \gamma}       \right] \nonumber \\
		& ~~~ + \tfrac 1{16} D^2 \left( (\bar \sigma_{ab})^{\dbeta \dgamma} \bm R_{\dbeta \dgamma \ul{nm}}   - \tfrac 13 (\sigma_{ab})^{\beta \gamma} \bm R_{\beta \gamma \ul{nm}}   \right) +
		\tfrac 14 (\sigma_a \bar \sigma_b)^{\gamma \beta} D_\beta T_{[\ul{nm}]_{14}, \gamma}~.
 	\end{align}
\end{subequations}
We note that the component $\bm R_{ab, \um \un} = \bm R_{\um \un, ab}$ is related through the algebraic Bianchi identity to the antisymmetric part in $ab$ of $\bm R_{\un a, b \um}$ given above. This leads to another relation which we quote in the next paragraph (\ref{E:dim2-reln4}f).

Next, we consider $\bm R_{\un c, \ul{pq}}$, a curvature component with three internal indices. As a first step, we decompose the last two (antisymmetric) internal indices into a $\bm 7$ and a $\bm{14}$,
\begin{align}
	\bm R_{\un c, \ul{pq}} &= \tfrac 16 \varphi_{\ul{pq}} {}^\ur \mathcal R_{\un c, \ur} + R_{\un c, [\ul{pq}]_{14}}~,
\end{align}
where $\mathcal R_{\un c, \um}$ is in $\bm 7 \times \bm 7 = \bm 1 + \bm{27} + \bm 7 + \bm{14}$ of $G_2$:
\begin{subequations}
	\begin{align}
		\mathcal R_{\un c, \um} &=: \tfrac 17 \delta_{\ul{nm}} \mathcal R_c + \tilde{\mathcal R}_{c \ul{nm}} + \tfrac 16 \varphi_{\ul{nm}} {}^\up \mathcal R_{c \up} + \mathcal R_{c [\ul{nm}]_{14}}  \\
		\tilde{\mathcal R}_{\ul{nm}} &= \tilde{\mathcal R}_{\ul{mn}}~, \qquad
		 \delta^{\ul{nm}} \mathcal R_{\un c, \um} = \mathcal R_c ~.
	\end{align}
\end{subequations}
These superfields are determined completely by Bianchi identities:
\begin{subequations}
	\begin{align}
		\mathcal R_c &= - \pa_c \tilde{\bar R} + \tfrac i{12} (\sigma_c)_{\alpha \dbeta} D^{\alpha} \bar D^{\dbeta} \left(  \tilde{\bar R} - \tilde R  \right) - 2 D^2 X_c + i \pa^\un \left( 5 \bar S_{\un c} + 3 S_{\un c} \right) \nonumber \\
		& ~~~~ - \tfrac 18 (\bar \sigma_c)^{\dgamma \gamma } \pa^\un \left(  D_{\gamma} \bar \lambda_{\un \dgamma} - \bar D_\dgamma \lambda_{\un \gamma}  \right)  \\
		\tilde{\mathcal R}_{c \ul{nm}} &= - \pa_c \tilde{\bar R}_{\ul{nm}} + \tfrac i{12} (\sigma_c)_{\alpha \dbeta} D^\alpha \bar D^\dbeta \left(  \tilde{\bar R}_{\ul{nm}} - \tilde R_{\ul{nm}}  \right) - \tfrac 43 D^2 \tilde X_{c \ul{nm}}  \nonumber \\
		& ~~ + \left[   i \pa_{\un} \left( 5 \bar S_{\um c} + 3 S_{\um c} \right) - \tfrac 18 (\bar \sigma_c)^{\dgamma \gamma}  \pa_\un \left( D_{\gamma} \bar \lambda_{\um \dgamma} - \bar D_\dgamma \lambda_{\um \gamma}  \right) \right]_{(\ul{nm})_{\textrm{traceless}}} \\
		\mathcal R_{\alpha \dbeta \um} &= (\sigma^a)_{\alpha \dbeta} \mathcal R_{a \um}  \nonumber \\
		&= 2 \pa_{\alpha \dbeta} \bar R_\um - \tfrac {3i}2 D_\alpha \bar \rho_{\um \dbeta} - 2 D_\alpha \bar D_\dbeta S_\um + \tfrac 52 D^2 S_{\um \alpha \dbeta}  \nonumber \\
		&~~~ + \tfrac i{24} D_\alpha \bar D^2 \bar \lambda_{\um \dbeta} + \tfrac i6 D^2 \bar D_\dbeta \lambda_{\um \alpha} + \tfrac i{12} D_\alpha \bar D_\dbeta D^\beta \lambda_{\um \beta} \nonumber \\
		& ~~~ + \varphi_\um {}^{\ul{np}} \pa_\un \left[ 5i \bar S_{\up \alpha \dbeta} + 3i S_{\up \alpha \dbeta} + \tfrac 14 \left(  D_\alpha \bar \lambda_{\up \dbeta} - \bar D_\dbeta \lambda_{\up \alpha}  \right)  \right] \\ 
	    \mathcal R_{c [\ul{nm}]_{14}} &= 2 \pa_c L_{[\ul{nm}]_{14}} + \pa_{[\un} \left( 5i \bar S + 3i S  \right)_{\um]_{14} c} + (\bar \sigma_c)^{\dgamma \gamma} D_\gamma \left[ \tfrac 12  T_{[\ul{nm}]_{14} , \dgamma} +  \tfrac 1{24} \bar D^{\dot \rho } \bm R_{\dot \rho \dgamma \ul{nm}}   \right]   \nonumber \\
	    & ~~  + (\bar \sigma_c)^{\dgamma \gamma} \left[ \tfrac 18 D^\rho \bar D_\dgamma \bm R_{\rho \gamma \ul{nm}}  - \tfrac 18 \pa_{[\un} \left(  D_{|\gamma|} \bar \lambda_{\um ]_{14} \dgamma} - \bar D_{| \dgamma | } \lambda_{\um ]_{14}  \gamma}  \right)   \right]
	\end{align}
\end{subequations}
The remaining part of $\bm R_{\un c, \ul{pq}}$, namely $R_{\un c, [\ul{pq}]_{14}}$, is in $\bm 7 \times \bm{14} = \bm 7 + \bm{14} + \bm{64}$ of $G_2$. A separate dimension 2 Bianchi identity determines the component $\bm R_{\ul{pq}, \un c}$ fully, from which, using pair exchange symmetry, one can determine $R_{\un c, [\ul{pq}]_{14}}$, and an additional derivative relation\footnote{since $R_{\un c, [\ul{pq}]_7}$ gets determined in two different ways from the Bianchi identities, they must be set equal.}. We give the expression for $R_{\un c, [\ul{pq}]_{14}}$ here, and quote the relation in the next paragraph (\ref{E:dim2-reln4}e).
\begin{align}
	R_{\un c, [\ul{pq}]_{14}} &= R_{[\ul{pq}]_{14}, \un c} \nonumber \\
	&= - (\bar \sigma_c)^{\dalpha \beta} \Big[ D_\beta \bm T_{\ul{pq}, \un \dalpha} - 2i \pa_{[\up} \left( \tilde X_{|\beta \dalpha \un | \uq]} + \tfrac 17 \delta_{\uq] \un} \tilde X_{\beta \dalpha}  \right)  \nonumber \\
	& \qquad \qquad \qquad + i \pa_{[ \up} \varphi_{\uq] \un} {}^\ur \left( \tfrac i{12} D_\beta \bar \lambda_{\ur \dalpha} + \bar S_{\ur \beta \dalpha}  \right) \Big]_{[\ul{pq}]_{14}} 
\end{align}

Finally, we consider the purely internal component $\bm R_{\ul{pq}, \ul{mn}}$. Pair exchange symmetry implies that
\begin{subequations}
	\begin{align}
		\bm R_{ [\ul{mn}]_7, [\ul{pq}]_7 } &= \bm R_{[\ul{pq}]_7, [\ul{mn}]_7} \in (\bm 7 \times \bm 7)_{\textrm{symmetric}} = \bm 1 + \bm{27} \\
		\bm R_{[\ul{mn}]_7, [\ul{pq}]_{14}} &= \bm R_{[\ul{pq}]_{14}, [\ul{mn}]_7} \in \bm 7 \times \bm{14} = \bm 7 + \bm{27} + \bm{64} \\
		\bm R_{[\ul{mn}]_{14}, [\ul{pq}]_{14}} &= \bm R_{[\ul{pq}]_{14}, [\ul{mn}]_{14}} \in (\bm{14} \times \bm{14})_{\textrm{symmetric}} = \bm 1 + \bm{27} + \bm{77'}
	\end{align}
\end{subequations}
We find that everything except the $\bm{77'}$ piece of $\bm R_{\ul{mn}, \ul{pq}}$ gets determined by dimension 2 Bianchi identities. To illustrate this, we first decompose the last pair of antisymmetric indices into $\bm 7 + \bm{14}$:
\begin{align}
	\bm R_{\ul{pq}, \ul{mn}} &= \tfrac 16 \varphi_{\ul{mn}} {}^\ur \mathcal R_{\ul{pq}, \ur} + R_{\ul{pq}, [\ul{mn}]_{14}} 
\end{align}
Clearly, $\mathcal R_{\ul{np}, \um} \in (\bm 7 + \bm{14}) \times \bm 7$, so it can be further $G_2$ decomposed. The same goes for $ R_{\ul{mn}, [\ul{pq}]_{14}} \in (\bm 7 + \bm{14}) \times \bm{14}$. Even without doing these decompositions explicitly, we find that the entire $\mathcal R_{\ul{np}, \um}$ is determined by a dimension 2 Bianchi identity:
\begin{align}
		\mathcal R_{\ul{np}, \um} &= 4 \pa_{[\un} \left[  \tfrac 12 \tilde{\bar R}_{\up ] \um}  + \tfrac 17 \delta_{\up] \um} \tilde{\bar R} + \tfrac 16 \varphi_{\up] \ul{mq}} \bar R^{\uq} + L_{\up] \um_{14}}   \right]     \nonumber  \\
		& - 2D^\alpha \bigg[ \tfrac 1{42} \varphi_{\ul{npm}} + \tfrac 1{24} \psi_{\ul{npmq}} V^\uq {}_\alpha + \tfrac 34 \varphi^\uq {}_{[\ul{np}} V_{\um] \uq \alpha}     \nonumber \\
		& ~~~~~~~~~ + \varphi_{\ul{np}} {}^\uq J_{\ul{qm} \alpha} - \tfrac 12 \varphi_{\ul{pm}} {}^\uq J_{\ul{qn} \alpha} + \tfrac 12 \varphi_{\ul{nm}} {}^\uq J_{\ul{qp} \alpha}   \nonumber  \\
		& ~~~~~~~~~ + \varphi_{\ul{np}} {}^\uq I_{\ul{qm} \alpha} - \tfrac 12 \varphi_{\ul{pm}} {}^\uq I_{\ul{qn} \alpha} + \tfrac 12 \varphi_{\ul{nm}} {}^\uq I_{\ul{qp} \alpha}   \nonumber \\
		&~~~~~~~~~ + Z_{\ul{np}| \um \alpha}  \bigg]~,
\end{align}
from which the irreducible pieces in $\mathcal R_{\ul{np}, \um}$ can be extracted. The remaining piece $R_{\ul{pq}, [\um \un]_{14}}$ does not participate in any Bianchi identities in $4|4+7$ except for the algebraic one, namely $\bm R_{[\ul{mn}, \up] \uq} = 0$. Hence, it can only be (possibly partially) determined in terms of $\mathcal R_{\ul{np}, \um}$ by using the algebraic identity. Its first two (form) indices can again be split into a $\bm 7$ and a $\bm{14}$, 
\begin{align}
	R_{\ul{pq}, [\ul{mn}]_{14}} &= R_{[\ul{pq}]_7, [\ul{mn}]_{14}} + R_{[\ul{pq}]_{14}, [\ul{mn}]_{14}}
\end{align}
The first term $R_{[\ul{pq}]_7, [\ul{mn}]_{14}}$ is determined in terms of $\mathcal R_{\ul{np}, \um}$ using pair exchange:
\begin{align}
	R_{[\ul{pq}]_7, [\ul{mn}]_{14}} &= R_{[\ul{mn}]_{14}, [\ul{pq}]_7} = \tfrac 16 \varphi_{\ul{pq}} {}^\ur \mathcal R_{[\ul{mn}]_{14}, \ur}~.
\end{align}
The second term is $ R_{[\ul{pq}]_{14}, [\ul{mn}]_{14}} \in (\bm{14} \times \bm{14})_{\textrm{symmetric}} = \bm 1 + \bm{27} + \bm{77'}$. We can use the algebraic Bianchi identity to determine the $\bm 1$ and the $\bm{27}$ pieces in the following way. Projecting the identity onto $\bm 1$,
\begin{align}
	\psi^{\ul{mnpq}} \bm R_{\ul{mn}, \ul{pq}} &= 0 \qquad 
	\Rightarrow \qquad 
	R_{[\ul{mn}]_{14},} {}^{[\ul{mn}]_{14}}  = \tfrac 13 \varphi^{\ul{mnp}} \mathcal R_{\ul{mn}, \up} ~,
\end{align}
meaning that the singlet in $R_{[\ul{pq}]_{14}, [\ul{mn}]_{14}}$ is proportional to $\varphi^{\ul{mnp}} \mathcal R_{\ul{mn}, \up}$.
Next, projecting the algebraic Bianchi identity onto $(\bm 7 \times \bm 7)_{\textrm{symmetric}} = \bm 1 + \bm{27} $ (and subtracting the $\bm 1$) gives the $\bm{27}$ piece:
\begin{subequations}
	\begin{align}
		\psi_{(\ur} {}^{\ul{mnp}} \bm R_{|\ul{mn}, \up| \uq)} &= 0  \\
		\Rightarrow \left(  \bm R_{(14 \times 14)|_{27}} \right)_{\ul{qr}} &=  - \tfrac 1{21} \delta_{\ul{rq}} \varphi^{\ul{mnp}} \mathcal R_{\ul{mn}, \up} + \tfrac 16 \psi_{(\ur} {}^{\ul{mnp}} \varphi_{\uq) \up} {}^\us \mathcal R_{\ul{mn}, \us}   \nonumber \\
		& ~~~ - \tfrac 23 \varphi_{(\ur} {}^{\ul{ps}} ~~ \Pi^{\boldsymbol{14}} {}_{\uq) \up} {}^{\ul{ij}} \mathcal R_{\ul{ij}, \us}
	\end{align}
\end{subequations}
It is not possible to determine the $\boldsymbol{77'}$  piece in terms of $\mathcal R_{\ul{mn}, \up}$ using the algebraic Bianchi identity simply because $\mathcal R_{\ul{mn}, \up} \in (\bm 1 + \bm 7 + \bm{14} + \bm{27}) + (\bm 7+ \bm{27} + \bm{64})$. It is a piece of the curvature component that is completely unconstrained.

\paragraph{Relations.} 
We again state some derivative relations implied by Bianchi identities. Some of these arise from the fact that certain components of the Riemann tensor that are equal by pair exchange symmetry are determined by different Bianchi identities to be different stuff. So these different stuff must be set equal. Other relations arise from projecting Bianchi identities into their symmetric/antisymmetric pieces with respect to two dotted/undotted spinor indices. The symmetric pieces determine Riemann components (which went in the paragraph above) and the anti-symmetric pieces give rise to derivative relations between torsion components.
\begin{subequations}
	\begin{align}
		D_\delta \bar W_{\dgamma \dbeta \dalpha} &= 0 = \bar D_\ddelta W_{\gamma \beta \alpha} \\
		D^\alpha W_{\alpha \beta \gamma} &= \tfrac 12 \bar D_\dgamma D_{(\beta} G_{\gamma)} {}^\dgamma
	\end{align}
\end{subequations}
\begin{subequations}
	\begin{align}
		\bar D_\dgamma D_{(\gamma} \bar S^\um {}_{\beta) \dalpha} &= 0 \\
		2i D^2 \bar S_{\um \dgamma \dalpha} &= - D^\beta \bar D_{( \dgamma} \bar S_{|\um \beta| \dalpha)} - \tfrac 12 \bar D_{(\dgamma} D^\beta \bar S_{|\um \beta| \dalpha)} \\
		8i \pa_\um R^\dagger &= - D^2 S_\um - \left( D^\beta \bar D^\dalpha + \tfrac 12 \bar D^\dalpha D^\beta  \right) \bar S_{\um \beta \dalpha} \\
		 4i \varphi_\um {}^{\ul{np}} \partial_\un S_\up &= - \tfrac {3i}2 D^\alpha \rho_{\um \alpha} - D^2 S_\um + D^\alpha \bar D^\dbeta \bar S_{\um \alpha \dbeta} + \tfrac i{24} (D^\alpha \bar D^2 \lambda_{\um \alpha} + \bar D_\dbeta D^2 \bar \lambda_\um {}^\dbeta) \\
		D^\alpha T_{[\ul{mn}]_{14}, \alpha} &= 4i \pa_{[\um} S_{\un]_{14}} \\
		2i \varphi_\um {}^{\ul{np}} \partial_\un S_{\up \alpha \dbeta} &= - \tfrac {3i}2 \bar D_\dbeta \rho_{\um \alpha} - \bar D_\dbeta D_\alpha S_\um + \tfrac 12 \bar D^2 \bar S_{\um \alpha \dbeta} - \tfrac i{48} \bar D_\dbeta D^2 \lambda_{\um \alpha} + \tfrac i{48} \bar D^2 D_\alpha \bar \lambda_{\um \dbeta} \nonumber \\
		& \qquad + \tfrac 1{12} \bar D_\dbeta \pa_\alpha {}^\dgamma \bar \lambda_{\um \dgamma}    \\
		\bar D_\dbeta T_{[\ul{mn}]_{14}, \alpha} &= 2i \partial_{[\um} S_{\un]_{14} \alpha \dbeta}
	\end{align}
\end{subequations}

\begin{subequations}
	\begin{align}
		3i D_{(\alpha} \rho_{|\um| \beta)} + 3i \pa_{(\alpha} {}^\dgamma  S_{|\um| \beta) \dgamma} - &\, \tfrac 12 \left( 7 D_{(\alpha} \bar D^\dgamma + 3 \bar D^\dgamma D_{(\alpha}  \right) \bar S_{|\um| \beta) \dgamma} - 4 \varphi_\um {}^{\ul{np}} \pa_\un S_{\um \alpha \beta}  \nonumber  \\
		&= \tfrac i{24} D_{(\alpha} \bar D^\dgamma D_{\beta)} \bar \lambda_{\um \dgamma} - \tfrac i{12} \left( 2D_{(\alpha} \bar D^2 - 3\bar D^2 D_{(\alpha}  \right) \lambda_{|\um| \beta)} \\
		(\bar \sigma_c)^{\dalpha \beta} \left( 2 \pa_{[b} \bm T_{a] \beta, \um \dalpha} + D_\beta \bm T_{ba, \um \dalpha} \right) &=  (\sigma_{ba})^{\beta \alpha} \bm R_{\um c, \beta \alpha} + (\bar \sigma_{ba})^{\dbeta \dalpha} \bm R_{\um c, \dbeta \dalpha} \nonumber \\
		\Rightarrow \pa_{(\alpha} {}^\dbeta \Big[ \epsilon_{\beta) \gamma} \left( i \epsilon_{\dbeta \dgamma} S_\um + 3 \bar S_{\um \dbeta \dgamma} \right) + & \, S_{|\um| \beta) \gamma} \epsilon_{\dbeta \dgamma} \Big] - \tfrac 12 \epsilon_{\gamma (\alpha} D^2 \bar S_{|\um| \beta) \dgamma} + 2i D_\gamma \bar D_\dgamma S_{\um \alpha \beta} \nonumber \\
		&= R_{\um, \gamma \dgamma, \alpha \beta}~,  \qquad \textrm{use (\ref{E:R_{mc,ba}}c) here.}
	\end{align}
\end{subequations}

\begin{subequations}
\label{E:dim2-reln4}
	\begin{align}
		0 &= i \pa^\un \bar S_{\un \alpha \dbeta} - \tfrac 38 D^2 X_{\alpha \dbeta} + \tfrac i{24} D_\alpha \bar D_\dbeta \left( \tilde{\bar R} - \tilde R  \right)  \\
		0 &= i \pa_{ (\un} \bar S_{\um)_{\textrm{traceless}} \alpha \dbeta} - \tfrac 16 D^2 \tilde X_{\alpha \dbeta \ul{nm}} + \tfrac i{24} D_\alpha \bar D_\dbeta \left(  \tilde{\bar R}_{\ul{nm}} - \tilde R_{\ul{nm}}  \right)  \\
		0 &= \varphi_\um {}^{\ul{np}} \pa_\un \bar S_{\up \alpha \dbeta} + \tfrac 38 D_\alpha \bar \rho_{\um \dbeta} - \tfrac i2 D_\alpha \bar D_\dbeta S_\um - \tfrac i8 D^2 S_{\um \alpha \dbeta} \nonumber \\
		&~~~~ + \tfrac 1{48} \left[ D^2 \bar D_\dbeta \lambda_{\um \alpha} - D_\alpha \bar D_\dbeta D^\beta \lambda_{\um 
		\beta} - \tfrac 12 D_\alpha \bar D^2 \bar \lambda_{\um \dbeta}  \right]  \\
		0 &= 3i \pa_{[\un} \bar S_{\um]_{14} \alpha \dbeta}  + \tfrac 34 D_\alpha T_{[\ul{nm}]_{14}, \dbeta} \nonumber  \\
		& \qquad - \tfrac 1{16} \left[ 3 D^\gamma \bar D_\dbeta + 2 \bar D_\dbeta D^\gamma \right] \bm R_{\alpha \gamma \ul{nm}} + \tfrac 1{16} D_\alpha \bar D^\dgamma \bm R_{\dgamma \dbeta \ul{nm}} \\
		 \mathcal R_{\um, c \un} &=  - (\bar \sigma_c)^{\dalpha \beta} \varphi_\un {}^{\ul{pq}} \Big[ D_\beta \bm T_{\ul{pq}, \um \dalpha} - 2i \pa_\up \left( \tilde X_{\beta \dalpha \ul{mq}} + \tfrac 17 \delta_{\ul{mq}} \tilde X_{\beta \dalpha}  \right)  \nonumber \\
	& \qquad \qquad \qquad  ~~~ + i  \varphi_{\ul{qm}} {}^\ur \pa_\up \left( \tfrac i{12} D_\beta \bar \lambda_{\ur \dalpha} + \bar S_{\ur \beta \dalpha}  \right) \Big] \\
		 2 \bm R_{\um [a, b] \un} &= (\sigma_{ab})^{\beta \alpha} \bm R_{\ul{mn}, \beta \alpha} + (\bar \sigma_{ab})^{\dbeta \dalpha} \bm R_{\ul{mn}, \dbeta \dalpha} 
	\end{align}
\end{subequations}

\begin{align}
	\left[2 \partial_{[\un} \bm T_{\up ] \beta, \um \alpha}  + D_\beta \bm T_{\ul{np}, \um \alpha}  \right]_{(\alpha \beta)} &= 0
\end{align}


\subsubsection{Remaining Bianchi identities}
We mention for completeness that there are dimension 2 Bianchi identities satisfied by the 4-form field strength:
\begin{align}
	\pa_{[\he} \bm G_{\hd \hc \hb \ha]} &= 0~.
\end{align} 
There are also dimension $\tfrac 52$ torsion Bianchi identities, only containing bosonic curls like the one above:
\begin{subequations}
	\begin{align}
		\pa_{[\um} \bm T_{cb]} {}^{\ul \alpha} &= \pa_{[\un} \bm T_{\um b]} {}^{\ul \alpha} = \pa_{[\un} \bm T_{\ul{mp}]} {}^{\ul \alpha} = 0 \\
		\pa_{[d} \bm T_{cb]} {}^{\ul{m \alpha}} & = \pa_{[\un} \bm T_{cb]} {}^{\ul{m \alpha}} = \pa_{[ \un} \bm T_{\up b]} {}^{\ul{m \alpha}} = \pa_{[\un} \bm T_{\ul{pq}]} {}^{\ul{m \alpha}} = 0
	\end{align}
\end{subequations}

\acknowledgments
We thank William Linch for discussions, and Artem Bolshov and Nathan Brady for collaboration at an early stage of this work. This work is partially supported by the NSF under grants NSF-2112859, and the Mitchell Institute for Fundamental Physics and Astronomy at Texas A\&M University.

\appendix


\section{$\Gamma$ matrices in 4, 7, and 11 dimensions}
\label{A:Gamma-matrices}

The defining postulates for matrices $B$ and $C$ which relate a representation of the Clifford algebra with its complex conjugate, and transpose representations respectively, are
\begin{align}
	\label{E:define_eta_epsilon_for_gammas}
	\Gamma_{\mu}^{*} = - \eta (-1)^t B \Gamma_{\mu} B^{-1}~, \qquad
	\Gamma_{\mu}^{T} = - \eta C \Gamma_{\mu} C^{-1}
\end{align}
where $t$ is the number of time-like directions. All $\Gamma$'s are chosen to be unitary. The $B$ and $C$ matrices can be related as $C = B^T A$, where $A$ is the product of all time-like $\Gamma$ matrices. This implies
\begin{subequations}
	\label{E:B_C_properties}
	\begin{align}
	B^{T} = \epsilon \eta^t (-1)^{\tfrac{t(t-1)}{2}} B~, \qquad
	C^{T} = -\epsilon C~, \qquad
	\epsilon = - \sqrt 2 \cos(\tfrac{\pi}{4}(1+\eta D))
	\end{align}
\end{subequations}
In even dimensions, both $\eta=1$ and $\eta = -1$ are allowed. In odd dimensions, the first $(D-1)$ $\Gamma$'s are borrowed from the preceding even dimension, and the last one is obtained by choosing one from two possibilities: $\Gamma_D = \pm i^{t+\tfrac{D(D-1)}{2}}\Gamma_1  \Gamma_2 \ldots \Gamma_{D-1}$. This last $\Gamma$ satisfies the same complex conjugation rule as the first $D-1$ $\Gamma$'s only for one of the two values of $\eta$ in $(D-1)$ dimensions. The rule is
$-\eta = (-1)^{\tfrac{D(D-1)}{2}}$. Further details may be found e.g. in \cite{VanProeyen:1999ni}.

For example, in $1+10$ dimensions, we must have $\eta =1$, $\epsilon =1$. Therefore, any set of 11D gamma matrices, which for later convenience we denote $\hat \Gamma_{\ha}$, must satisfy
\begin{gather}
	\label{E:11d_gammas}
	\hat{\Gamma}_{\ha}^* = \hB \hat{\Gamma}_{\ha} \hB^{-1}~, \qquad
	\hat{\Gamma}_{\hat a}^T = - \hC \hat{\Gamma}_\ha \hC^{-1} \\
	\hB^{\dagger} \hB = \boldsymbol{1} = \hB^* \hB  ~ \Rightarrow \hB^T = \hB~, \qquad
	\hC^T = - \hC ~, \qquad
	\hB = - \hC \hat{\Gamma}_0~.
\end{gather}
The index structures are $\hat C \sim \hat C^{\hat \alpha \hat \beta}$, and $\hat C^{-1} \sim \hat C_{\hat \alpha \hat \beta}$. Spinor indices are raised and lowered following the conventions
\begin{align}
	\label{E:raise/lower}
	A_\halpha = - \hC_{\halpha \hbeta} A^\hbeta ~, \qquad
	A^\halpha = - \hC^{\halpha \hbeta} A_\hbeta~.
\end{align}
In practice, we will build our 11D gamma matrices by taking tensor products of 4D and 7D gamma matrices.

\subsection{4D gamma matrices}
Our conventions are similar to \cite{Wess:1992cp}. We introduce the $\sigma^a$ matrices
\begin{subequations}
	\label{4d-Pauli}
	\begin{align}
	\sigma^0 &= \begin{bmatrix}
	-1& 0\\
	0& -1
	\end{bmatrix}, ~~
	\sigma^1 = \begin{bmatrix}
	0& 1\\
	1& 0
	\end{bmatrix}, ~~
	\sigma^2 = \begin{bmatrix}
	0& -i\\
	i& 0
	\end{bmatrix}, ~~
	\sigma^3 = \begin{bmatrix}
	1& 0\\
	0& -1
	\end{bmatrix} \\
	(\bar \sigma^a)^{\dalpha \alpha} &= \epsilon^{\dalpha \dbeta} \epsilon^{\alpha \beta} (\sigma^a)_{\beta \dbeta}
	\end{align}
\end{subequations}
and use these to build a Weyl representation for 4D $\gamma$ matrices:
\begin{align}
	\gamma^a &= \begin{bmatrix}
	\boldsymbol 0_2& i \sigma^a \\
	i \bar{\sigma}^a& \boldsymbol 0_2
	\end{bmatrix}
\end{align}
As matrices, $\gamma^0 = i \sigma^1 \otimes \sigma^0$, and $\gamma^{1,2,3} = - \sigma^2 \otimes \sigma^{1,2,3}$.
We choose $\eta = \epsilon = 1 $ with
\begin{align}
C_{4 \textrm{D}} &= -i \sigma^3 \otimes \sigma^2 
=\begin{bmatrix}
-\epsilon^{\alpha \beta}& \boldsymbol 0_2 \\
\boldsymbol 0_2 & - \epsilon_{\dalpha \dbeta}
\end{bmatrix}~, \\
B_{4 \textrm{D}} &= \sigma^2 \otimes \sigma^2 
	= \begin{bmatrix}
	\boldsymbol  0_2 & -\epsilon^{\alpha \beta} \\
	- \epsilon_{\dalpha \dbeta}& \boldsymbol  0_2
	\end{bmatrix}~.
\end{align}
We take the chiral $\gamma_5$ matrix to be
\begin{align}
	\gamma_5 &:= i \gamma^0 \gamma^1 \gamma^2 \gamma^3 = - \sigma^3 \otimes \sigma^0 = \begin{bmatrix}
	\delta_\alpha {}^\beta & \boldsymbol  0_2\\
	\boldsymbol  0_2 & - \delta^\dalpha  {}_\dbeta
	\end{bmatrix}
\end{align}

\subsection{7D gamma matrices}
Euclidean $SO(7)$ $\Gamma$ matrices obey $\{\Gamma^{\ul a}, \Gamma^{\ul b}\} = 2\,\delta^{\ul a \ul b} \boldsymbol{1}_8 $. Let $\Gamma^{\ul 1}, \ldots, \Gamma^{\ul 6}$ supply the (unique up to similarity transformations) unitary irrep of dimension $8$ of $SO(6)$. We choose
$\Gamma^{\ul 7} = i\Gamma^{\ul 1}\Gamma^{\ul 2}\ldots\Gamma^{\ul 6}$. The dimension being odd, we have only one option for $\eta$ and $\epsilon$, in this case $\eta = 1$ and $\epsilon = -1$. Unitarity of the gamma matrices and Euclidean signature implies $\Gamma^{\ul a}$ are Hermitian. It also follows that
\begin{align}
	\Gamma^{[\ul a_1} \ldots \Gamma^{\ul a_7]} =: \Gamma^{\ul{a_1 \ldots a_7}} = -i \epsilon^{\ul{a_1 \ldots a_7}} ~, \qquad
	\epsilon^{\ul{12 \ldots 7}} = 1~.
\end{align}
The $C$ and $B$ matrices obey $C_{7 \textrm{D}}^T = C_{7 \textrm{D}}$,
$B_{7 \textrm{D}}^T = B_{7 \textrm{D}}$, and since all $\Gamma$'s are spacelike, $C_{7 \textrm{D}} = B_{7 \textrm{D}}$. A Majorana basis can be chosen in which $B_{7\textrm{D}} = C_{7 \textrm{D}}$ is the identity matrix. The proof goes as follows. 
Under a unitary change of basis, $\Gamma^{' \ul a} = U^{-1} \Gamma^{\ul a} U$, $C_{7 \textrm{D}}$ transforms as $C_{7 \textrm{D}}^{'} = U^T C_{7 \textrm{D}} U$. We invoke the Autonne-Takagi factorization theorem which states that, since $C_{7 \textrm{D}}$ is a complex symmetric matrix, then there exists a unitary matrix $U$ such that $C'_{7 \textrm{D}}$ is a real diagonal matrix with non-negative entries. Being unitary, the eigenvalues of $C'_{7 \textrm{D}}$ must be pure phases. These two facts mean we can choose $C'_{7 \textrm{D}} = B'_{7 \textrm{D}} = \boldsymbol{1}_8$ (and henceforth dropping the primes). This means that
$\Gamma^{\ul a}$ are antisymmetric and purely imaginary,
	\begin{align}
	\label{E:7d_B-C-properties}
	(\Gamma^{\ul a})^T &= - \Gamma^{\ul a}, ~~ (\Gamma^{\ul a})^* = - \Gamma^{\ul a} \eol
	\Rightarrow 
	(\Gamma^{\ul a})_I {}^J &= - (\Gamma^{\ul a})_J {}^I = (\Gamma^{\ul a})_{IJ} = - (\Gamma^{\ul a})_{JI} = -(\Gamma^{\ul a})^*_{IJ} ~~\textrm{etc.}
	\end{align}

\subsection{Explicit 11D gamma matrices}
We choose the following 11D $\Gamma$ matrices:
\begin{subequations}
	\label{E:11d_gammas_as_4dx7d}
\begin{align}
\hat \Gamma^a := \gamma^a \otimes \boldsymbol{1}_8 \implies
    (\hat \Gamma^a)_\halpha {}^\hbeta &= \begin{bmatrix}
	\boldsymbol 0_{16}& i (\sigma^a)_{\alpha \dbeta} \delta_{IJ}\\
	i (\bar \sigma^a)^{\dalpha \beta} \delta^{IJ}& \boldsymbol 0_{16}
	\end{bmatrix} \\
\hat \Gamma^{\ul a} := - \gamma_5 \otimes \Gamma^{\ul a} \implies
	(\hat \Gamma^{\ul a})_\halpha {}^\hbeta &=  \begin{bmatrix}
	- \delta_\alpha {}^\beta (\Gamma^{\ul a})_I {}^{J}& \boldsymbol 0_{16} \\
	\boldsymbol 0_{16}& \delta^\dalpha {}_\dbeta (\Gamma^{\ul a})^I {}_{J}
	\end{bmatrix}
	\end{align}
\end{subequations}
The charge conjugation matrix is
\begin{subequations}
	\label{E:11d_C-as-C4_x_C7}
	\begin{align}
	\hC &:= C_{4 \textrm{D}} \otimes C_{7 \textrm{D}}~~ \Rightarrow ~~\hat C^{\halpha \hbeta} = \begin{bmatrix}
	- \epsilon^{\alpha \beta} \delta^{IJ}& \boldsymbol 0_{16} \\
	\boldsymbol 0_{16}& - \epsilon_{\dalpha \dbeta} \delta_{IJ}
	\end{bmatrix} \\
	\hC^{-1} &= - \hC ~~ \Rightarrow ~~ \hC_{\halpha \hbeta} = \begin{bmatrix}
	- \epsilon_{\alpha \beta} \delta_{IJ}& \boldsymbol 0_{16} \\
	\boldsymbol 0_{16}& - \epsilon^{\dalpha \dbeta} \delta^{IJ}
	\end{bmatrix}
	\end{align}
\end{subequations}
The matrices $\hC \hat \Gamma$, for $\hat \Gamma$ of any rank, have both spinor indices upstairs, and $\Gamma\hC^{-1}$ have both indices downstairs, and have the following symmetry properties:
\begin{align}
\begin{array}{rl}
	\textrm{Symmetric ranks:} &  1,2,5,6,9,10 \\
	\textrm{Antisymmetric ranks:}& 3,4,7,8
\end{array}
\end{align}
Other useful results include
\begin{subequations}
 	\label{E:Rank2_Gammas}
 	\begin{align}
 		\hat \Gamma^{ab} &= \gamma^{ab} \otimes \boldsymbol 1_8 = -2 \begin{bmatrix}
 			\sigma^{ab}& \boldsymbol 0_2\\
 			\boldsymbol 0_2& \bar{\sigma}^{ab}
 		\end{bmatrix} \otimes \boldsymbol 1_8 \\
 	\hat \Gamma^{a \ul b} &= \tfrac 12 [\gamma_5, \gamma^a] \otimes \Gamma^{\ul b} = \begin{bmatrix}
 		\boldsymbol 0_2 & i \sigma^a \\
 		-i \bar{\sigma}^a & \boldsymbol 0_2
 	\end{bmatrix} \otimes \Gamma^{\underline b} \\
 	\hat \Gamma^{\ul{ab}} &= \boldsymbol 1_4 \otimes \Gamma^{\ul{ab}} = \begin{bmatrix}
 		\delta_\alpha {}^\beta & \boldsymbol 0_2 \\
 		\boldsymbol 0_2 & \delta^\dalpha {}_\dbeta
 	\end{bmatrix} \otimes \Gamma^{\underline{ab}}
 	\end{align}
 \end{subequations}
where $\sigma^{ab} = \tfrac 14 (\sigma^a \bar \sigma^b - \sigma^b \bar \sigma^a)$
and $\bar \sigma^{ab} = \tfrac 14 (\bar \sigma^a \sigma^b - \bar \sigma^b \sigma^a)$.


\section{Engineering dimensions}
\label{AppendixB}
We define the mass (engineering) dimensions of various components of connections and curvatures. Superspace coordinates, and coordinate differentials have
\begin{subequations}
	\begin{align}
		[x^{\hat m}] &= [d \hat x^{\hat m}] = -1, ~~~~
		[\theta^{\hat \mu}] = [d \theta^{\hat \mu}] = - \tfrac 12
	\end{align}
\end{subequations}
The frame basis has the same mass dimension as the coordinate basis:
\begin{subequations}
	\begin{align}
[\hat E^\ha] &= -1 \quad\implies \quad
[\hat E_{\hat m} {}^\ha] = 0~, \quad
[\hat E_{\hat \mu} {}^\ha] = - \tfrac 12 \\
[\hat E^\halpha] &= - \tfrac 12 \quad \implies\quad
    [\hat E_{\hat m} {}^\halpha] = \tfrac 12, \quad [\hat E_{\hat \mu} {}^\halpha] = 0
\end{align}
\end{subequations}
Exterior differentiation doesn't change mass dimension. The mass dimensions of an arbitrary $(p,q)$ super-tensor, and those of its components in any basis $\{\hat e_{\hat M_i}\}, \{\hat e^{\hat N_j}\}$,
\begin{equation}
	\hat \Sigma = \hat e_{\hat M_1} \otimes \ldots \otimes \hat e_{\hat M_p} \otimes \hat e^{\hat N_1} \otimes \ldots \otimes \hat e^{\hat N_q} ~ \hat \Sigma^{\hat M_1 \ldots \hat M_p} {}_{\hat N_1 \ldots \hat N_q}
\end{equation}
are therefore related as
\begin{align}
		[\hat \Sigma] &= [\hat \Sigma^{\hat M_1 \ldots \hat M_p} {}_{\hat N_1 \ldots \hat N_q}] - n_v - \tfrac 12 n_s + m_v + \tfrac 12 m_s \eol
		n_v &= \textrm{number of vector } \hat N\textrm{'s}, ~~ n_s = \textrm{number of spinor } \hat N\textrm{'s} \eol
		m_v &= \textrm{number of vector } \hat M\textrm{'s}, ~~ m_s = \textrm{number of spinor } \hat M\textrm{'s}
\end{align}
The engineering dimension of the spin connection, torsion, and Riemann tensors are
\begin{subequations}
\begin{alignat}{5}
[\Omega_\hb{}^\ha] &= 0 &~~ &\implies &\quad
[\hat \Omega_{\hc \hb} {}^\ha] &= 1, &~~ 
[\hat \Omega_{\hgamma \hb} {}^\ha] &= \tfrac 12~, \\
[\hat T^\ha] &= -1 &~~ &\implies &\quad
		[\hat T_{\hb \hc} {}^\ha] &= 1, &~~~~
		[\hat T_{\hb \hgamma} {}^\ha]  &= \tfrac 12,&~~~~
		[\hat T_{\hbeta \hgamma} {}^\ha]  &= 0~, \\
[\hat T^\halpha] &= - \tfrac 12 &\quad &\implies &\quad
		[\hat T_{\hb \hc} {}^\halpha] &= \tfrac 32, &~~~~
		[\hat T_{\hb \hgamma} {}^\halpha]  &= 1, &~~~~
		[\hat T_{\hbeta \hgamma} {}^\halpha]  &= \tfrac 12~, \\
[\hat R_\hb{}^\ha] &= 0 &\quad &\implies &\quad
    [\hat R_{\hd \hc \hb} {}^\ha] &= 2, &~~ 
		[\hat R_{ \hd \hgamma \hb} {}^\ha] &= \tfrac 32, &~~ 
		[\hat R_{ \hdelta \hgamma \hb} {}^\ha] &= 1~.
\end{alignat}
\end{subequations}
From the component action of 11D supergravity, the mass dimension of 3-form must be
\begin{alignat}{7}
[\hC] &= -3 &\quad &\implies &\quad
    [\hC_{\ha \hb \hc}] &= 0, &~~ 
    [\hC_{\ha \hb \hgamma}] &= - \tfrac 12, &~~ 
    [\hC_{\ha \hbeta \hgamma}] &= -1, &~~ 
    [\hC_{\halpha \hbeta \hgamma}] &= - \tfrac 32 \\\nonumber
[\hat G] &= -3 &\quad &\implies &\quad
    [\hat G_{\ha \hb \hc \hd}] &= 1, &~~ 
    [\hat G_{\ha \hb \hc \hdelta}] &= \tfrac 12, &~~ 
    [\hat G_{\ha \hb \hgamma \hdelta}] &= 0, &~~ 
    [\hat G_{\ha \hbeta \hgamma \hdelta}] &= - \tfrac 12, &~~ 
    [\hat G_{\halpha \hbeta \hgamma \hdelta}] &= -1~.
\end{alignat}
The prepotential superfields necessarily have dimensions
\begin{align}
	\label{E:prepot-dim}
	[X] &= -1, ~~~~ [\Sigma_{\alpha \um}] = - \tfrac 12, ~~~~ [V_{\ul{mn}}] = -1, ~~~~ [\Phi_{\ul{mnp}}] = 0, ~~~ [\cV^\um] = -1,\nonumber \\
	[H_{\alpha \dalpha}] &= -1, ~~~~ [\Psi_{\um \alpha}] = - \tfrac 12
\end{align}


\section{Decomposing 11D spinor indices}
\label{A:AppendixC}

When restricting the $11|32$ superspace to $4|4+7$, the 32-component spinor index $\halpha$ can be decomposed by expanding a generic 11D: spinor $\Psi$ as
\begin{equation}
		\psi \otimes \eta, ~~\psi_\um \otimes (i \Gamma^\um \eta)
\end{equation}
where $\psi$ is a spinor of $SO(3,1)$ and $\eta$ is a real commuting spinor of $SO(7)$,
which we normalize as $\eta^T \eta = 1$. The spinors $\eta$ and $i \Gamma^\um \eta$ are linearly independent and provide a basis for $8$-component real spinors of $SO(7)$. The spinor $\eta$ is the spinor associated to the $G_2$-structure,
\begin{equation}
\label{E:eta<=>varphi}
	\varphi_{\ul{mnp}} = i \eta^T \Gamma_{\ul{mnp}} \eta~.
\end{equation}
Therefore, $\eta$ and $i \Gamma^\um \eta$ are singlets under $G_2$ and parametrize the decomposition of a generic $SO(7)$ spinor $\boldsymbol{8}_{SO(7)} = \boldsymbol{1}_{G_2} \oplus \boldsymbol{7}_{G_2}$. An 11D spinor $\Psi$ can then be explicitly decomposed as
\begin{align}
\Psi_\halpha = \begin{bmatrix}
			\Psi_{\alpha I} \\
			\Psi^{\dalpha I}
\end{bmatrix}~, \qquad
	\Psi_{\alpha I} = \eta_I \Psi_\alpha + i (\Gamma^\um \eta)_I \Psi_{\um \alpha}~, \quad
	\Psi^{\dalpha I} = \eta^I \Psi^\dalpha + i (\Gamma_\um \eta)^I \Psi^{\um \dalpha}~.
\end{align}
Such a decomposition makes it transparent how our 11D gamma matrices act.

At this stage, let us record a few useful results for the $G_2$ spinors:
\begin{subequations}
	\begin{align}
		\eta^T \Gamma^\um \eta &= 0 \\
		\eta^T \Gamma^\um \Gamma^\un \eta &= \delta^{\ul{mn}} \\
		\eta^T \Gamma^\um \Gamma^\un \Gamma^\up \eta &= \eta^T \Gamma^{\ul{mnp}} \eta = -i \varphi^{\ul{mnp}} \\
		\eta^T \Gamma^{\ul{mnpq}} \eta &=  \psi^{\ul{mnpq}} = \tfrac 1{3!} \epsilon^{\ul{mnpqrst}} \varphi_{\ul{rst}} = (\star \varphi)^{\ul{mnpq}} \\
		\eta^T \Gamma^\um \Gamma^\un \Gamma^\up \Gamma^\uq \eta &= \psi^{\ul{mnpq}} + \delta^{\ul{mn}} \delta^{\ul{pq}} - \delta^{\ul{mp}} \delta^{\ul{nq}} +\delta^{\ul{mq}} \delta^{\ul{np}}
	\end{align}
\end{subequations}
In order to extract $\Psi_\alpha$, $\Psi_{\um \alpha}$ etc. from $\Psi_{\alpha I}$, one can use the projection relations:
\begin{subequations}
	\begin{alignat}{2}
		\Psi_\alpha &= \eta^I \Psi_{\alpha I}~, &\quad
		\Psi_{\um \alpha} & = i (\Gamma_\um \eta)^I \Psi_{\alpha I}~, \\
		\Psi^\dalpha &= \eta_I \Psi^{\dalpha I} ~, &\quad
		\Psi^{\um \dalpha} & = i (\Gamma^\um \eta)_I \Psi^{\dalpha I}~.
	\end{alignat}
\end{subequations}
Contractions of 11D spinors decompose in the following way:
\begin{equation}
	A^\halpha B_\halpha := - A^\halpha \hC_{\halpha \hbeta} B^\hbeta = A^\alpha B_\alpha + A_\dalpha B^\dalpha + A^{\um \alpha} B_{\um \alpha} + A_{\um \dalpha} B^{\um \dalpha}~.
\end{equation}


\section{Background torsion and curvature tensors}
In a Minkowski background, the only non-zero components of the torsion are
$\mathring T_{\hgamma \hbeta}{}^\ha = 2 (\hat \Gamma^\ha)_{\hgamma \hbeta}$,
which decompose as
\begin{alignat}{2}
	\mathring T_{\alpha \dbeta} {}^a &= 2i \,(\sigma^a)_{\alpha \dbeta}~, &\qquad
	\mathring T_{\um \alpha, \un \dbeta} {}^a &= 2i \,\delta_{\ul{mn}} (\sigma^a)_{\alpha \dbeta}~, \\
    \mathring T_{\alpha , \un \beta} {}^{\um} &= 2i\, \delta^\um_\un \epsilon_{\alpha \beta}, &\qquad
    \mathring T^{\dalpha , \un \dbeta , \um} &= -2i \,\delta^{\ul{mn}} \epsilon^{\dalpha \dbeta}\\
    \mathring T_{\un \beta , \up \gamma} {}^\um &= 2i\, \varphi^\um {}_{\ul{np}} \epsilon_{\beta \gamma}~, &\qquad
		\mathring T^{\un \dbeta , \up \dgamma , \um} &= -2i \,\varphi^{\ul{mnp}} \epsilon^{\dbeta \dgamma}
\end{alignat}
In particular, all components of $\mathring T^\halpha$ vanish. All components of the Riemann tensor vanish. The only non-zero components of the four-form flux are
$\mathring G_{\ha \hb \hgamma \hdelta} = 2 (\hat \Gamma_{\ha \hb})_{\hgamma \hdelta}$,
which decompose as
\begin{subequations}
\begin{alignat}{2}
\mathring G_{ab \gamma } {}^\delta &= -4\, (\sigma_{ab})_{\gamma} {}^\delta~, &\quad ~~~
\mathring G_{ab} {}^\dgamma {}_\ddelta &= -4\, (\bar \sigma_{ab})^\dgamma  {}_\ddelta \\
\mathring G_{ab, \un \gamma} {}^{\um \delta} &= -4\, \delta_\un^\um (\sigma_{ab})_{\gamma} {}^{\delta}~, &\qquad
\mathring G_{ab} {}^{\un \dgamma} {}_{\um \ddelta} &= -4\, \delta^\un_\um (\bar \sigma_{ab})^\dgamma  {}_\ddelta~,\\[3ex]
\mathring G_{a \um , \gamma , \un \ddelta} &= -2\, (\sigma_a)_{\gamma \ddelta} \delta_{\ul{mn}}~, &\qquad
\mathring G_{a \um , \un \gamma , \ddelta} &= 2\, (\sigma_a)_{\gamma \ddelta} \delta_{\ul{mn}} \\
\mathring G_{a \um} {}^{\dgamma , \un \delta} &= 2\, (\bar \sigma_a)^{\dgamma \delta} \delta^\un_\um~, &\qquad
\mathring G_{a \um} {}^{\un \dgamma , \delta} &= -2\, (\bar \sigma_a)^{\dgamma \delta} \delta^\un_\um \\
\mathring G_{a \um, \un \gamma , \up \ddelta} &= - 2\, (\sigma_a)_{\gamma \ddelta} \varphi_{\ul{mnp}}~, &\qquad 
\mathring G_{a \um} {}^{\un \dgamma , \up \delta} &= 2\, (\bar \sigma_a)^{\dgamma \delta} \varphi_\um {}^{\ul{np}}~, \\[3ex]
\mathring G_{\ul{mn}, \gamma} {}^{\up \delta}	&= 2\, \delta_{\gamma} {}^{\delta} \varphi_{\ul{mn}} {}^\up~, &\qquad 
\mathring G_{\ul{mn}, \up \gamma} {}^\delta &= -2\, \delta_\gamma {}^\delta \varphi_{\ul{mnp}} \\
\mathring G_{\ul{mn}} {}^\dgamma {}_{\up \ddelta} &= 2\, \delta^\dgamma {}_\ddelta \varphi_{\ul{mnp}}~,  &\qquad 
\mathring G_{\ul{mn}} {}^{\up \dgamma} {}_\ddelta &= -2\, \delta^\dgamma {}_\ddelta \varphi_{\ul{mn}} {}^\up \\
\mathring G_{\ul{mn} , \up \gamma} {}^{\uq \delta } &= 2\, \delta_\gamma {}^\delta \left[ \psi_{\ul{mnp}} {}^\uq + 2\, \delta_{\up [\um} \delta_{\un]}^{\uq} \right]~, &\quad 
\mathring G_{\underline{mn}} {}^{\up \dgamma} {}_{\uq \ddelta} &= 2\, \delta^\dgamma  {}_\ddelta \left[  \psi_{\ul{mn}} {}^\up {}_\uq + 2\, \delta^\up_{[\um} \delta_{\un ] \uq}   \right]
\end{alignat}
\end{subequations}



\begin{thebibliography}{10}

\bibitem{Berkovits:2000fe}
N.~Berkovits,
``Super-Poincar\'{e} covariant quantization of the superstring,''
JHEP \textbf{04}, 018 (2000)
[arXiv:hep-th/0001035].


\bibitem{Cederwall:2010tn}
M.~Cederwall,
``D=11 supergravity with manifest supersymmetry,''
Mod. Phys. Lett. A \textbf{25}, 3201-3212 (2010)
[arXiv:1001.0112 [hep-th]].


\bibitem{Berkovits:2018gbq}
N.~Berkovits and M.~Guillen,
``Equations of motion from Cederwall\textquoteright{}s pure spinor superspace actions,''
JHEP \textbf{08}, 033 (2018)
[arXiv:1804.06979 [hep-th]].


\bibitem{Cederwall:2004cg}
M.~Cederwall, U.~Gran, B.~E.~W.~Nilsson and D.~Tsimpis,
``Supersymmetric corrections to eleven-dimensional supergravity,''
JHEP \textbf{05}, 052 (2005)
[arXiv:hep-th/0409107.


\bibitem{Hyakutake:2006aq}
Y.~Hyakutake and S.~Ogushi,
``Higher derivative corrections to eleven dimensional supergravity via local supersymmetry,''
JHEP \textbf{02}, 068 (2006)
[arXiv:hep-th/0601092].

\bibitem{Marcus:1983wb}
N.~Marcus, A.~Sagnotti and W.~Siegel,
``Ten-dimensional Supersymmetric {Yang-Mills} Theory in Terms of Four-dimensional Superfields,''
Nucl. Phys. B \textbf{224}, 159 (1983).




\bibitem{Becker:2016xgv}
K.~Becker, M.~Becker, W.~D.~Linch and D.~Robbins,
``Abelian tensor hierarchy in 4D, N = 1 superspace,''
JHEP \textbf{03}, 052 (2016)
[arXiv:1601.03066 [hep-th]].


\bibitem{Becker:2016rku}
K.~Becker, M.~Becker, W.~D.~Linch and D.~Robbins,
``Chern-Simons actions and their gaugings in 4D, $N =$ 1 superspace,''
JHEP \textbf{06}, 097 (2016)
[arXiv:1603.07362 [hep-th]].


\bibitem{Becker:2016edk}
K.~Becker, M.~Becker, S.~Guha, W.~D.~Linch and D.~Robbins,
``M-theory potential from the $G_{2}$ Hitchin functional in superspace,''
JHEP \textbf{12}, 085 (2016)
[arXiv:1611.03098 [hep-th]].

\bibitem{Becker:2017zwe}
K.~Becker, M.~Becker, D.~Butter, S.~Guha, W.~D.~Linch and D.~Robbins,
``Eleven-dimensional supergravity in 4D, $N = 1$ superspace,''
JHEP \textbf{11}, 199 (2017)
[arXiv:1709.07024 [hep-th]].


\bibitem{Becker:2018phr}
K.~Becker, M.~Becker, D.~Butter and W.~D.~Linch,
``$N=1$ supercurrents of eleven-dimensional supergravity,''
JHEP \textbf{05}, 128 (2018)
[arXiv:1803.00050 [hep-th]].

\bibitem{Becker:2020hym}
K.~Becker and D.~Butter,
``4D $N=1$ Kaluza-Klein superspace,''
JHEP \textbf{09}, 091 (2020)
[arXiv:2003.01790 [hep-th]].



\bibitem{Becker:2021oiz}
K.~Becker, D.~Butter, W.~D.~Linch and A.~Sengupta,
``Components of eleven-dimensional supergravity with four off-shell supersymmetries,''
JHEP \textbf{07}, 032 (2021)
[arXiv:2101.11671 [hep-th]].


\bibitem{Brink:1980az}
L.~Brink and P.~S.~Howe,
``Eleven-Dimensional Supergravity on the Mass-Shell in Superspace,''
Phys. Lett. B \textbf{91}, 384-386 (1980).


\bibitem{Cremmer:1980ru}
E.~Cremmer and S.~Ferrara,
``Formulation of Eleven-Dimensional Supergravity in Superspace,''
Phys. Lett. B \textbf{91}, 61-66 (1980).

\bibitem{Wess:1992cp}
J.~Wess and J.~Bagger,
{\it Supersymmetry and supergravity},
Princeton University Press, Princeton, 1992.

\bibitem{Gates:1979gv}
S.~J.~Gates, Jr. and W.~Siegel,
``(3/2, 1) Superfield of O(2) Supergravity,''
Nucl. Phys. B \textbf{164}, 484-494 (1980).

\bibitem{Binetruy:2000zx}
P.~Binetruy, G.~Girardi and R.~Grimm,
``Supergravity couplings: A Geometric formulation,''
Phys. Rept. \textbf{343}, 255-462 (2001)
[arXiv:hep-th/0005225].




\bibitem{VanProeyen:1999ni}
A.~Van Proeyen,
``Tools for supersymmetry,''
Ann. U. Craiova Phys. \textbf{9}, no.I, 1-48 (1999)
[arXiv:hep-th/9910030].



%
%

\end{thebibliography}
\end{document}